\documentclass{jfm}

\usepackage{amsmath}
\usepackage{color}
\usepackage{amssymb}
\usepackage{graphicx}
\usepackage{upgreek}

\newcommand{\od}[2]{\frac{d #1}{d #2}}
\newcommand{\ods}[2]{\frac{d^2 #1}{d #2^2}}
\newcommand{\pd}[2]{\frac{\partial #1}{\partial #2}}

\newcommand{\ie}{\textit{i.e.}}
\newcommand{\bs}{\boldsymbol}

\graphicspath{ {./} }

\usepackage{dcolumn}% Align table columns on decimal point
\usepackage{bm}% bold math
\usepackage{amsmath}
\usepackage{amssymb}
\usepackage{scalerel}[2014/03/10]
\usepackage{stackengine}

\usepackage[euler]{textgreek}

\usepackage{multirow}

\title[ ]{Detailed numerical investigation of the drop aerobreakup {in the bag breakup regime}}

\author{
 Y. Ling\aff{1,2}  \corresp{\email{stanley\_ling@sc.edu}}  \and T. Mahmood\aff{2}
}

\affiliation{
\aff{1}Department of Mechanical Engineering, University of South Carolina, Columbia, SC 29208, USA
\aff{2}Department of Mechanical Engineering, Baylor University, Waco, TX 76798, USA
}

\begin{document}

\maketitle

\begin{abstract}
Aerobreakup of drops is a fundamental two-phase flow problem that is essential to many spray applications. A parametric numerical study was performed by varying the gas stream velocity, focusing on the regime of moderate Weber numbers, in which the drop deforms to a forward bag. When the bag is unstable, it inflates and disintegrates into small droplets. Detailed numerical simulations were conducted using the volume-of-fluid method on an adaptive octree mesh to investigate the aerobreakup dynamics. Grid-refinement studies show that converged 3D simulation results for drop deformation and bag formation are achieved by the refinement level equivalent to 512 cells across the initial drop diameter. To resolve the thin liquid sheet when the bag inflates, the mesh is further refined to 2048 cells across the initial drop diameter. The simulation results for the drop length and radius were validated against previous experiments, and a good agreement was achieved. The high-resolution results of drop morphological evolution were used to identify the different phases in the aerobreakup process, characterize the distinct flow features and dominant mechanisms in each phase. In the early time, the drop deformation and velocity are independent of the Weber number, and a new internal-flow deformation model, which respects this asymptotic limit, has been developed. The pressure and velocity fields around the drop were shown to better understand the internal flow and interfacial instability that dictate the drop deformation. Finally, the impact of drop deformation on the drop dynamics was discussed.
\end{abstract}

%\pacs{Valid PACS appear here}% PACS, the Physics and Astronomy
                             % Classification Scheme.
%Use showkeys class option if keyword
                              %display desired
\maketitle

\section{Introduction}
When a drop is subjected to a high-speed gas stream, the drop can experience significant deformation or even breakup. If the stabilizing forces, including surface tension and viscous forces, are not sufficient to overcome the destabilizing inertia force, the drop will break into a collection of children droplets of different sizes \citep{Taylor_1949a, Guildenbecher_2009a, Theofanous_2011a}. The aerodynamic breakup (or aerobreakup) of drops is essential to many spray applications, such as liquid fuel injection and spray cooling, and has therefore been extensively studied in the past decades.

Though the interaction between the drop and surrounding gas in practical aerobreakup applications usually involves many complex factors, drop aerobreakup is typically formulated in a simplified configuration,\ie, an initially stationary and spherical drop is suddenly exposed to an unbounded uniform gas stream \citep{Theofanous_2008a, Marcotte_2019a, Jain_2019a}. Then, the two-phase interfacial flows that dictate the drop deformation and dynamics are fully determined by the densities and viscosities of the drop liquid and the gas, $\rho_l$, $\mu_l$, $\rho_g$, and $\mu_g$, the surface tension $\sigma$, the initial drop diameter $d_0$, and the gas stream velocity $U_0$. The subscripts $g$ and $l$ are used to denote the properties for the gas and liquid, respectively, while the subscript $0$ is used to represent the initial state. The drop is considered to be far away from the free stream boundary. Furthermore, it is assumed that the Mach number of the gas stream is significantly lower than the critical Mach number, so the compressibility effect can be neglected. Aerobreakup of drops in a supersonic gas stream \citep{Theofanous_2004a, Theofanous_2007a, Sharma_2021c} is outside the scope of the present study. We have also considered the liquid as a Newtonian fluid and excluded the non-Newtonian effect of the drop liquids on the process \citep{Joseph_1999a,Theofanous_2013a}. As a result, the present problem can be fully characterized by four independent dimensionless parameters: the Weber number $\text{We} = \rho_g U_0^2 D_0/\sigma$, the Reynolds number $\text{Re} = \rho_g U_0 D_0/\mu_g$, the Ohnesorge number $\text{Oh} = \mu_l/\sqrt{\rho_l D_0 \sigma}$, and the gas-to-liquid density ratio $\textsl{r}=\rho_g/\rho_l$ \citep{Pilch_1987a, Hsiang_1992a, Joseph_1999a, Guildenbecher_2009a}. Alternative dimensionless parameters can be defined based on the above four parameters, such as the gas-to-liquid viscosity ratio $\textsl{m}=\mu_g/\mu_l$ \citep{Guildenbecher_2009a}.

The present study focuses on the regime of millimeter water drops in an air stream, following the recent work by \cite{Jackiw_2021a}. In this regime, the small values of $\textsl{r}\sim O(10^{-3})$ and $\text{Oh}\sim O(10^{-3})$ and the large value of $\text{Re}\sim O(10^3)$ indicate that surface tension plays the dominant role in resisting drop breakup. The key parameter characterizing drop dynamics and morphology evolution is $\text{We}\sim O(10^2)$, which is moderate. Previous experiments using shock tubes \citep{Hinze_1955a,Hsiang_1995a, Joseph_1999a,Theofanous_2004a} and continuous jets \citep{Liu_1993a, Flock_2012a, Zhao_2013a, Opfer_2014a, Guildenbecher_2017a, Jackiw_2021a, Jackiw_2022a} have identified different breakup modes for low-$\text{Oh}$ drops. The critical Weber number for breakup to occur is $\text{We}_{cr}\approx 11 \pm 2$. When $\text{We}<\text{We}_{cr}$, the drop will oscillate without breakup or break into a small number of large children drops due to large-amplitude nonlinear oscillations (also referred to as the \emph{vibration} breakup mode). As $\text{We}>\text{We}_{cr}$, the drop will break into a large number of children droplets. The drop morphology upon breakup varies significantly with $\text{We}$, from \emph{bag}, \emph{bag-stem} (or multi-mode), \emph{multi-bag}, and \emph{stripping} (or sheet-thinning mode) \citep{Hsiang_1995a, Liu_1997a, Guildenbecher_2009a, Jackiw_2021a}. When $\text{We}$ is sufficiently large, such as $\text{We}>10^3$, it has been claimed in previous studies that the drop breaks in the catastrophic mode \citep{Joseph_1999a}. However, more recent high-resolution experimental diagnostics suggest that such a breakup mode does not exist \citep{Theofanous_2008a}. Different classifications of the breakup regimes have also been proposed, based on the dominant breakup mechanisms \citep{Theofanous_2004a}. Since the bag, bag-stem, and multi-bag breakup modes all involve the formation and inflation of bags, which is driven by Rayleigh-Taylor instability, they have been reclassified as the Rayleigh-Taylor piercing (RTP) mode. On the other hand, the sheet-thinning mode has been renamed the shear-induced entrainment (SIE) mode, since it is mainly driven by the shear on the interface periphery.

The impulsive gas acceleration and continuous non-zero relative velocity between the gas and the drop induce both viscous and inviscid, steady and unsteady drag forces on the drop \citep{Maxey_1983a, Mei_1991a, Ling_2013b}, resulting in a streamwise acceleration of the drop. As the drop deforms, the gas flow will be modulated, which in turn influences the drop deformation. Consequently, although the surrounding gas flow in the far field is steady, the drop deformation and the gas flow around the drop are highly unsteady. According to the scaling analysis \citep{Ling_2013b}, the ratios between the unsteady forces and the quasi-steady drag are proportional to $1/\textsl{r}$. For the present cases with $\textsl{r}\ll 1$, the unsteady forces will have a small contribution to the overall drop acceleration. The drop dynamics are dictated by the quasi-steady drag, and the drop velocity will increase following the gas viscous time scale. For cases with a large $\textsl{r}$, such as in a pressurized gas environment, the contribution of the unsteady forces to the drop acceleration cannot be ignored \citep{Marcotte_2019a, Jain_2019a}.

Aerodynamic breakup of an isolated drop involves rich physics, and the investigation of the subject is challenging for both experiments and simulations. Due to the highly unsteady flow and complex topology change, an analytical approach is generally inviable, except for limiting cases \citep{Vanden_1980a, Miksis_1981a}. Previous studies of drop aerobreakup are based on experiments. Thanks to advancements in high-speed imaging, the temporal evolution of the drop morphology can be clearly captured \citep{Theofanous_2008a, Opfer_2014a, Flock_2012a, Jackiw_2021a}. With more sophisticated diagnostics, such as digital in-line holography, the velocity and size of the children droplets can also be measured \citep{Gao_2013a, Guildenbecher_2017a}. However, it is still challenging to measure high-level details of the drop deformation and gas flows due to the small length and time scales involved. For example, for the bag breakup mode, it is difficult to visualize the velocity and pressure fields inside the bag and to directly measure the thickness and velocity in the bag sheet \citep{Opfer_2014a}. Therefore, high-fidelity interface-resolved numerical simulations that can provide these high-level details are essential to investigate drop aerobreakup \citep{Jain_2019a, Marcotte_2019a}.

Fully-resolved simulations of drop aerobreakup are expensive if one aims to resolve all the length scales. For example, for a millimeter-sized drop that breaks in the bag mode, the bag sheet thickness reduces to tens of nanometers before the bag sheet ruptures \citep{Williams_1982b, Burelbach_1988s}. Previous numerical studies could afford to use a high mesh resolution for 2D axisymmetric simulations only \citep{Han_1999a, Han_2001a, Jing_2010a, Chang_2013a, Jalaal_2014a, Strotos_2016a, Stefanitsis_2017a, Marcotte_2019a, Jain_2019a}. However, as will be shown later, 2D axisymmetric simulations will \emph{not} correctly capture the bag formation in the high $\text{Re}$ regime since the turbulent wake and its influence on the drop deformation are not faithfully resolved. That explains why even converged 2D simulations fail to reproduce the bag morphology observed in experiments \citep{Marcotte_2019a}. The mesh resolutions in previous 3D simulations are generally much lower and insufficient to capture the turbulent wake and the bag development \citep{Kekesi_2014a, Jain_2015a, Yang_2016a, Jain_2019a}. There remains a significant gap between the existing simulation results and experimental data.

%(XXX literature survey for existing models for breakup time, drop size, velocity, and spatial distributions. XXX)
An important motivation for the study of drop aerobreakup is to develop sub-scale models that can be used in simulations of sprays consisting of a huge number of droplets, for which it is intractable to resolve the interface of each individual drop. The cell size used for simulations of sprays in practical scales is typically much larger than the size of individual drops, and therefore, the drops are represented by Lagrangian point-particles \citep{Apte_2003a, Pai_2006a, Balachandar_2009a}. The drop motion, shape deformation, topology change, and heat and mass transfer between the drop and the surrounding flow need to be captured by subgrid models \citep{ORourke_1987a, Hsiang_1992a}. When drop breakup occurs, one also needs to estimate the starting time and end time \citep{Dai_2001a, Chou_1998a} and the statistics of the children droplets produced \citep{ORourke_1987a, Hsiang_1992a, Wert_1995a, Zhao_2013a}. For the moderate $\text{We}$ regime, the Taylor analogy breakup (TAB) model \citep{ORourke_1987a} and its subsequent variants \citep{Tanner_1997a, Park_2002a} have been widely adopted to predict the drop deformation (the time evolution of the drop lateral radius) and the breakup time. The TAB models are simply based on the mass-spring-damper analogy and there exist coefficients that need to be calibrated based on experimental data. Models that incorporate the flow physics have also been proposed. In the model of \citet{Villermaux_2009a}, the pressure distribution induced by the stagnation flow on the windward surface is considered as the driving force for the lateral expansion of a thin liquid disk, although the model was also extended to predict the overall drop deformation \citep{Kulkarni_2014a, Stefanitsis_2019a, Rimbert_2020o, Jackiw_2021a}. Since the Rayleigh-Taylor Instability (RTI) plays a significant role in the formation and development of bags, models based on linear RTI analysis have also been developed to predict the critical condition for the bag formation and also the number of bags to be formed in the multi-bag mode \citep{Harper_1972a, Joseph_1999a, Theofanous_2004a}. For very high $\text{We}$, the drop breaks in the stripping mode, and the corresponding breakup models are based on the shear instability \citep{Ranger_1969a} and the rupture of the thinning liquid sheets \citep{Liu_1997a, Lee_2001a}. Comprehensive reviews of these models have been given in previous studies \citep{Guildenbecher_2009a, Jackiw_2021a}, thus are not repeated here.
 
%RTI 
Though significant progress has been made in understanding the physics that drive the deformation and breakup of drops at moderate $\text{We}$, there remain important fundamental questions unanswered. When $\text{We}$ is just above the critical value, a single bag is formed, and the breakup starts when the bag sheet ruptures. When $\text{We}$ is very large, liquid sheets are formed at the equator and break into small droplets before a bag gets a chance to form near the center. Between these two asymptotic limits, when $\text{We}$ is moderate, the drop can deform to a variety of shapes before breakup occurs \citep{Jackiw_2021a}. The variation in breakup morphology is very sensitive to $\text{We}$ in the moderate range.
{
As the drop is accelerated by the free stream, it experiences baroclinic torque and Rayleigh-Taylor instability (RTI) on its windward side. Several models based on the linear stability analysis of RTI on a liquid layer of finite thickness \citep{Mikaelian_1990a, Mikaelian_1996i} have been proposed to predict the critical Weber number \citep{Joseph_1999a, Theofanous_2004a}. However, the significance of RTI and whether the linear stability model captures the nonlinear development of the bag and the bag morphology change from a simple axisymmetric bag to more complex shapes remain uncertain, despite its presence on the interface \citep{Guildenbecher_2009a}.
} 
It has been suggested that the Laplace pressure on the edge of the drop/disk contributes to hindering the bag formation or development \citep{Guildenbecher_2009a}, yet solid evidence to support the argument remains absent.

%(XXX Goal of the present study: what will be done in this paper and what are the key questions to be addressed? XXX) 
The goal of the present study is to investigate the aerobreakup of an isolated drop at moderate Weber numbers through detailed interface-resolved 3D numerical simulation. By using advanced numerical techniques, including the mass-momentum consistent volume-of-fluid method \citep{Zhang_2020a, Arrufat_2020a}, balanced-force surface tension discretization \citep{Popinet_2009a}, and adaptive mesh refinement, we aim to fully resolve the two-phase turbulent interfacial flows involved in aerobreakup through large-scale parallel computations. The drop's initial diameter and fluid properties are kept unchanged, and the free-stream gas velocity is varied for a parametric study. Although the lateral radius of the drop typically increases monotonically in time \citep{Flock_2012a, Opfer_2014a, Jackiw_2021a}, the overall deformation is a complex process consisting of different phases with distinct features. The high-level details revealed by the simulation results will be used to identify the dominant mechanisms that dictate the deformation in each phase and also to characterize the effect of $\text{We}$.

For the range of $\text{We}$ considered, the drop will deform from its initial spherical shape to a \emph{forward} bag with the opening facing the upstream direction. Although it is of great interest to study the rupture of thin bag sheets and the statistics of the droplets generated by the bag and rim breakups \citep{Guildenbecher_2017a, Jackiw_2022a}, the mesh resolution required to fully resolve the sheet rupture and hole formation is beyond the current available computational capability. If a liquid sheet breaks due to molecular forces, the thickness must be reduced to tens of nanometers. For simulations of millimeter drops, a fully-resolved simulation would need to cover over six orders of magnitude in length scales, which is obviously impractical even with today's computer power. Therefore, the present study will mainly focus on the drop deformation and bag formation and development before the bag sheet rupture occurs. The present simulations can be considered fully resolved and mesh-independent up to the early stage of the bag inflation, but not for the sheet rupture and children drop statistics.
{
Nevertheless, the simulation results still provide important insights into the bag rupture mechanisms, such as the hole-hole interaction, which remain not fully understood. 
}
Since the occurrence of holes in the liquid sheet is random, ensemble averaging and statistics of multiple realizations (e.g., hole appearing locations and the number of holes) will be required to fully characterize the rupture dynamics and the statistics of the droplets formed {\citep{Mostert_2022g, Tang_2022h}}, though such studies are outside the scope of the present work.

The rest of the paper will be organized as follows. Section \ref{sec:equations} introduces the governing equations and section \ref{sec:methods} describes the numerical methods used in the present simulations. The physical parameters and simulation setup are presented in section \ref{sec:setup}. {Before discussing the simulation results, section \ref{sec:char_length_time} defines the characteristic length and time scales used to characterize the drop shape and gas flows. Grid-refinement and validation studies are presented in section \ref{sec:validation}. The simulation results identify different deformation phases, which are discussed in sequence in sections \ref{sec:phases_overall}. The effect of drop deformation on the aerodynamic drag and lift exerted on the drop is discussed in section \ref{sec:dynamics}.} Finally, section \ref{sec:conclusions} concludes the key findings of the present study.

\section{Simulation methods}
\label{sec:sim}
\subsection{Governing equations}
\label{sec:equations}
 The two-phase interfacial flows are governed by the incompressible Naviers-Stokes equations with surface tension, 
\begin{align}
	\rho \left(\pd{ u_i}{t} + u_i \pd{u_j}{x_j}\right) & = -\pd{p}{x_i} + \pd{}{x_j}\left[ \mu\left( \pd{u_i}{x_j} + \pd{u_j}{x_i} \right) \right] + \sigma \kappa\delta_s n_i\, ,	
	\label{eq:mom}\\
	\pd{u_i}{x_i} & =0\, .	
	\label{eq:cont}
\end{align}
where $\rho, u_i, p, \mu$ represent density, velocity vector, pressure and viscosity, respectively.  The Dirac distribution function $\delta_s$ is localized on the interface. The surface tension coefficient $\sigma$ is constant and $\kappa$ and $n_i$ represent the curvature and normal vector of the interface. 

The gas and liquid phases are distinguished by the liquid volume fraction $c$, the evolution of which follows the advection equation: 
\begin{align}
  \pd{c}{t} + u_i \pd{c }{x_i} & = 0 \, .
  \label{eq:adv}
\end{align}
After spatial discretization, the cells with only liquid or gas will exhibit $c=1$ and 0, respectively, while for cells containing the gas-liquid interface, $c$ is a fractional number. The density and viscosity are both defined based on the arithmetic mean
\begin{align}
  \rho & = \rho_l c + \rho_g (1-c)  \, , \\
  \mu & = \mu_l c + \mu_g (1-c)\, .
  \label{eq:mean}
\end{align}

\subsection{Numerical methods}
\label{sec:methods}
The governing equations (Eqs.\ \eqref{eq:mom}-\eqref{eq:adv}) are solved using the open-source, multiphase flow solver \textit{Basilisk}. The \textit{Basilisk} solver uses a finite-volume method for spatial discretization. The projection method is used to incorporate the incompressibility condition. The advection equation (Eq.\ \eqref{eq:adv}) is solved via a geometrical Volume-of-fluid (VOF) method \citep{Scardovelli_1999a}. The advection of momentum across the interface is conducted consistently with the VOF advection \citep{Arrufat_2020a, Zhang_2020a}. The mass-momentum consistency is essential for multiphase flows with large density contrast \citep{Zhang_2020a}. The balanced-force method is used to discretize the singular surface tension term in the momentum equation \citep{Popinet_2009a}. The interface curvature required to calculate surface tension is computed based on the height-function (HF) method \citep{Popinet_2009a}. The staggered-in-time discretization of the volume fraction/density and pressure leads to a formally second-order-accurate time discretization \citep{Popinet_2009a}. An adaptive quadtree/octree mesh is used to discretize the computational domain, which allows adaptive mesh refinement (AMR) in user-defined regions. The mesh adaptation is based on the wavelet estimate of the discretization errors of the liquid volume fraction and velocity \citep{Hooft_2018a}. The parallelization of the solver is done through a tree decomposition to guarantee high parallel performance even if a large number of levels of octree cells are used. Numerous validation studies for the \emph{Basilisk} solver can be found on the \textit{Basilisk} website and also in our previous studies \citep{Zhang_2020a, Sakakeeny_2020a, Sakakeeny_2021a, Sakakeeny_2021b}.

\subsection{Simulation setup}
\label{sec:setup}
\begin{figure}
\centering
\includegraphics[trim={0cm 0cm 0cm 0cm},clip,width=0.95\textwidth]{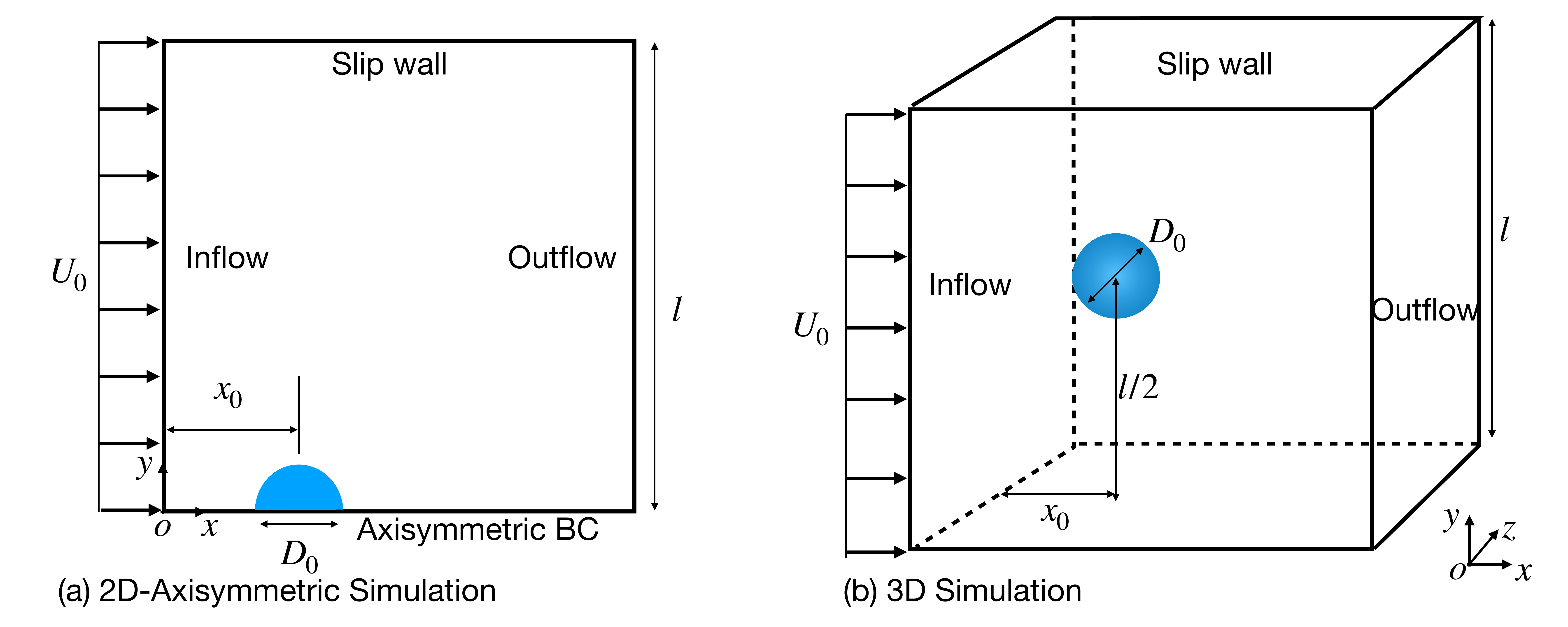}
\caption{Schematics of the computational domains for (a) 2D-axisymmetric and (b) 3D simulations. }
\label{fig:setup}
\end{figure}

The spherical drop is initially placed in a computational domain filled with stationary gas. We have considered both 2D-axisymmetric and 3D domains, as shown in figures \ref{fig:setup}(a) and \ref{fig:setup}(b), respectively. At $t=0^+$, a uniform velocity boundary condition is imposed on the left boundary of the domain, while the pressure outflow boundary condition is applied to the right boundary. Due to the incompressibility condition, the gas is suddenly accelerated to $U_0$ in an infinitesimal time (one time step in the simulation).

All lateral boundaries in the 3D domain are considered as slip walls. For the 2D domain, the bottom is the axis and the top is a slip wall. The 2D domain is a square with an edge length of $l/d_0=16$, while the 3D domain is a cube with an edge length of $l/d_0=32$. The domain size is sufficient to resolve the wake behind the drop during the time considered. For all 2D and 3D simulations, the drop is initially placed $x_0/d_0=3$ away from the inlet. In the 3D simulations, the initial location of the drop is at the center of the $y$-$z$ cross-section.

The physical parameters used in the present simulations are chosen to be similar to the experiment of \cite{Jackiw_2021a}. The liquid and gas are water and air at room temperature, and their fluid properties are given in table~\ref{tab:phy_para1}. The initial drop diameter is fixed at $d_0=1.9$ mm. The corresponding gas-to-liquid density and viscosity ratios are thus small, {$\textsl{r}=0.0012$} and $\textsl{m}=0.018$, respectively. The Ohnesorge number $\text{Oh}=0.0026$ is very small, indicating that the effect of liquid viscosity on drop breakup is small compared to that of surface tension, and the latter is the dominant force to resist the drop deformation/breakup induced by interaction with the gas stream. Therefore, the Weber number $\text{We}$ is the key dimensionless parameter that characterizes the aerobreakup dynamics. A parametric study of We is performed here by varying $U_0$ from 18.7 to 24.0 m/s. The Reynolds number will vary from about 2400 to 3000 correspondingly. More cases were simulated, but we will focus the discussion on the five cases shown in table~\ref{tab:cases}.

\begin{table}
\setlength{\tabcolsep}{3.5pt}
  \caption{Fluid properties for simulation cases.}
\centering
  \begin{tabular}{ccccccccc} 
 $\rho_{l}$      & $\rho_{g}$     & $\mu_{l}$ & $\mu_{g}$ & $\sigma$  & $d_0$ &\textsl{r} & \textsl{m} & Oh	\\ 
    (kg/m$^{3}$) & (kg/m$^{3}$) & (Pa\,s) & (Pa\,s)  & (N/m) & (mm) & $\rho_g/\rho_l$ & $\mu_g/\mu_l$ & $\mu_l/\sqrt{\rho_l D_0 \sigma}$\\
$1000$             & $1.2$                  & 0.001 & $0.000018$ & $1.9$ & $0.0483$ &  $0.0012$ & $0.018$ & 0.00269 \\ 
  \end{tabular}
  \label{tab:phy_para1}
\end{table}

\begin{table*}
\setlength{\tabcolsep}{5pt}
     \caption{Simulation cases and key parameters. }
   \centering
    \begin{tabular}{c c c c c ccc}
\hline
Cases         	 & $U_0$	& We & Re  	& 3D Mesh & 2D Mesh\\
         & (m/s)  & $\rho_gU_0^2d_0/\sigma$ & $\rho_g U_0 d_0/\mu_g$	& $N=d_0/\Delta_{\min}$ & $N$\\
\hline
  			& 18.7  & 10.9 	& 2369  			& 256	& \\
   			& 19.2  & 11.5	& 2432  			& 256 - 2048	&  \\
 			& 19.6  & 12.0	& 2483  			& 256 - 2048 	& 256 - 2048\\
 			& 22.1  & 15.3 	& 2800  			& 128 - 2048 	& \\
 			& 24.0  & 18.0 	& 3040  			& 256 - 2048	&  \\
%  			& 27.9  & 25.4	& 3538  			& 1.22	& 4.9		& 256, 512(2048)& 256 - 1024\\
% 7 			& 32.6  & 33.2	& 4129  			& 1.43	& 5.8		& 256(2048)\\
%  			& 36.7  & 42.0	& 4641  			& 1.61	& 6.5		& 256, 512(1024)& 256 - 2048\\
% 9 			& 48.0  & 72.0	& 6077  			& 2.11	& 8.5		& 256, 512(1024)& 256 - 2048\\
\hline
    \end{tabular}
\label{tab:cases}
\end{table*}

The 2D/3D domains are discretized by quadtree/octree meshes, which are dynamically adapted based on the wavelet estimates of the discretization errors of the liquid volume fraction and all velocity components \citep{Hooft_2018a}. The minimum cell size is controlled by the maximum refinement level $\mathcal{L}$, \ie, $\Delta_{\min}=l/2^\mathcal{L}$. For 2D simulations, $\mathcal{L}$ has been varied from 12 to 15, corresponding to 256 to 2048 minimum quadtree cells across the initial drop diameter, \ie, $N=d_0/\Delta_{\min}=256$ to 2048. Due to the higher computational cost for 3D simulations, we have used a different mesh adaptation strategy. As will be shown later, the mesh resolution required to capture the early deformation of the drop is significantly lower than that required to resolve the thin liquid sheet in the bag. Therefore, we have used a maximum refinement level of {$\mathcal{L}=12$} to 14, which is equivalent to {$N=128$} or 512 to start the simulations, for the purpose of confirming that the resolution is enough to capture the drop deformation until the bag is formed. As the bag grows and the sheet thickness decreases, $\mathcal{L}$ is then increased until 16 ($N=2048$). For the case $\text{We}=10.9$, since a thin bag sheet is never formed, $\mathcal{L}=12$ ($N=256$) is used throughout the simulation.

According to the knowledge of the authors, the current mesh resolution is significantly higher than all previous numerical studies, including both 2D and 3D simulations in the literature, while the domain is large enough to resolve the wake \citep{Marcotte_2019a, Jain_2019a}. For the 3D domain, the mesh for $N=2048$ corresponds to 16 levels of adaptively refined octree cells, which is equivalent to $2.8\times 10^{14}$ uniform Cartesian cells. The total number of octree cells generally increases over time, as the drop surface area increases and the wake develops. For the case $\text{We}=15.3$, the number of octree cells reaches about 500 million. The high-performance tree-decomposition parallelization technique in the \emph{Basilisk} solver is essential to guarantee efficient simulations using such large numbers of levels (up to 16), computational cores (up to 3584 cores), and octree cells (up to 500 million).

For the present study, the drop diameter is 1.9 mm, and the minimum cell size for $N=2048$ is about $\Delta_{\min} =$ 0.9 \textmu m, which is still significantly larger than the physical cutoff length scale dictated by {van der Waals forces} ($\sim O(10\ \text{nm})$). Therefore, the interfaces' pinching and hole formation induced by {van der Waals forces} will not be captured in the simulation. The hole nucleation in the liquid sheet here is due to the numerical cutoff length, \ie, the minimum cell size $\Delta_{\min}$. When the sheet thickness is smaller than about $2\Delta_{\min}$, the flow and pressure in the liquid cannot be well resolved, and the thin region of the liquid sheet represented by the VOF field will be perturbed, and numerical pinching of the interfaces will occur. The numerical breakup leads to earlier hole formation, and as a result, the bag will break at a size smaller than what was observed in experiments. Nevertheless, as indicated in previous experimental studies, holes are observed in the liquid sheet with thickness larger than the length for active molecular forces \citep{Lhuissier_2012a, Opfer_2014a}, probably due to tiny bubbles in the liquid sheet. A more detailed discussion of the effect of mesh resolution on the bag breakup dynamics will be given later in the results section.

The 2D simulations were run on the Baylor University cluster \emph{Kodiak} using 4 to 36 cores (Intel Xeon Gold 6140 processor). The most refined simulation ($N=2048$) takes about 20 days using 36 cores. The 3D simulations were run on the TACC-Stampede2 (Intel Xeon Platinum 8160) and TACC-Frontera (Intel Xeon Platinum 8280) machines, using up to 3584 cores. The simulation time varies with cases, and the longest one takes about 50 days.

\section{Parameters to characterize drop aerobreakup}
\label{sec:char_length_time}
Before we present and discuss the results, we will first introduce the parameters to be measured in the simulation to characterize the drop shape, dynamics, and turbulent flows involved in the drop aerobreakup process. 

\subsection{Drop {morphology}}
\label{sec:drop_length}
\begin{figure}
\centering
\includegraphics[trim={0cm 0cm 0cm 0cm},clip,width=0.99\textwidth]{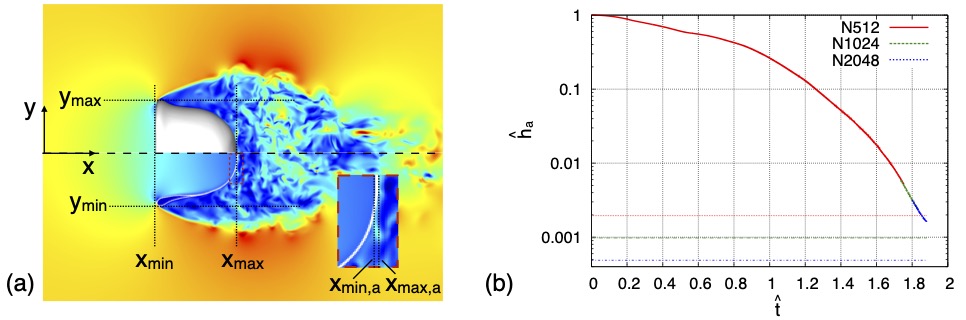}
\caption{(a) Representative snapshot of the drop surface and velocity magnitude on the central $x$-$y$ plane, with annotations showing the characteristic length scales for the drop shape. (b) Time evolution of $h_a=x_{\max,a}-x_{\min,a}$ for $\text{We}=15.3$, indicating the time ranges resolved by different mesh resolutions. The horizontal lines indicate the minimum cell size for each mesh. The dimensionless variables are denoted by $\hat{}$, defined in Eqs.\eqref{eq:dim_x} and \eqref{eq:dim_t}. }
\label{fig:notation_mesh}
\end{figure}

To characterize the drop shape at a given time, the maximum and minimum interfacial positions along the Cartesian coordinates were measured, \ie, $x_{\max}$, $x_{\min}$, $y_{\max}$, $y_{\min}$, $z_{\max}$, $z_{\min}$, as shown in figure~\ref{fig:notation_mesh}(a). Furthermore, we also measured the maximum and minimum interfacial positions along the $x$-axis (along the streamwise direction), $x_{\max,a}$, $x_{\min,a}$. Based on these measurements, several characteristic length scales can be obtained:
\begin{itemize}
\item \emph{Length of the drop}, $L=x_{\max}- x_{\min}$.
\item \emph{Lateral radius of the drop}, $R=(y_{\max}-y_{\min}+z_{\max}-z_{\min})/4$. When the drop shape is axisymmetric, the maximum lateral radius $R=y_{\max}=-y_{\min}=z_{\max}=-z_{\min}$; otherwise, $R$ is the average for $y$ and $z$ directions.
\item \emph{Thickness of the sheet along the $x$-axis}: $h_{a}=x_{\max,a}$-$x_{\min,a}$. At early times, when the drop shape is axisymmetric and both windward and leeward sides are convex, $h_a=L$. At later times, when the drop deforms to a bag, $h_a$ represents the sheet thickness at the center of the bag, as shown in figure~\ref{fig:notation_mesh}(b).
\item \emph{Length of the forward bag}: $L_{fb}=x_{\min,a}- x_{\min}$. Here, the term \emph{forward bag} refers to the bag with the opening facing the upstream direction.
\end{itemize}

\subsection{{Turbulent gas flow and drop dynamics}}
For the range of $\text{Re}$ considered, the gas flow in the wake of the drop lies in the chaotic vortex shedding or the subcritical turbulent wake regimes, according to the regime map for a solid sphere \citep{Tiwari_2020b}. The gas enstrophy is then used to characterize the gas turbulence, which is calculated as
\begin{equation}
\Omega_g = \frac{1}{\mathbb{V}}\int (1-c)|\bs{\omega}|^2 dV \,,
\end{equation}
where $\bs{\omega}$ is the vorticity vector and $\mathbb{V}$ is the volume of the computational domain. The normalized gas enstrophy is defined as $\hat{\Omega}_g=\Omega_g d_0/U_0^2$.

The aerodynamic drag exerted on the drop will cause the drop to accelerate along the streamwise direction. The mean velocity of the drop in the $x$ direction is calculated by integrating over all the liquid cells in the domain,
\begin{equation}
u_d = \frac{\int c\, u dV}{\int c\, dV} \,.
\end{equation}
Note that this definition is only valid before breakup occurs. Then the drag coefficient can be defined as
\begin{equation}
C_D = \frac{2 m_d}{\rho_g (U_0-u_d)^2 \pi R^2} \od{u_d}{t}\,,
\label{eq:Cd}
\end{equation}
where $m_d$ is the mass of the drop. Here, $C_D$ is defined based on the instantaneous relative velocity ($U_0-u_d$), and the drop frontal area estimated by the lateral radius ($\pi R^2$). As a result, the variation of $C_D$ is generally not influenced by the time variation of $u_d$ and $R$.

The asymmetric drop deformation and the turbulent wake will introduce lift on the drop. The lift coefficients in the transverse directions are defined as,
\begin{equation}
C_{y} = \frac{2 m_d}{\rho_g (U_0-u_d)^2 \pi R^2} \od{v_d}{t}\,, \quad
C_{z} = \frac{2 m_d}{\rho_g (U_0-u_d)^2 \pi R^2} \od{w_d}{t}\,,
\end{equation}
where $v_d$ and $w_d$ are the mean drop velocities in $y$ and $z$ directions, respectively, similar to $u_d$.

\subsection{Characteristic time scales for drop deformation}
\label{sec:time_scales}
The early-time deformation of the drop is driven by the pressure variation in the radial direction, which is in turn induced by the stagnation gas flow on the windward surface of the drop. The viscous and surface tension effects are secondary to the early-time deformation. Then simple scaling analysis shows that the time scale governing the drop deformation is \citep{Ranger_1969a, Villermaux_2011a}
\begin{equation}
	\tau_d = \frac{d_0}{U_0 \sqrt{\textsl{r}}}\,. 
\end{equation}
The scaling $\tau_d\sim \textsl{r}^{-1/2}$ has been confirmed by the parametric study of \citet{Marcotte_2019a}. It was shown that the evolutions of $R$ for different $\textsl{r}$ collapse approximately if $t$ is scaled by $\tau_d$. This time scale $\tau_d$ was first introduced by \citet{Ranger_1969a} based on the drag on the drop and the resulting acceleration. However, it was shown by \citet{Marcotte_2019a} that the time scale $\tau_d$ actually does not collapse the drop velocity  for different $\textsl{r}$. In the present study, $\textsl{r}$ and $d_0$ are kept as constant, so $\tau_d$ varies due to $U_0$. 

The surface tension and liquid viscosity resist the drop deformation, and the corresponding time scales are 
\begin{align}
	\tau_s & = \sqrt{ \frac{\rho_l d_0^3}{\sigma}}\,, \\
	\tau_{vl} & =\rho_l d_0^2/\mu_l\, .
\end{align}
The time scale ratios are related to the key dimensionless parameters \ie, $\tau_s/\tau_{vl}=\text{Oh}$ and ${\tau_s}/{\tau_d} = \sqrt{\text{We}}$. For the cases considered here, $\text{Oh}\sim O(10^{-3})$ is very small and $\text{We}\sim O(10)$. As a result, $\tau_{vl} \gg \tau_s > \tau_d$. Therefore, $\tau_{vl}$ is too large to be material and surface tension is the dominant resisting force. Furthermore, since $\tau_s > \tau_d$, we expect the early-time drop deformation will be insensitive to $\text{We}$. 

\subsection{Characteristic time scales for drop acceleration}
When a stationary drop is suddenly exposed to the free stream, it experiences an impulsive gas acceleration and a continuous non-zero relative velocity. The resulting aerodynamic drag drives the drop to move along the streamwise direction. This treatment can be considered as an approximation of the passage of a shock wave with the post-shock velocity equal to $U_0$, neglecting the compressibility effect, as discussed by \citet{Marcotte_2019a}.

The overall drag can be divided into quasi-steady, pressure-gradient, added-mass, and Basset history forces \citep{Maxey_1983a,Ling_2013b}. The last three are unsteady contributions, which are mainly triggered by the impulsive acceleration. The pressure-gradient and added-mass forces, often referred to as the inviscid unsteady forces, are only active at $t=0^+$, when the surrounding gas is suddenly accelerated from 0 to $U_0$. In the simulation, this impulse spans one time step, and the integration of the impulse is $U_0$. Therefore, the drop velocity jumps over the first time step due to the inviscid unsteady forces (including added-mass and pressure-gradient forces) as follows \citep{Ling_2011a, Ling_2013b}:
\begin{equation}
u_d(0^+)= \frac{3\textsl{r}}{2} U_0.
\label{eq:drop_vel_jump_fiu}
\end{equation}
{
Similar estimates of the velocity jump have been proposed by \citet{Marcotte_2019a} and \citet{Hadj-Achour_2021z}. \citet{Marcotte_2019a} found that the velocity jump scales with $\textsl{r}$, but the scaling coefficient of 1/2 is different from the 3/2 in Eq.~\eqref{eq:drop_vel_jump_fiu} since they did not include the pressure gradient force.
}
The viscous-unsteady (Basset history) force will last for a finite period of time, characterized by the viscous-unsteady time scale $\tau_{vu}$ \citep{Ling_2013b}. Due to the small value of $\textsl{r}$ considered here, the contributions of the unsteady forces to the increase in drop velocity are generally small \citep{Ling_2013b}. For cases with larger $\textsl{r}$, additional attention to the unsteady forces will be required. For the present cases, the quasi-steady force is the dominant force that accelerates the drop. The time scale for the quasi-steady drag is referred to as the response time, $\tau_{qs}$. For $\textsl{r}\ll 1$ \citep{Ling_2013b},
\begin{equation}
\tau_{qs} = \frac{\rho_g d_0^2}{36\mu_g}\Phi,
\end{equation}
where $\nu_g=\mu_g/\rho_g$, and $\Phi$ is the correction factor to the Stokes drag, accounting for the effect of Re and the drop shape. In the limit of zero Re and We, the drop acts like a solid sphere in the Stokes limit, and $\Phi=1$. For the range of Re considered here, $\tau_{qs}\sim O(10^{-1}\ \textrm{s})$. As a result, drop breakup occurs on a time scale much smaller than $\tau_{qs}$ when the drop velocity $u_d$ remains significantly lower than the free-stream velocity $U_0$.

Finally, we define dimensionless variables using $d_0$, $U_0$, and $\rho_g$ as characteristic scales, namely 
\begin{equation}
	\hat{x}=x/d_0\,, \hat{L}=L/d_0\,, \hat{R}=R/d_0\,, \hat{u} = u/U_0\,, \hat{p} = p/(\rho_g U_0^2)\, . 
	\label{eq:dim_x}
\end{equation}
Due to the focus on the drop deformation, $\tau_d$ is selected as the characteristic time scale. The dimensionless time is then defined as
\begin{equation}
	\hat{t} = t/\tau_d\, .
	\label{eq:dim_t}
\end{equation}

\section{Verification and Validation}
\label{sec:validation}
\subsection{Grid-refinement studies}
\begin{figure}
\centering
\includegraphics[trim={0cm 0cm 0cm 0cm},clip,width=0.99\textwidth]{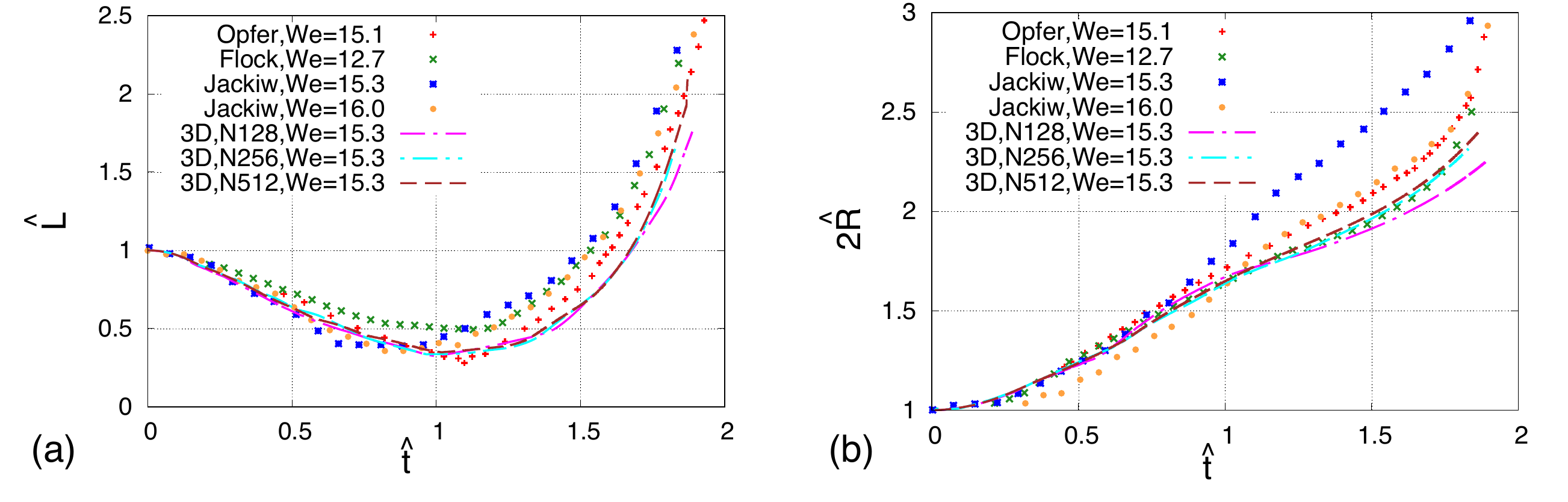}
\caption{Time evolutions of (a) the drop length $\hat{L}$ and (b) the drop radius $\hat{R}$ for We=15.3 {obtained from the 3D simulations}. The experimental results by \citet{Opfer_2014a}, \citet{Flock_2012a}, and \citet{Jackiw_2021a} are shown for comparison. }
\label{fig:validation}
\end{figure}

The time evolutions of the drop length $\hat{L}$ and radius $\hat{R}$ for different mesh resolutions are shown in figure~\ref{fig:validation}. The initial mesh resolutions are $N=128$, $256$, and $512$. The same mesh refinement strategy has been used for all three meshes, \ie, when the drop deforms to a bag and the bag sheet thickness decreases to about 4 cells, then refinement level will be increased until $N=2048$, as shown in figure~\ref{fig:notation_mesh}(b). It can be seen from figure~\ref{fig:validation} that the results for $N=256$ and $512$ match very well. The results for $N=128$ are also similar in general, though small deviations are observed for $\hat{t}>1.3$ when the forward bag forms. Therefore, one can conclude that $N=512$ is sufficient to yield converged results for the drop deformation until the bag ruptures and thus is sufficient to determine the breakup criteria.

\begin{figure}
\centering
\includegraphics[trim={0cm 0cm 0cm 0cm},clip,width=0.99\textwidth]{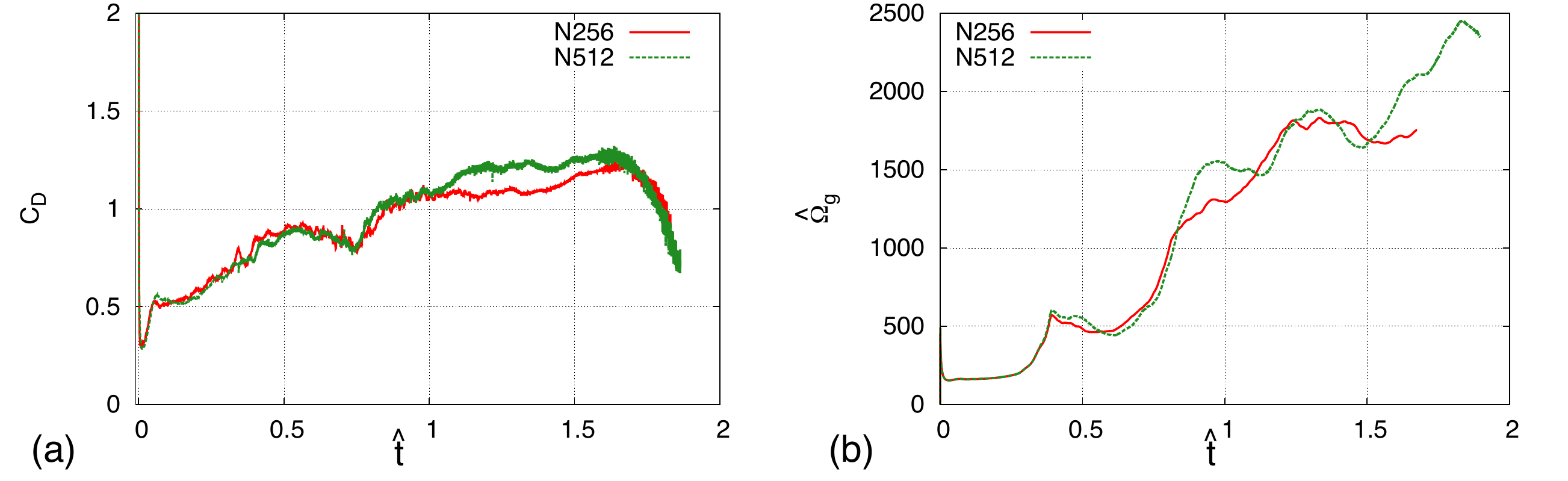}
\caption{Time evolutions (a) drag coefficient $C_D$ and (b) gas enstrophy $\hat{\Omega}_g=\Omega_g d_0/U_0^2$ for $\text{We}=15.3$. {The 3D simulation results for two different mesh resolutions $N=256$ and 512 are shown.} In (b) the results for $N=256$ were obtained by the mesh refinement $\mathcal{L}=14$ without further increase as in other cases.  }
\label{fig:validation_gas}
\end{figure}

In order to assess the adequacy of the mesh resolution in resolving the turbulent gas flow and its effect on the drag force, we examine the evolution of the drag coefficient $C_D$ and the gas enstrophy (see figure~\ref{fig:validation_gas}). Both $C_D$ and $\Omega_g$ exhibit non-monotonic behavior due to the development of the turbulent wake and the drop deformation. The results obtained with meshes of $N=256$ and $512$ show good agreement, with only minor discrepancies for $\hat{t}\gtrsim 1$. Given the inherently chaotic nature of the gas flow, this level of agreement is considered satisfactory.

\subsection{Experimental validation}
The simulation results obtained for the case $\text{We}=15.3$ are validated against available experimental data for the drop length $\hat{L}$ and radius $\hat{R}$. Specifically, comparison is made with the experiments of \citet{Jackiw_2021a} (water, $\text{We}=15.3$ and 16.0), \citet{Opfer_2014a} (Ethylene glycol, $\text{We}=15.1$), and \citet{Flock_2012a} (Ethyl Alcohol, $\text{We}=12.7$). The parameters employed for the present simulations are chosen to be similar to those adopted in the experiments conducted by \citet{Jackiw_2021a}. Although the dimensionless parameters for the two other experiments are not identical to those in the present study, the small values of $\text{Oh}$ and $\textsl{r}$ and the large value of $\text{Re}$ in all cases suggest that the aerobreakup dynamics for similar $\text{We}$ is expected to be comparable to the present cases. Thus, the experimental data serve as an important validation benchmark for the present numerical simulations.

{
It can be observed from figure~\ref{fig:validation} that the simulation results for both $\hat{L}$ and $\hat{R}$ exhibit remarkable agreement with the experimental data of JA and \citet{Opfer_2014a} at early times. The deviations from the experimental results of \citet{Flock_2012a} are relatively small, given that the experimental parameters, such as $\text{We}=12.7$, do not match the present simulations exactly. The simulation results for $\hat{R}$ deviate from the JA experimental data for the same $\text{We}=15.3$ at around $\hat{t}=0.7$, but they match quite well with the JA experimental results for a slightly larger $\text{We}=16$ and the other two experiments. The discrepancy may be due to the high degree of variation inherent in such experiments. Unlike the results of \citet{Opfer_2014a} and \citet{Flock_2012a}, which are averaged data for multiple experiments, those of JA correspond to single experimental runs and thus exhibit higher uncertainty. Moreover, in JA's experiment, the drop is suspended from a needle, which may significantly influence the development of the drop radius, as discussed in the appendix of their paper.
}

\subsection{Limitations of 2D axisymmetric simulations}
{
Due to the high computational cost of 3D simulations, 2D axisymmetric simulations have been widely used in numerical studies of drop aerobreakup \citep{Marcotte_2019a, Jain_2019a, Stefanitsis_2019b}. However, previous studies for bubbles \citep{Blanco_1995v, Magnaudet_2007d} and drops \citep{Rimbert_2020o} have also indicated that 2D axisymmetric simulations may not be sufficient to resolve the asymmetric wake when $\text{Re}$ and $\text{We}$ are not small. It remains unclear to what extent (such as time duration and ranges of $\text{Re}$ and $\text{We}$) a 2D axisymmetric simulation is valid in resolving drop aerobreakups.
}
In the present study, we have performed both 2D and 3D simulations using identical numerical methods and initial/boundary conditions. Therefore, the difference between the results characterize the effect of the artificial axisymmetric constraint in the 2D simulations on the drop. 

\begin{figure}
\centering
\includegraphics[trim={0cm 0cm 0cm 0cm},clip,width=0.99\textwidth]{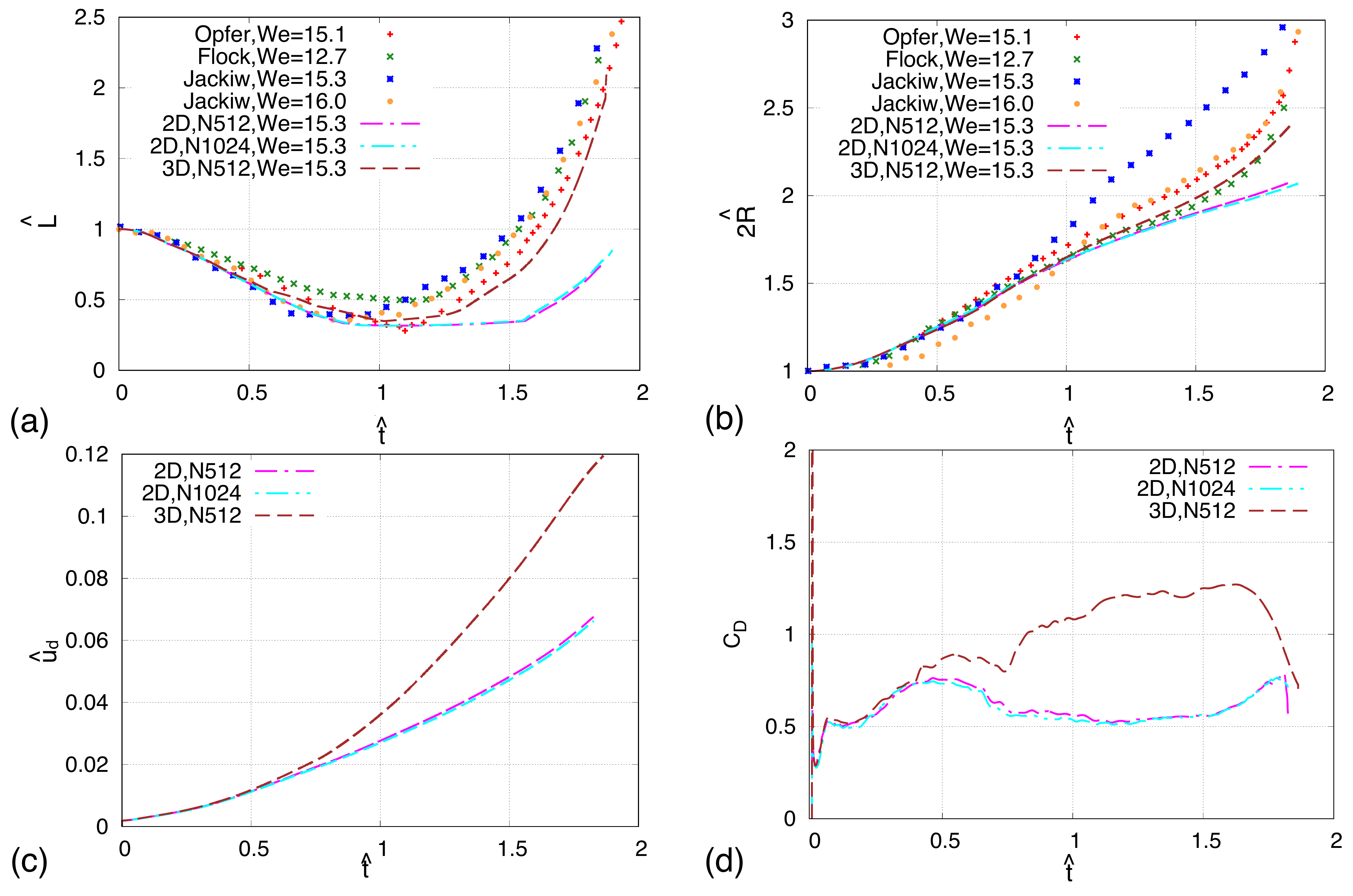}
\caption{Comparison between 2D-axisymmetric and 3D simulation results for We=15.3, including (a)  drop length $\hat{L}$, (b) drop radius $\hat{R}$, (c) drop velocity $u_d$, and (d) drag coefficient $C_D$. {The experimental results by \citet{Opfer_2014a}, \citet{Flock_2012a}, and \citet{Jackiw_2021a} are shown in (a) and (b) for comparison.} }
\label{fig:Compare_2D_We15}
\end{figure}

The temporal evolutions of $\hat{L}$ and $\hat{R}$ for both 2D and 3D simulations are shown in figure~\ref{fig:Compare_2D_We15}. The 2D simulation results for two different mesh resolutions, $N=512$ and $N=1024$, agree remarkably well, indicating that the 2D results presented here are mesh independent. The discrepancy between the 2D and 3D results is not related to the mesh resolution. The 2D and 3D results match very well for $\hat{t}\lesssim 0.5$, but the deviation grows over time. In general, the 3D results agree much better with the experimental data. The 2D simulation significantly underestimates the drop length compared to the 3D simulation and experimental results, as shown in figure~\ref{fig:Compare_2D_We15}(a). This indicates that the 2D simulations fail to capture the bag development. Significant discrepancies can also be observed in the drop velocity and drag coefficient, as shown in figures \ref{fig:Compare_2D_We15}(c) and (d). For $1\lesssim \hat{t} \lesssim 1.6$, the value of $C_D$ predicted by the 2D simulation is only about half of that obtained from the 3D simulation.

\begin{figure}
\centering
\includegraphics[trim={0cm 0cm 0cm 0cm},clip,width=0.99\textwidth]{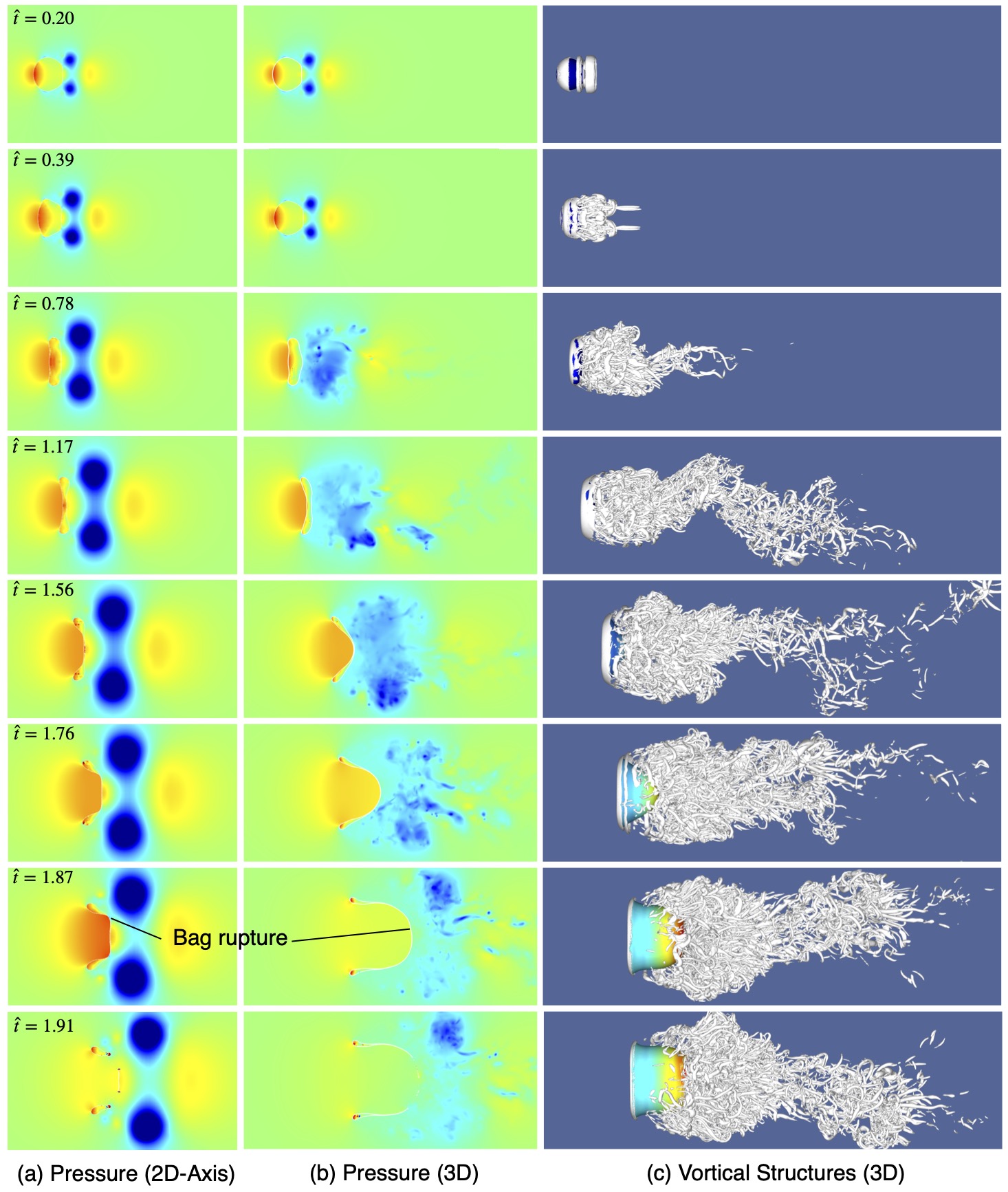}
\caption{Time evolutions of the pressure field on the $x$-$y$ plane using (a) 2D-axisymmetric and (b) 3D simulations, and (c) the vortical structures using 3D simulations. The results are for the case $\text{We}=15.3$. The color on the drop surface in (c) represents the velocity magnitude. }
\label{fig:wake_We15}
\end{figure}

The poor prediction of the 2D simulation is due to the fact that it cannot accurately resolve the turbulent wake. The pressure fields obtained from both 2D and 3D simulations are shown in figures~\ref{fig:wake_We15}(a) and (b), respectively. The vortical structures at different times for the 3D simulation, visualized by the $\lambda_2$ criterion, are shown in figure~\ref{fig:wake_We15}(c) to illustrate the development of the wake. The wake is initially approximately axisymmetric and later transitions to being fully 3D and turbulent. The artificial axisymmetric constraint in the 2D simulation limits the azimuthal development of the wake, leading to high pressure near the leeward pole and a low-pressure region in the symmetric vortex ring, as seen in figure~\ref{fig:wake_We15}(a) at $\hat{t}=0.78$. In contrast, the gas pressure near the leeward pole is lower and more uniform in the turbulent wake for the 3D results. The different pressure distributions result in different drags on the drop. The high pressure on the leeward pole in the 2D simulation also hinders the development of the bag, which is the reason for the discrepancy for $\hat{L}$ observed in figure~\ref{fig:Compare_2D_We15}(a). Actually, the 2D simulation fails to capture the correct shape of the bag, as seen at $\hat{t}=1.76$ in figures~\ref{fig:wake_We15}(a) and (b). While the 3D simulation shows a bag curved forward shape, the 2D bag bends upstream near the center, showing a ``bag-stem" shape. The maximum bag curvature and minimum sheet thickness in the 2D bag are near the 45-degree corner, where the rupture of the bag and the hole will first appear. This is not consistent with the experiment and 3D simulation results and is purely a numerical artifact.

In summary, for drops with high $\text{Re}$ the 2D simulation can only capture the early-time drop dynamics and deformation ($\hat{t}\lesssim 0.7$),  when the wake remains approximately axisymmetric. Even though the drop shape for $\text{We}=15.3$ remains approximately axisymmetric for a longer time, it will not be correctly captured by the 2D axisymmetric simulation. A full 3D simulation, though expensive, is necessary to resolve the drop aerobreakup. 

\section{Different phases in drop morphological evolution}
\label{sec:phases_overall}
\begin{figure}
\centering
\includegraphics[trim={0cm 0cm 0cm 0cm},clip,width=0.99\textwidth]{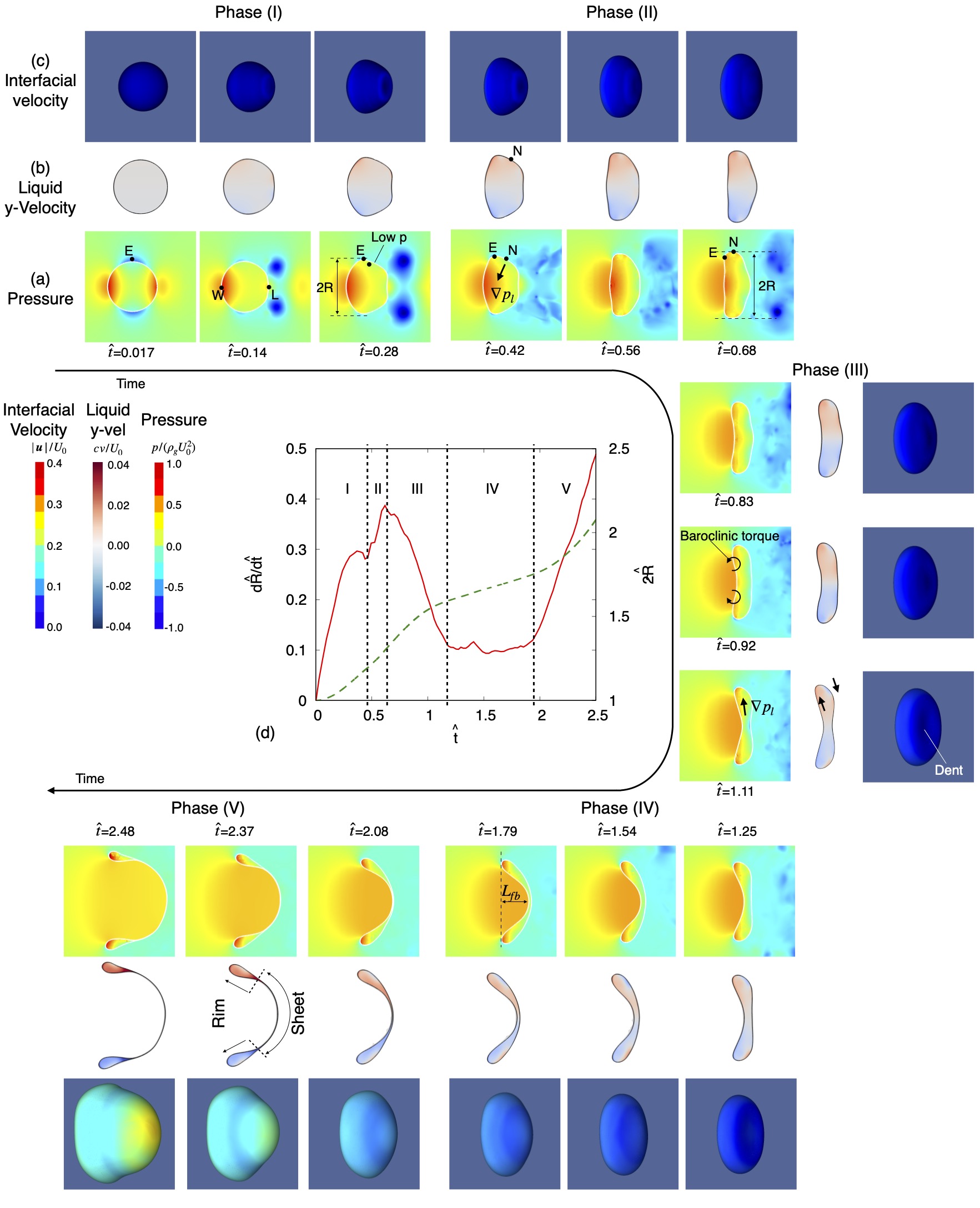}
\caption{Temporal evolutions of (a) pressure $p$ and (b) $y$-velocity in the drop on the central $x$-$y$ plane, (c) drop surface colored with velocity magnitude, (d) the drop lateral radius $\hat{R}$ and its rate of change $d\hat{R}/d\hat{t}$ for $\text{We}=12.0$. Different phases of deformation are defined and indicated. }
\label{fig:phase_def}
\end{figure}

The shape deformation of a drop is a complex process consisting of multiple phases with distinct features and dominant physics. We will use the case $\text{We}=12.0$ to illustrate the different deformation phases, as shown in figure~\ref{fig:phase_def}. In figures~\ref{fig:phase_def}(a) and (b), we have presented the temporal evolution of pressure $\hat{p}$ and y-velocity $\hat{u}_y$ inside the drop on the central $x$-$y$ plane. The corresponding drop surfaces, colored with the velocity magnitude, are shown in figure~\ref{fig:phase_def}(c). The drop radius $\hat{R}$ has been commonly used in previous studies to describe drop deformation, and it increases monotonically over time. However, its rate of change, $d\hat{R}/d\hat{t}$, exhibits a non-monotonic variation over time, as shown in figure~\ref{fig:phase_def}(d), from which different phases can be identified:
\begin{enumerate}
\item \emph{Phase I, ellipsoid deformation}: $d\hat{R}/d\hat{t}$ increases over time with a decreasing rate and gradually approaches a plateau. The deformation is driven by the stagnation pressure on the windward surface. The drop shape is similar to an \emph{ellipsoid}, and the thickness in the streamwise direction decreases over time.
\item \emph{Phase II, disk formation}: $d\hat{R}/d\hat{t}$ rises quickly again as the drop extends laterally, forming a circular \emph{disk}.
\item \emph{Phase III, disk deformation}: $d\hat{R}/d\hat{t}$ decreases over time. The deceleration of the drop edge and equator is due to surface tension at the drop periphery.
\item \emph{Phase IV, bag development}: $d\hat{R}/d\hat{t}$ is approximately constant, indicating that forces reach equilibrium at the edge rim. The sheet thickness near the center continues to decrease, and the thin region of the bag starts to curve toward downstream, \emph{forming a forward bag}.
\item \emph{Phase V, bag inflation}: $d\hat{R}/d\hat{t}$ increases rapidly due to the \emph{inflation of the forward bag}.
\end{enumerate}
After the above phases, there is an additional phase, \emph{Phase VI}, for the \emph{bag rupture}, which is not shown in figure~\ref{fig:phase_def}. These phases have not been systematically addressed in previous studies, probably because little attention has been paid to the evolution of $d\hat{R}/d\hat{t}$. The key features of drop deformation in each phase and the effect of $\text{We}$ will be discussed in subsequent sections.

\subsection {Phase I: Ellipsoid deformation}
In Phase I, the windward surface remains convex, so $L_{fb}=0$. Stagnation flows are formed near the windward and leeward poles (see points W and L in figure~\ref{fig:phase_def}(a)). On the windward surface, the gas pressure is high near the stagnation point and decreases along the surface in the lateral direction. If the viscous effect is ignored, the gas pressure near the stagnation point is given as
\begin{equation}
p_g(r) = p_g(0)-\rho_g \frac{a^2U^2}{8d_0^2}r^2,
\label{eq:press_stag}
\end{equation}
where $r$ is the radial coordinate from the $x$-axis. The stretching rate of the stagnation flow, $a$, varies with the shape of the windward surface geometry, e.g., $a=6$ for a spherical surface and $a=\pi/4$ for a flat disk \citep{Villermaux_2009a}.
The liquid pressure inside the drop can be estimated as the sum of the gas pressure and the Laplace pressure,
\begin{equation}
p_l(r)=p_g(r)+\sigma\kappa,
\label{eq:press_liq}
\end{equation}
which also exhibits a similar variation in $r$ as $p_g$. The gradient of $p_l$ drives the extension of the drop in the radial direction, as shown in figure~\ref{fig:phase_def}(b) at $\hat{t}=0.14$. If the liquid flow is modeled as 1D and inviscid (due to the low $\text{Oh}$ for the present cases), the momentum equation can be expressed as
\begin{equation}
\rho_l \left(\frac{\partial u_r}{\partial t} +u_r \frac{\partial u_r}{\partial r}\right)=-\pd{p_l}{r},
\label{eq:NS_cyl}
\end{equation}
which can be integrated from $r=0$ to $R$ to yield
\begin{equation}
R \ods{R}{t} = \frac{2(p_l(0)-p_l(R))}{\rho_l}.
\label{eq:NS_int}
\end{equation}
Combining Eq.~\eqref{eq:press_liq}, an evolution equation for $\hat{R}$ can be obtained \citep{Villermaux_2009a, Kulkarni_2014a, Jackiw_2021a}:
\begin{equation}
\frac{d^2 \hat{R}}{d\hat{t}^2}\frac{1}{\hat{R}}=\left(\frac{a}{2}\right)^2\left[1-\frac{8}{a^2\text{We}}\frac{(\hat{\kappa}_E-\hat{\kappa}_W)}{\hat{R}^2}\right],
\label{eq:R_evol}
\end{equation}
where $\hat{\kappa}_W$ and $\hat{\kappa}_E$ are the curvatures at the windward pole ($r=0$, point W) and the lateral edge ($r=\hat{R}$, point E).

\subsubsection{Asymptotic early-time dynamics}
At $t=0$, $\hat{\kappa}_W=\hat{\kappa}_E=1/\hat{R}0$, so Eq.~\eqref{eq:R_evol} reduces to
\begin{equation}
\left(\frac{d^2 \hat{R}}{d\hat{t}^2}\right)_{t=0}=\frac{a^2}{8}, ,
\label{eq:R_evol_t0}
\end{equation}
indicating that the drop deformation in the asymptotic limit is independent of $\text{We}$ and is purely determined by $a$. This universal early-time behavior is consistent with the time scale analysis above (\ie, $\tau_{vl} \gg \tau_s > \tau_d$) and is also confirmed by the simulation results, see figure~\ref{fig:early-time}. {It is observed that $d\hat{R}/d\hat{t}$ for all cases collapse very well at early times ($\hat{t}\lesssim0.15$) for the range of $\text{We}$ considered ($10.9\le\text{We}\le 18$).}

Similarly, the results for the drop velocity $u_d$ also collapse in the early time ($\hat{t}\lesssim 0.3$), see figure~\ref{fig:early-time}(b). The initial jump of $u_d$ at $t=0^+$ is observed, which is due to the impulsive gas acceleration and the resulting inviscid-unsteady forces. Therefore, the velocity jump is also independent of $\text{We}$ and $\text{Re}$. The velocity jump, $\hat{u}(t=0^+)\approx0.002$, agrees well with the prediction given by Eq.~\eqref{eq:drop_vel_jump_fiu}. The results from the inviscid compressible flow simulation of \citet{Meng_2018a} are also shown for comparison. In their simulations, the stationary water drop is hit by a planar air shock of Mach number $M_s=1.47$. Due to the compressibility effect, the drop velocity ``jump" occurs in a finite period of time ($\hat{t}\lesssim 0.04$). Nevertheless, the subsequent evolution of $u_d$ agrees well with the present simulations assuming incompressible flows. {The good agreement affirms that the early-time drop dynamics for the range of $\text{We}$ and $\text{Oh}$ considered are independent of viscous and surface tension effects.}

\subsubsection{Effect of $\text{We}$}
The inviscid results of \citet{Meng_2018a} deviate from the present cases at later times $\hat{t}\gtrsim 0.3$ due to the effect of surface tension. Additional 3D simulations were performed for higher $\text{We}$, and it is shown that the present simulation results for $\text{We}=42$ and 72 agree well with the results of the inviscid simulation results for the whole time range shown. This indicates that the compressibility effect on drop breakup induced by weak shocks on the drop velocity is secondary, and the incompressible flow simulations can yield a reasonable estimate for shock-induced drop aerobreakup.

\begin{figure}
\centering
\includegraphics[trim={0cm 0cm 0cm 0cm},clip,width=0.99\textwidth]{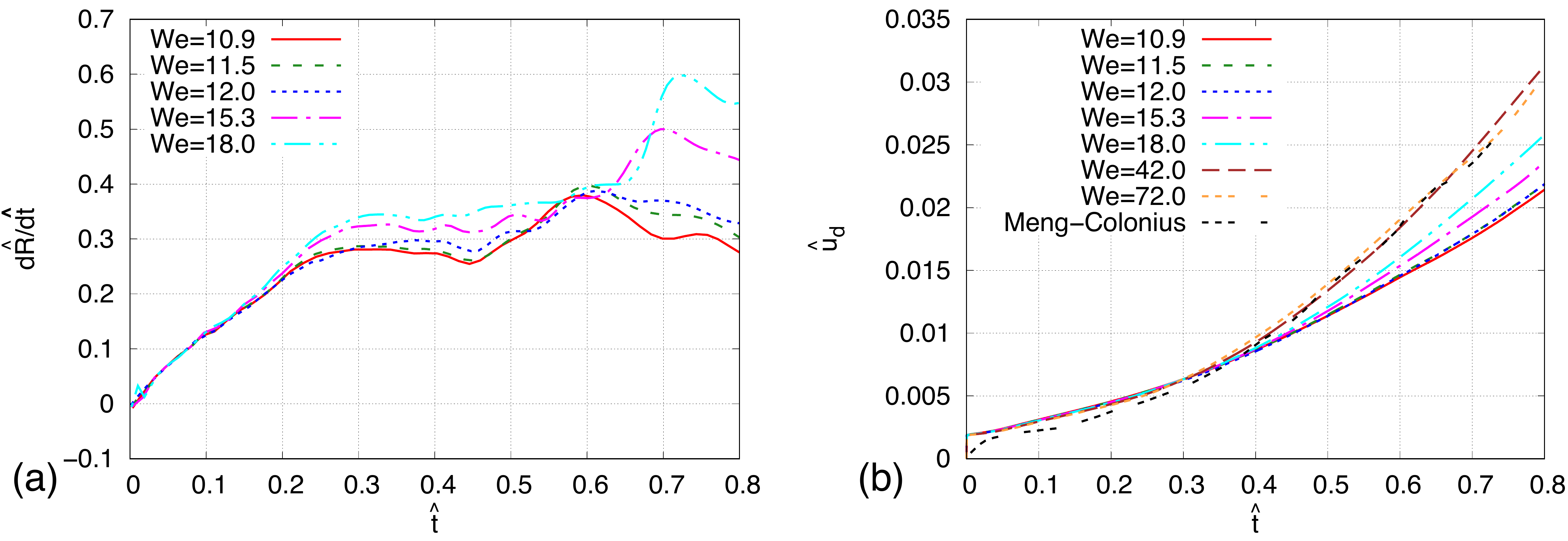}
\caption{Early-time evolution of (a) rate of change of drop radius $d\hat{R}/d\hat{t}$ and (b) drop velocity $\hat{u}_d$ for different $\text{We}$. {The inviscid compressible flow simulation results for shock-droplet interaction by \citet{Meng_2018a} are shown in (b) for comparison.}}
\label{fig:early-time}
\end{figure}

Deviations among different cases for $d\hat{R}/d\hat{t}$ arise after about $\hat{t}=0.2$, as shown in figure~\ref{fig:early-time}(a). It can be observed that $d\hat{R}/d\hat{t}$ for different $\text{We}$ eventually reaches different plateau values. As the curvatures of the windward surface decrease, the stagnation flow is modified, and the pressure gradient that drives the lateral expansion is reduced. Simultaneously, the curvature at the lateral edge increases, and as a result, the retraction force due to surface tension increases. The balance between the surface tension and stagnation pressure brings $d\hat{R}/d\hat{t}$ to a constant. For cases with larger $\text{We}$, the retraction force from the lateral edge is smaller, and thus it takes slightly longer for $d\hat{R}/d\hat{t}$ to reach the plateau, and the plateau values of $d\hat{R}/d\hat{t}$ are also higher.

The above discussions mainly focus on the windward surface. Due to flow separation, the pressure field on the leeward side is different from the windward counterpart, as shown in figure~\ref{fig:phase_def}(a). Generally, the pressure at the leeward pole is lower. Nevertheless, the pressure decreases radially even faster due to the vortex ring formed behind the separation point. As a result, the leeward surface curvature decreases faster and turns from convex to concave earlier ($\hat{t}=0.28$) than the windward counterpart.

\subsubsection{{Modeling early-time} evolution of drop radius $R$}
The evolution of the drop radius $R$ is mainly dictated by windward surface deformation. The shape of the drop can be approximated as ellipsoidal. Then $\kappa_W$ and $\kappa_E$ can be expressed in terms of $R$ as $\kappa_W = h_a/R^2$ and $\kappa_E=4R/h_a^2 + 1/R$, where $h_a=d_0^3/4R^2$ is the thickness of the drop for an ellipsoid. By substituting the estimates of curvatures into Eq.~\eqref{eq:R_evol}, we obtain:
\begin{equation}
    \frac{d^2 \hat{R}}{d\hat{t}^2}\frac{1}{\hat{R}}=\left(\frac{a}{2}\right)^2\left[1-\frac{64}{a^2\text{We}}\left( 8 \hat{R}^3 +\frac{1}{8 \hat{R}^3} -\frac{1}{32 \hat{R}^6} \right) \right]\, .
    \label{eq:R_evol2}
\end{equation}
Similar models have been developed by \citet{Villermaux_2009a} (VB) and were then extended by \citet{Kulkarni_2014a} (KS) and \citet{Jackiw_2021a} (JA). The difference between the present model (Eq.\eqref{eq:R_evol2}) and the previous ones lies in the different estimates of $\kappa_W$ and $\kappa_E$. In the VB and KS models, the drop is assumed to be a cylindrical disk with a rounded edge, so the windward surface is flat ($\kappa_W=0$), and only the principal curvature on the $x$-$y$ plane is considered for the edge, namely $\kappa_E=2/h_a$, where $h_a=d_0^3/6R^2$ for a cylinder. As a result, Eq.\eqref{eq:R_evol} becomes
\begin{equation}
    \left(\frac{d^2 \hat{R}}{d\hat{t}^2}\frac{1}{\hat{R}}\right)_{KS}=\left(\frac{a}{2}\right)^2\left[1-\frac{96}{a^2\text{We}} \right]\, .
    \label{eq:R_evol_KS}
\end{equation}
Note that the above expression holds for both the VB and KS models, the difference between the two lies in the different values of $a$. In the JA model, the drop was first considered as a disk, so $\kappa_W=0$ and $\kappa_E=2/h_a+1/R$, with the second principle curvature added compared to the KS model. Nevertheless, when relating $h_a$ with $R$, they have used the relation for an ellipsoid, so $h_a=d_0^3/4R^2$. Eventually, they obtained:
\begin{equation}
    \left(\frac{d^2 \hat{R}}{d\hat{t}^2}\frac{1}{\hat{R}}\right)_{JA}=\left(\frac{a}{2}\right)^2\left[1-\frac{64}{a^2\text{We}} \left( 1+ \frac{1}{8\hat{R}^3}\right) \right]\, .
    \label{eq:R_evol_JA}
\end{equation}
Based on experimental observations, they argued that for a short time $\hat{t}<\hat{t}_{bal}=1/8$, the drop radius $R$ remains unchanged, and then $R$ increases linearly. Therefore, Eq.~\eqref{eq:R_evol_JA} can be further simplified as:
\begin{equation}
    \label{eq:R_evol_JA2}
%\[
    \left(\frac{d \hat{R}}{d\hat{t}}\right)_{JA2}= 
\begin{cases}
     0\, ,& \text{if } \hat{t}\le \hat{t}_{bal}, \\
      \left(\frac{a}{2}\right)^2\left[1-\frac{128}{a^2\text{We}} \right] \frac{\hat{t}_{bal}}{2},              & \text{if } \hat{t}> \hat{t}_{bal}, 
\end{cases}
%\]
\end{equation}
which is referred to as JA2 model here, to distinguish it from Eq.~\eqref{eq:R_evol_JA}. 
In the present model, we have used the geometric features of an ellipsoid to estimate $\kappa_W$, $\kappa_E$, and the relation between $h_a$ and $R$ consistently. This treatment leads to an important feature, which is the \text{We}-independent asymptotic limit at $t=0$ (see Eq.~\eqref{eq:R_evol_t0}), being correctly captured. It can be easily shown that none of the above models guarantee this feature.

\begin{figure}
\centering
\includegraphics[trim={0cm 0cm 0cm 0cm},clip,width=0.99\textwidth]{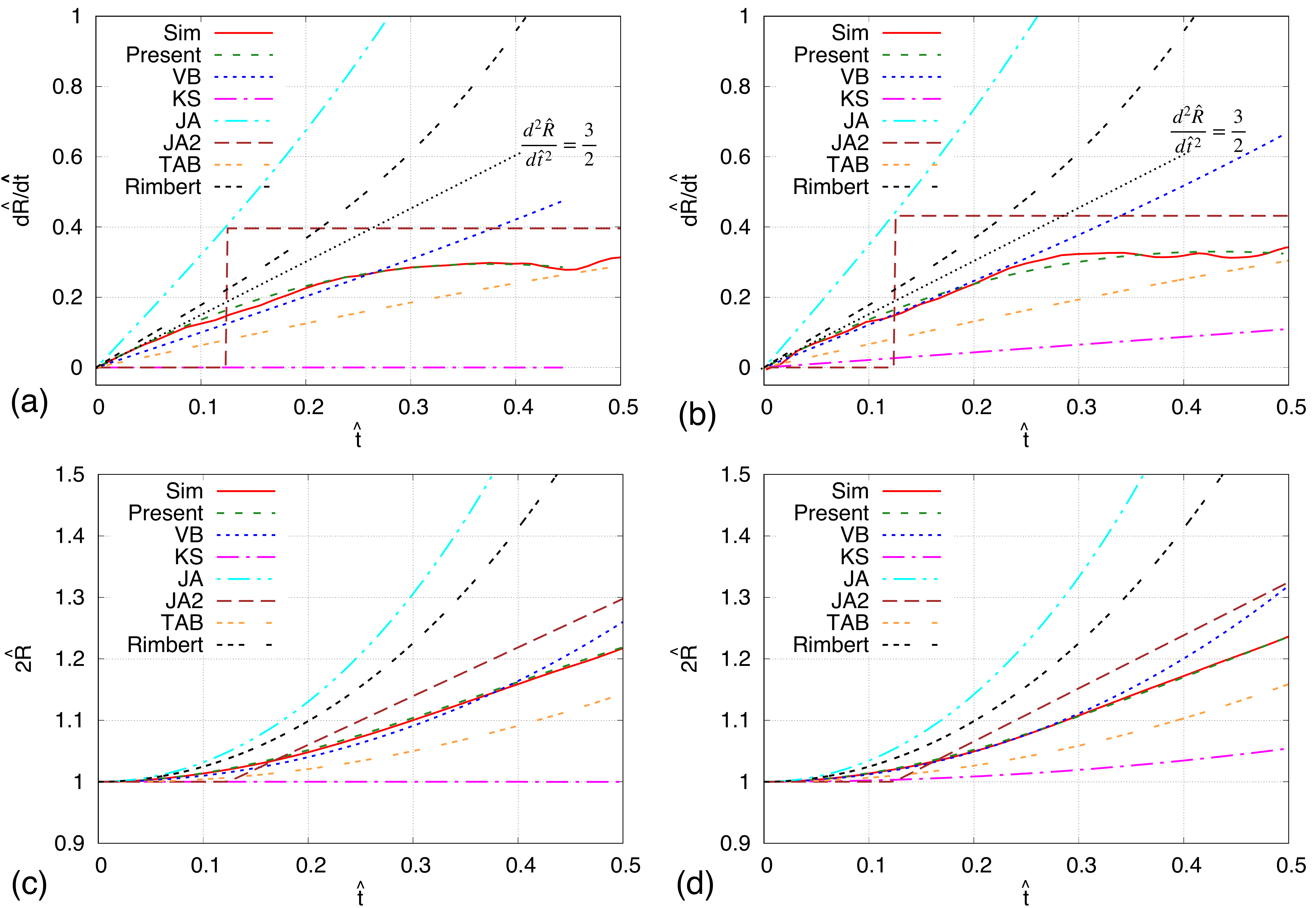}
\caption{Comparison between the present simulation and model results for $d\hat{R}/d\hat{t}$ and $\hat{R}$ with predictions of other models for (a,c) $\text{We}=12.0$ and (b,d) $\text{We}=15.3$. {The results for other models are shown as well: VB, KS, TAB, Rimbert represent the models of \citet{Villermaux_2009a}, \citet{Kulkarni_2014a}, \citet{ORourke_1987a}, \cite{Rimbert_2020o}, while JA and JA2 represent the two models of   \citet{Jackiw_2021a} (Eqs.~\eqref{eq:R_evol_JA} and \eqref{eq:R_evol_JA2}). }}
\label{fig:compare_model}
\end{figure}

To close the above models, the stretching rate of the stagnation flow, $a$, remains to be determined. In the VB model, $a$ is taken to be 4, which is an arbitrary intermediate chosen between the two limits $a=6$ and $\pi/4$ for sphere and cylinder. Both VB and KS proposed that the condition $d^2R/dt^2 =0$ can be used to determine the critical Weber number, $\text{We}_{cr}$, \ie, when $\text{We}<\text{We}_{cr}$, $d^2R/dt^2<0$, then $R$ will not increase over time and the drop will be stable. The value $a=4$ used by VB will lead to $\text{We}_{cr}=6$, which does not agree with experimental observations for drop aerobreakup. That is why KS proposed $a=2\sqrt{2}$, so that according to their model (Eq.~\eqref{eq:R_evol_KS}) $d^2R/dt^2 =0$ will yield $\text{We}_{cr}=12$, which agrees with the experimental observations. Nevertheless, this way to estimate $a$ is problematic since both experimental and simulation results show that $d^2R/dt^2>0$ is not a necessary condition for the drop to be unstable. A good counterexample is the case $\text{We}=10.9$, for which $d^2R/dt^2>0$ at an early time. However, the drop is eventually stable. Generally, the KS model significantly under-predicts the temporal growth of $R$ at an early time, see figures~\ref{fig:compare_model}(b) and (d). In particular, for the case $\text{We}=12.0$, the KS model will yield $d^2R/dt^2=0$, and as a result, $R$ would not grow in time, see figures~\ref{fig:compare_model}(a) and (c), which is obviously incorrect.

In the JA model, the value of $a$ for a sphere was used, \ie, $a=6$. In contrast to the KS model, the JA model significantly overpredicts the evolution of $\hat{R}$ and $d\hat{R}/d\hat{t}$. 
{
Among the VB, KS, and JA models, the VB model actually matches the best with the present simulation results for the initial development of $\hat{R}$ and $d\hat{R}/d\hat{t}$ for $\hat{t}\lesssim 0.25$. Nevertheless, a common issue of all these models is that they fail to capture the trend that $d\hat{R}/d\hat{t}$ gradually reaches a plateau around $\hat{t}\approx 0.3$.
The constant $d\hat{R}/d\hat{t}$ was observed in the experiments of JA, which motivated them to add this assumption in the original JA model, leading to the simplified JA2 model that approximates $d\hat{R}/d\hat{t}$ as a piece-wise constant function of time, see Eq.~\eqref{eq:R_evol_JA2}. This simplification reduces the discrepancy between the model predictions and the simulation results. However, as shown in figures\ref{fig:compare_model}(a-b), the evolution of $d\hat{R}/d\hat{t}$ in reality does not change abruptly.
}

{
In a recent study by \citet{Rimbert_2020o}, the potential flow solution around a spheroid was used to estimate the gas pressure on the drop surface. This approach does not require a priori knowledge of the stretching rate of the stagnation flow, $a$, and the time-varying pressure work on the drop is expressed as a function of $\hat{R}$. By introducing a small perturbation at the initial spherical state, $\hat{\epsilon}=\hat{R}-0.5$, the linearized model is given as follows:
\begin{equation}
\ods{\hat{\epsilon}}{\hat{t}}+20 \frac{\text{Oh}}{\sqrt{\text{We}}} \od{\hat{\epsilon}}{\hat{t}} + \left(\frac{32}{\text{We}} -\frac{342}{35} \right)\hat{\epsilon} = \frac{3}{2}.
\end{equation}
The constant $3/2$ on the right-hand side is from the pressure work of the gas flow, estimated from the potential flow solution. At time zero, $\hat{\epsilon}=d\hat{\epsilon}/d\hat{t}=0$, the above equation reduces to
\begin{equation}
\left(\ods{\hat{R}}{t}\right)_{t=0} = \left(\ods{\epsilon}{t}\right)_{t=0} = \frac{3}{2},
\label{eq:Rimbert_t0}
\end{equation}
which gives the initial slope of $d\hat{R}/d\hat{t}$ that is independent of $\text{We}$ and $\text{Oh}$. It can be observed from figures~\ref{fig:compare_model}(a) and (b) that the estimated slope matches very well with the present simulation results. However, the model of \citet{Rimbert_2020o} overpredicts the subsequent evolution of $d\hat{R}/d\hat{t}$ and $\hat{R}$, which may be related to their assumption that the drop shape is a spheroid.
}

{To predict the early-time evolution of $\hat{R}$, a simple model is proposed here by incorporating a time varying $a$. First of all, by combining Eqs.~\eqref{eq:R_evol_t0} and \eqref{eq:Rimbert_t0} we can determine the initial value of $a$, \ie, $a_0=a(t=0)=\sqrt{(d^2R/dt^2)_{t=0}}= \sqrt{12}$. The value is lower than 6 even though the initial drop shape is spherical, and the difference is related to the assumptions made in the model, such as the flow in the drop is only 1D in the radial direction. As the drop deforms and the windward surface curvature decreases, 
}
$a$ decreases over time and eventually reaches  $a_1$. The time variation of $a$ is simply modeled by an error function as, 
\begin{equation}
	a = a_0 -  (a_0-a_1) \mathrm{erf} (\hat{t}/\hat{\tau}_a)\, , 
	\label{eq:stretch_rate2}
\end{equation}
where $a_1=2.5$ is the asymptotic value of $a$ at the end of Phase I and $\tau_a=0.35$ is the transition time, found based on the simulation results. Though the present model (Eqs.~\eqref{eq:R_evol2} and \eqref{eq:stretch_rate2}) require simulation results to calibrate the model parameters, its predictions agree very well with the simulation results, see figure~\ref{fig:compare_model}. It is also worth noting that Eq.~\eqref{eq:stretch_rate2} is independent of $\text{We}$ and thus can be used predict the evolution of $R$ in Phase I for other values $\text{We}$ that are not simulated here. 

{
Finally, as the Taylor analogy breakup (TAB) model \citep{ORourke_1987a} is widely used in the literature, its results are also shown in figure~\ref{fig:compare_model} for comparison. It can be observed that the TAB model underpredicts the initial growths of $d\hat{R}/d\hat{t}$ and $\hat{R}$.
}

\subsection{Phase II: Disk formation}
A distinct feature of Phase II is that $d\hat{R}/d\hat{t}$ increases again after it reaches the plateau at the end of Phase I, as shown in figure~\ref{fig:phase_def}(d). As shown in figure~\ref{fig:phase_def}(a), as the pressure at point E increases, the growth of $d\hat{R}/d\hat{t}$ slows down at the end of Phase I. In Phase II, the liquid turns to flow toward a new low-pressure location, point N marked in the figure. The low pressure is related to small curvature at point N on the $x$-$y$ plane, which is in turn related to the deformation of the leeward surface in Phase I. It can be observed that the $y$-velocity at point N increases faster than that at point E, as shown for $\hat{t}=0.42$ to 0.68 in figure~\ref{fig:phase_def}(b). As a result, the lateral extension velocity of the drop increases again. At around $\hat{t}=0.56$, the point N passed the point R, becoming the new location determining $R$. As the drop deforms, the curvature at point N increases and the drop gradually deforms from an approximate ellipse to a disk with rounded edge. At the end of Phase II, the Laplace pressure at the new periphery N becomes comparable to the stagnation pressure near the windward pole. Then $d^2\hat{R}/d\hat{t}^2=0$, and $d\hat{R}/d\hat{t}$ reaches a local maximum.

It is worth mentioning that the model given in Eq.~\eqref{eq:R_evol} can also be used to predict the drop deformation in this phase, with certain adaptation. The additional complexity is to estimate the time variation of the principal curvature $\kappa_N$ on the $x$-$y$ plane since the shape of the drop is neither an ellipse nor a disk in the transition. In this study, we focus on discussions of the flow physics and will leave the model development for future work.

{
The appearance of a local maximum in $d\hat{R}/d\hat{t}$ is a distinctive feature of the bag breakup regime. As $\text{We}$ increases and the drop breakup mode transitions to the multi-mode regime, $d\hat{R}/d\hat{t}$ will monotonically increase. Here, we denote the time corresponding to $d^2\hat{R}/d\hat{t}^2=0$ in phase II as $\hat{t}^*$. The zero acceleration of $\hat{R}$ indicates a force balance at the drop periphery. When $\text{We}$ is too large, the surface tension will not be sufficient to establish such a balance. \citet{Marcotte_2019a} suggested that the force balance at the drop periphery is achieved when the rim is formed, which is consistent with the present results. They also suggested that rim formation is a critical feature in determining whether the drop will break in the bag modes (RTP mode) or in the shear modes (SIE mode). Nevertheless, the present results show that rim formation will not be achieved even in the multi-mode regime, and that $\text{We}$ does not need to reach high values in the shear breakup regime to prevent rim formation.
}

\begin{figure}
\centering
	\includegraphics[trim={0cm 0cm 0cm 0cm},clip,width=0.99\textwidth]{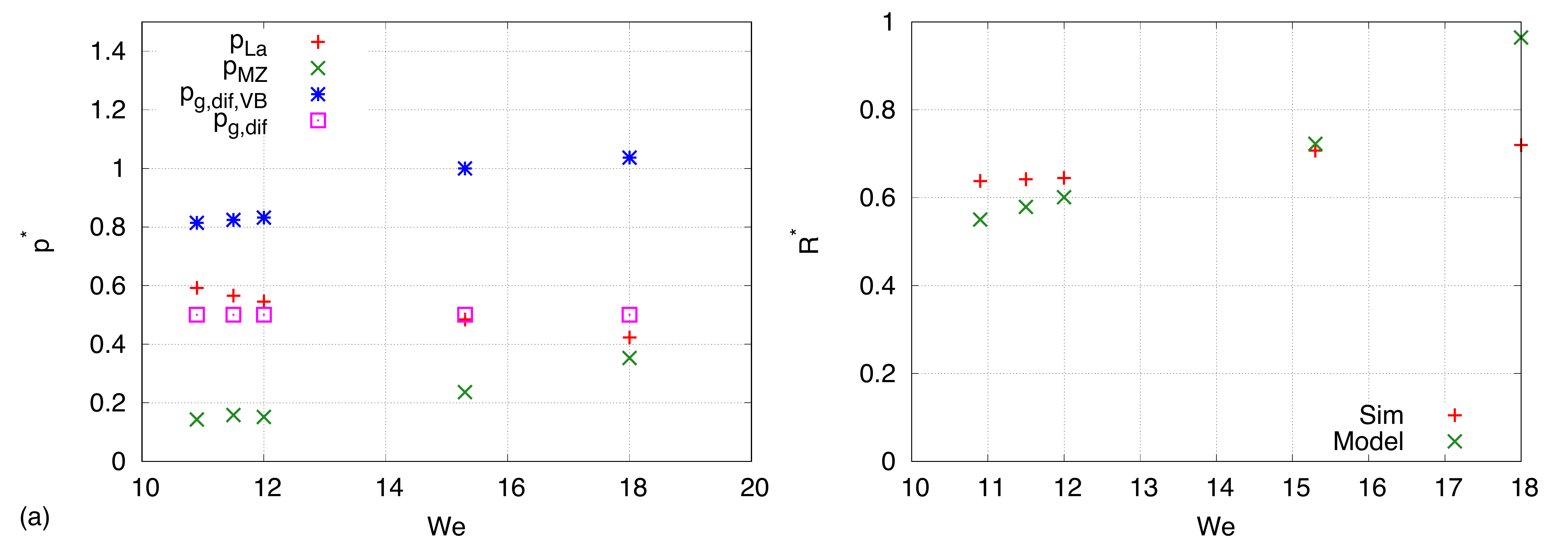}
\caption{(a) Comparison of Laplace pressure (Eq.~\eqref{eq:rim_p_La}) with liquid inertia (Eq.~\eqref{eq:rim_p_In}) and gas-pressure differences estimated by the VB (Eqs.~\eqref{eq:rim_p_gd_VB}) and present models (\eqref{eq:rim_p_gd}) as a function of $\text{We}$ when the rim is formed. {(b) Comparison of the drop lateral radius when the rim is formed, $\hat{R}^*$, measured from the present simulation and the model prediction (Eq.~\eqref{eq:drop_radius_rim_form}).}}
\label{fig:rim_formation}
\end{figure}

The Laplace pressure at the drop periphery at $\hat{t}^*$ can be estimated by assuming that the drop exhibits the shape of a disk:
\begin{equation}
\hat{p}_{La}^* = \frac{\sigma \kappa_N}{\rho_g U_0^2} = \frac{1}{\text{We}} \left(12\hat{R^*}^2 + \frac{1}{\hat{R^*}}\right),
\label{eq:rim_p_La}
\end{equation}
where $\hat{R}^*$ is the dimensionless drop radius for $\hat{t}=\hat{t}^*$. The values of $\hat{p}_{La}^*$ for different $\text{We}$ are shown in figure~\ref{fig:rim_formation}. \citet{Marcotte_2019a} suggested that the Laplace pressure when the rim is formed is in equilibrium with the liquid fluid inertia:
\begin{equation}
\hat{p}_{MZ}^* = \frac{\rho_l (dR/dt)^2}{\rho_g U_0^2} = \left(\od{\hat{R}}{\hat{t}}\right)^2.
\label{eq:rim_p_In}
\end{equation}
They further proposed to approximate $dR/dt$ by the Dimotakis speed, namely $U_0\sqrt{r}$. However, it can be observed in figure~\ref{fig:early-time}(a) that $U_0\sqrt{r}$ significantly over-predicts $d\hat{R}/d\hat{t}$ at $\hat{t}^*$. By using the corrected $(d\hat{R}/d\hat{t})$ obtained from simulations, it can be observed from figure~\ref{fig:rim_formation} that the liquid inertia is significantly lower than $\hat{p}_{La}^*$, and thus is unlikely to be the force that balances the Laplace pressure when the rim is formed.

According to the 1D inviscid model (Eq.\eqref{eq:NS_int}), when $d^2\hat{R}/d\hat{t}^2=0$, the liquid pressure at the disk center and the rim should be in equilibrium, $p_l(0)= p_l(R)$. Since the curvature at the windward pole is approximately zero at this time, $p_l(0)\approx p_g(0)$, see also figure~\ref{fig:phase_def}(a) at $\hat{t}=0.68$. The liquid pressure at the periphery, $p_l(R)=p_g(R)+p_{La}$. As a result, the Laplace pressure at the rim is actually balanced by the gas pressure difference $\hat{p}_{g,dif}=p_g(0)-p_g(R)$. \citet{Villermaux_2009a} have proposed estimating $p_g(R)$ using the stagnation pressure on the windward surface, namely Eq.\eqref{eq:press_stag},
\begin{equation}
\left(\hat{p}_{g,dif}^*\right)_{VB}= \frac{p_g(0)-p_g(R)}{\rho_g U_0^2} = \frac{a^2\hat{R}^2}{8},
\label{eq:rim_p_gd_VB}
\end{equation}
where $a=\pi/4$ for a disk. However, the computed pressure field (see figure~\ref{fig:phase_def}(a)) indicates that the gas pressure above the periphery of the drop, $p_g(R)$, is only dictated by stagnation flow at the windward surface when the drop exhibits an ellipsoidal shape in Phase I. When the drop becomes a disk in Phases II and III, the gas flows over the drop periphery without much influence from the windward surface, and as a result, $p_g(R) \approx 0$. Therefore, we estimate the gas pressure difference between the disk center and periphery as
\begin{equation}
\hat{p}_{g,dif}^*= \frac{p_g(0)}{\rho_g U_0^2} = \frac{1}{2}.
\label{eq:rim_p_gd}
\end{equation}

The results of $(\hat{p}_{g,dif}^*)_{VB}$ and $\hat{p}_{g,dif}^*$ are shown in figure~\ref{fig:rim_formation} and it can be observed that the present estimate $\hat{p}_{g,dif}^*$ agree better with $\hat{p}_{La}^*$, affirming that the rim formation and the maximal radial expansion velocity before inflation is established when the gas stagnation pressure at the windward pole is in equilibrium with the Laplace pressure at the periphery rim. Furthermore, the pressure balance $\hat{p}_{La}^*=\hat{p}_{g,dif}^*$ yields 
\begin{equation}
	 \frac{1}{\text{We}} \left(12\hat{R^*}^2 + \frac{1}{\hat{R^*}}\right)=  \frac{1}{2}\,, 
	 \label{eq:drop_radius_rim_form}
\end{equation}
{
which can be used to predict $\hat{R^*}$ as a function of $\text{We}$ for the bag breakup regime.
Figure~\ref{fig:rim_formation}(b) shows the comparison between the simulation results for $\hat{R^*}$ and the prediction from the model. It can be observed that the increasing trend of $\hat{R^*}$ with $\text{We}$ is captured. The agreement between the model predictions and the simulation measurements is generally good for $\text{We}=10.9$ to 15. However, the discrepancy is much larger for $\text{We}=18$, since this case is in the transition to the multi-mode regime.}

\subsection{Phase III Disk deformation}
After $d\hat{R}/d\hat{t}$ reaches a local maximum at the end of Phase II, it starts to decrease over time in Phase III, when the disk deforms, see figure~\ref{fig:phase_def}(d). {The Rayleigh-Taylor (RT) instability contributes to the subsequent deformation of the windward and leeward surfaces of the disk.} As the drop accelerates toward the right, the baroclinic torque induced by the misalignment between the pressure and density gradients destabilizes the windward surface of the disk and stabilizes the leeward counterpart, see $\hat{t}=0.92$ in figure~\ref{fig:phase_def}. The deformation of the two surfaces squeezes the liquid to move from the disk center towards the edge rim, see the distribution of $\hat{u}_y$ in figure~\ref{fig:phase_def}(b), resulting in a rapid decrease of disk thickness at the center $h_a$ and an increase in $\hat{R}$.

There are two mechanisms induced by surface tension that hinder the disk deformation, resulting in the deceleration of the disk edge in the radial direction ($d^2\hat{R}/d\hat{t}^2<0$). The first mechanism is the surface tension corresponding to the deformation of the windward surface and the resulting curvature on the $x$-$y$ plane. This surface force is accounted for in the stability analysis of RTI on a flat surface or a planar liquid layer with surface tension \citep{Mikaelian_1990a, Mikaelian_1996i}, which is used to predict the most unstable wavelengths on the windward surface \citep{Joseph_1999a,Theofanous_2004a}. The second mechanism is the surface force from the rounded edge of the disk. As the radius of curvature is smaller at the lateral edge, a higher pressure develops at the edge, and the resulting pressure gradient resists the disk's lateral expansion, as shown in figure~\ref{fig:phase_def} at $\hat{t}=1.11$. Similar to Phases I and II, the second mechanism is the dominant one, and the competition between this mechanism and the RT baroclinic torque {seems to} determine the drop deformation in this phase.

\begin{figure}
\centering
\includegraphics[trim={0cm 0cm 0cm 0cm},clip,width=0.99\textwidth]{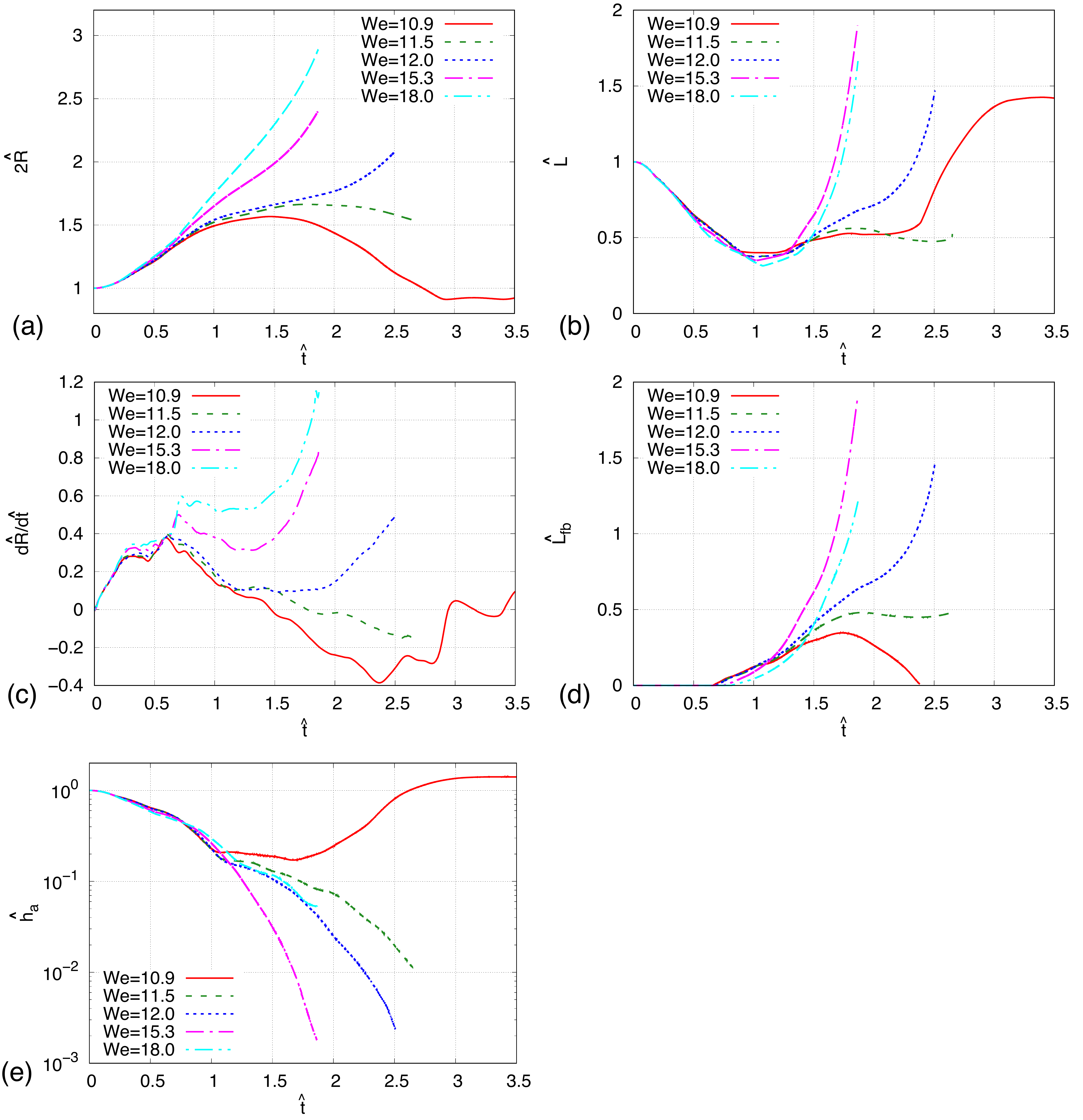}
\caption{Time evolutions of (a) the drop radius $\hat{R}$, (b) the drop length $\hat{L}$, (c) the rate of change of $\hat{R}$, (d) the bag length $L_{fb}$, and (e) the bag sheet thickness along $x$ axis, for different $\text{We}$. }
\label{fig:deform_bag}
\end{figure}

Though the effect of $\text{We}$ can be already seen from $d\hat{R}/d\hat{t}$ at the end of Phase II, the differences in the drop shapes do not become obvious until Phase III, see the temporal evolutions of the characteristic length scales in figure~\ref{fig:deform_bag}. Phase III is characterized by the decrease of $d\hat{R}/d\hat{t}$ in time, and it is observed that Phase III roughly starts around $\hat{t}= 0.7$ and ends around $\hat{t}=1.2$ for all cases except $\text{We}=10.9$. 

The evolutions of $\hat{L}_{fb}$ and $\hat{h}_a$ in Phase III  also showed distinct behaviors, see figure~\ref{fig:deform_bag}. When the drop windward/leeward surfaces becomes concave, it is hard to measure these two parameters in experiment, but which is easy in the simulation.  Due to the definition of $\hat{L}_{fb}$, it remains zero until the windward surface becomes concave in Phase III. 
{
Though the RTI contributes to the deformation of the windward surface, the growth of $\hat{L}_{fb}$ in the linear regime (when $\hat{L}_{fb}$ is small) is linear instead of exponential. While an exponential growth is expected for the linear RTI on an infinitely flat layer of finite thickness, a possible reason for not observing it here is the surface tension effect at the edge of the disk. The rapid growth of $\hat{L}_{fb}$ for $\text{We}=15.3$ and 18.0 at later times may appear to be exponential, but it is due to the bag inflation, which will be discussed later, and is not related to the linear stability development.
} 
It can be seen that $\hat{h}_a$ decreases monotonically in time, but the rate of decrease in Phase III is higher than that in Phase II. In addition, kinks are observed in the evolutions of $h_a$ for cases with $\text{We}\le 12$ at around $\hat{t}=1.1$, after which the rate of decrease is significantly reduced. For the case $\text{We}=10.9$, the slow decrease of $\hat{h}_a$ prevents it from reaching the threshold of about 0.2 (\ie, the disk thickness remains larger than 20\% of the original drop diameter). As will be shown later, the thick layer at the disk center hinders the bag growth, and a topology change never happens for this case. {It is noteworthy that even a slight increase in $\text{We}$, for instance, from $\text{We}=11.5$ to 12.0, can cause $\hat{h}_a$ to surpass the critical threshold, leading to the eventual piercing of the bag.}

\begin{figure}
\centering
\includegraphics[trim={0cm 0cm 0cm 0cm},clip,width=0.99\textwidth]{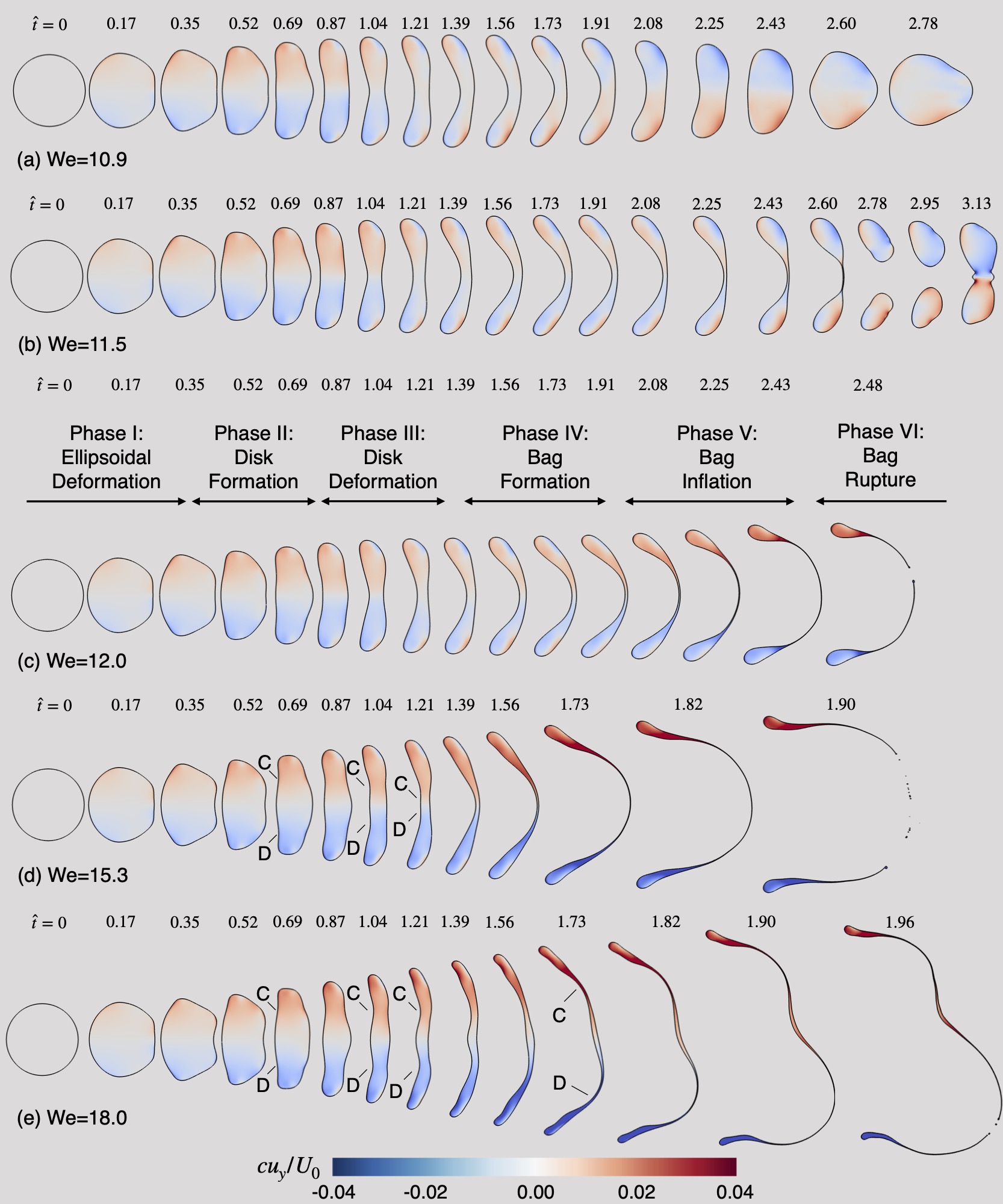}
\caption{Temporal evolution of the $y$-velocity ($u_y $) on the central $x$-$y$ plane inside the drop for different $\text{We}$. The points C and D marked in (d) and (e) indicates the concave locations on the perturbed windward surface due to drop deformation. Different deformation phases for $\text{We}=12.0$ are indicated in (c). }
\label{fig:uy_liq}
\end{figure}

To better illustrate the flow inside the drop and its impact on the drop deformation, the time evolutions of $\hat{u}_y$ inside the drop for different $\text{We}$ are also shown in figure~\ref{fig:uy_liq}. It is confirmed that the velocity distributions within the drop for different $\text{We}$ are generally similar for Phases I and II, and distinctions among different cases arise in Phase III. Due to the approximate flow symmetry, $\hat{u}_y$ is close to zero near the center of the disk ($y=0$) and is generally positive and negative in the top and bottom portions of the drop in Phases I and II for all cases, indicating the liquid flows outward from the disk center. The signs of $\hat{u}_y$ are generally similar in Phase III, but the radial flow is not driven by the liquid pressure gradient inside the drop as in Phases I and II, but by the RTI on the windward surface. As shown in figure~\ref{fig:phase_def}(a) at $\hat{t}=1.11$, the pressure actually increases along the radial direction and drives an inward flow back toward the center, as can be seen in the small regions of $\hat{u}_y<0$ near the right top corner of the disk for $\hat{t}=1.04$. For $\text{We}=10.9$ and 11.5, the inward flow gradually overcomes the outward counterpart, as the region for $\hat{u}_y<0$ in the upper half of the drop grows in time. Eventually, the lateral deformation of the disk changes from expansion to contraction. For $\text{We}\le 12$, the outward flow dominates, and the liquid continues to flow away from the center, leading to a continuous decrease of $h_a$, as shown in figure~\ref{fig:deform_bag}.

\subsection{Phase IV: Bag development}
The evolution of $d\hat{R}/d\hat{t}$ shown in figure~\ref{fig:phase_def}(d) also exhibits a phase for which $d\hat{R}/d\hat{t}$ is approximately constant, referred to as Phase IV. In this phase, the thin center of the disk starts to bend toward downstream, forming a forward bag. Here, we define a ``forward bag" as a curved liquid layer with the opening facing upstream. Based on this definition, a ``bag" is formed for all cases, see figure~\ref{fig:uy_liq}. The formation of the bag is due to the baroclinic torque on the unstable windward surface overcoming its counterpart on the stable leeward surface, which in turn is caused by the larger pressure difference on the windward surface. The deformation of the bag enhances liquid flow from the disk center to the edge, balancing the forces at the edge rim of the disk such that $d^2\hat{R}/d\hat{t}^2=0$. Phase IV ends when the bag inflates and $d\hat{R}/d\hat{t}$ increases rapidly again.

\subsubsection{Criteria for bag piercing}
The evolutions of the characteristic lengths and internal flows for different $\text{We}$ are shown in figures~\ref{fig:deform_bag} and \ref{fig:uy_liq}. It can be seen that the constant $d\hat{R}/d\hat{t}$ is only obvious for $\text{We}\le 12$ and thus is a near-$\text{We}_{cr}$ feature. For cases with higher $\text{We}$, the bags start to inflate right after they are formed. Among the three lower $\text{We}$ cases considered here, the bag for $\text{We}=10.9$ is stable, the edge of the bag retracts ($d\hat{R}/d\hat{t}<0$), and the thickness of the bag tunrs to increase in time ($d\hat{h}_a/d\hat{t}>0$). Topology change does not happen (vibration breakup may occur in a long time but which is out of scope of the study). In contrast, for $\text{We}=12.0$, the bag is unstable, $\hat{R}$ increases and $\hat{h}_a$ decreases, and eventually the bag inflates and is pierced through by the gas flow. The results for $\text{We}=11.5$ show {an interesting breakup mode that is less understood}, \ie, both $\hat{R}$ and $\hat{h}_a$ decrease over time. As a result, the bag will rupture similar to the unstable case $\text{We}=12.0$, but the remaining rim retracts back to form a big drop like the stable case $\text{We}=10.9$. Since the drop for $\text{We}=11.5$ will not fragment into a large number of small drops, we consider the case $\text{We}=12.0$ to represent the dynamics of aerobreakup at critical conditions and thus $\text{We}_{cr}=12.0$, which is consistent with previous experiments for water drops. Nevertheless, the results for $\text{We}=11.5$ show that bag piercing does not guarantee a complete fragmentation of the bag. The range of $\text{We}$ for such behavior is quite narrow for the density ratios and Reynolds numbers considered, and further investigation of {this breakup mode} will be relegated to future study.

The linear RTI model with surface tension and viscous effects for an accelerating infinite planar liquid sheet has been proposed to predict the critical condition, such as $\text{We}_{cr}$ \citep{Joseph_1999a, Theofanous_2004a, Theofanous_2012a}, based on the linear stability analysis \citep{Mikaelian_1990a, Mikaelian_1996i}. It was suggested that when the most unstable wavelength is smaller than the disk diameter, a bag will be formed and eventually pierced through by the gas. Although such models, along with other assumptions, reasonably capture the variation of $\text{We}_{cr}$ over $\text{Oh}$, the hypotheses of the model are not fully consistent with the flow physics. First of all, the effect of surface tension at the disk edge was not considered in the model. This simplification is acceptable for perturbations of small wavelengths but not for those with wavelengths comparable to the disk radius $R$. Furthermore, the previous models predict disk/bag stability based on the linear dynamics of RTI. However, a perturbation that grows in the linear regime (when its amplitude $\hat{L}_{fb}$ is small) does not guarantee its continuous nonlinear growth at later times, and the case $\text{We}=10.9$ explained above is a good example for that.

For cases with $\text{We}\le15.3$, the forward bag formed is approximately a symmetric spherical shell, and the tip of the bag is located near the central $x$-axis. Such bags are referred to as ``simple" bags here, and theoretical models have been proposed to predict its formation and development \cite{Villermaux_2009a, Kulkarni_2014a}. It is considered that the acceleration of the bag tip, which is perpendicular to the free stream, is driven by the pressure difference $(\Delta p)_a$ across the liquid sheet, 
\begin{align}
	\ods{L_{fb}}{t} = \frac{(\Delta p)_a}{\rho_l h_a}\,. 
	\label{eq:bag_tip}
\end{align}
In this simple model, surface tension and liquid viscosity are ignored. Extensions to incorporate these effects have been made by \citet{Kulkarni_2014a}. Since the pressure is hard to measure in experiments, \citet{Villermaux_2009a} approximates the pressure on the leeward side to be the free stream pressure, and that on the windward side to be the stagnation pressure, so $(\Delta p)_a \approx p_g(r=0)$, as given in Eq.~\eqref{eq:press_stag}. For cases with different $\text{We}$, $(\Delta p)_a$ is similar, so the different evolution of $L_{fb}$ is mainly due to the different $h_a$. The rate of decrease of $h_a$ generally increases with $\text{We}$. As a result, $h_a$ for low $\text{We}$ can be significantly larger than that for high $\text{We}$; for example, at $\hat{t}=1.5$, $h_a=0.2$ and 0.03 for $\text{We}=10.5$ and 15.3, respectively, and the former is almost an order of magnitude larger than the latter. If $h_a$ is too large, then the tip acceleration is slow compared to the capillary retraction of the edge rim, and $h_a$ will stop decreasing, and the bag will not be pierced through. For the present cases, the evolutions of $h_a$ for $\text{We}=10.5$ and 11.0 bifurcate at $\hat{h}_a=0.1-0.2$, which seems to indicate the $h_a$ threshold for bag piercing is about 0.2. The value may vary with other parameters that are kept constant in the study, and further investigations are still required to fully confirm this observation.

\subsubsection{Transition from simple to ring bags}
\begin{figure}
\centering
\includegraphics[trim={0cm 0cm 0cm 0cm},clip,width=0.89\textwidth]{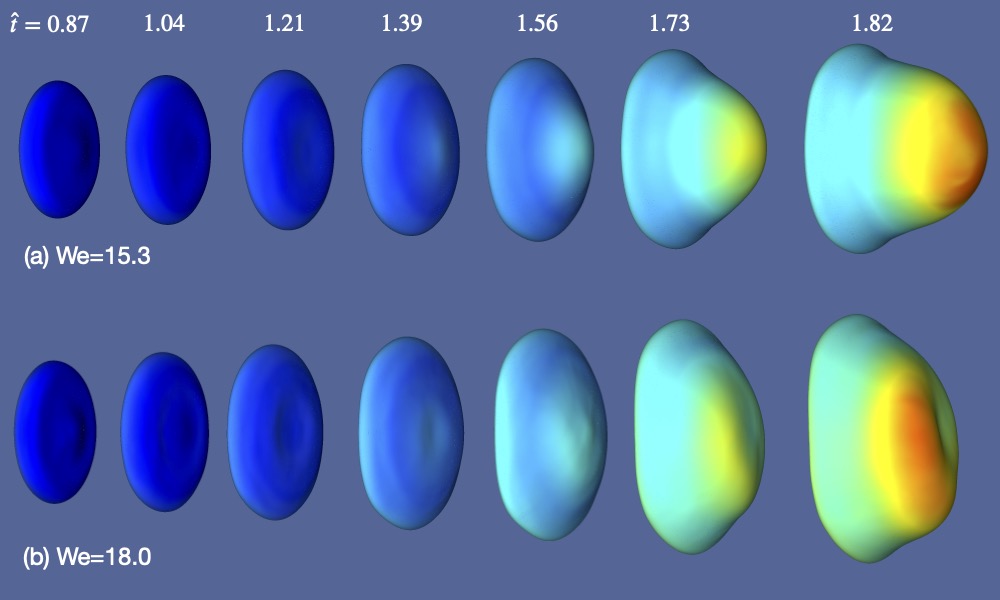}
\caption{Bag morphological evolutions for (a) $\text{We}=15.3$ (simple bag) and (b)18.0 (ring bag). The color represents the velocity magnitude and the color scale is the same as figure~\ref{fig:phase_def}. }
\label{fig:drop_deform_We18}
\end{figure}

The bag formed for $\text{We}=18.0$ exhibits a more complex shape compared to other cases, as shown in figure~\ref{fig:drop_deform_We18}. First, the bag loses the approximate symmetry observed for $\text{We}=15.3$. Furthermore, the minimum thickness of the disk is not located at the bag center but rather at a ring around it, as seen in figure~\ref{fig:uy_liq}(e). As the tip accelerates faster at the location with smaller sheet thickness, the interfacial velocity at $\hat{t}=1.82$ shown in figure~\ref{fig:drop_deform_We18}(b) results in the formation of a ``ring" bag with a dent at the center. As the ring bag grows, the liquid is squeezed to move away from the tip in two directions. The outward flowing liquid will gather at the rim, and the inward counterpart will gather at the center to form a small stem, as seen in figure~\ref{fig:uy_liq}(e). This drop morphology is also known as the ``bag-stem" mode. As the ring bag grows, transverse RTI may develop and introduce an azimuthal variation of the bag growth rate, leading to the ``multi-bag" mode. A detailed investigation of the bag-stem and multi-bag modes is outside the scope of this paper. However, the present results provide a good understanding of the transition from simple to ring bags and lay the foundation for future studies on the bag-stem and multi-bag modes.

A close look at the drop shapes at $\hat{t}=0.69$ in figure~\ref{fig:uy_liq} reveals wavy perturbations formed on the windward surface of the disk. The surface is convex at the center and concave about $\hat{R}/2$ away from the center. The concave locations are marked as points C and D in figures~\ref{fig:uy_liq}(d) and (e). The perturbation wavelength is about $R$, and such a feature holds for all $\text{We}$, making it an outcome of the early-time deformation and independent of $\text{We}$. As the drop for $\text{We}=18.0$ continues to deform, C and D become the locations for the minimum disk thickness where the ring bag is formed. In comparison, for $\text{We}\le 15.3$, C and D merge at the center, resulting in the minimum thickness being located at the center. RTI models based on the linear stability of an infinite liquid layer \citep{Theofanous_2004a} have been used to predict the transition from simple to ring bags. It is argued that the transition will occur when the most unstable RT wavelength $\lambda_{RT}$ satisfies $(2R)/\lambda_{RT}=3$. However, the simulation results presented here suggest a slightly different conclusion. When $\lambda_{RT}$ is smaller than $R$, the perturbation on the windward surface will grow and form the ring bag. When $\lambda_{RT}>R$, even if the location of minimum thickness is initially not at the center of the bag, it will eventually move back to the center because the mode with a larger wavelength grows faster.

\subsection{Phase V: Bag inflation}
As the thickness of the bag reduces to a certain level, $\hat{h}_a\lesssim 0.02$, the development of the bag enters Phase V, as shown in figure~\ref{fig:phase_def}. For simple bags, the bag in this phase consists of two regions, the rim and the bag sheet, as seen at $\hat{t}=2.37$ in figure~\ref{fig:phase_def}. The fraction of the liquid mass in the sheet increases with $\text{We}$. The ratio between the masses in the sheet and the rim determines the mass fractions of the children droplets generated by the rupture of the sheet and the rim. In Phase V, the sheet portion of the bag inflates. The rapid radial expansion of the sheet also enhances the lateral expansion of the edge rim, resulting in a rapid increase of $d\hat{R}/d\hat{t}$, which is the key feature to distinguish Phase V from Phase IV. This section will focus on the inflation of simple bags.

The sheet inflation is mainly driven by the pressure difference across the liquid sheet. It is shown in figure~\ref{fig:phase_def}(c) that $(\Delta p)_a$ varies over time. On the windward side, the gas pressure inside the bag is approximately uniform. The pressure in the bag is initially similar to the stagnation pressure of the free stream. However, as the bag inflates and the bag tip velocity increases, the pressure decreases. On the leeward side, due to flow separation on the lateral edge of the bag, the pressure in the wake is generally lower than the free stream pressure. Different from the pressure on the windward side of the bag, the pressure on the leeward side increases as the bag inflates due to the resulting delay of flow separation.

\begin{figure}
\centering
\includegraphics[trim={0cm 0cm 0cm 0cm},clip,width=0.99\textwidth]{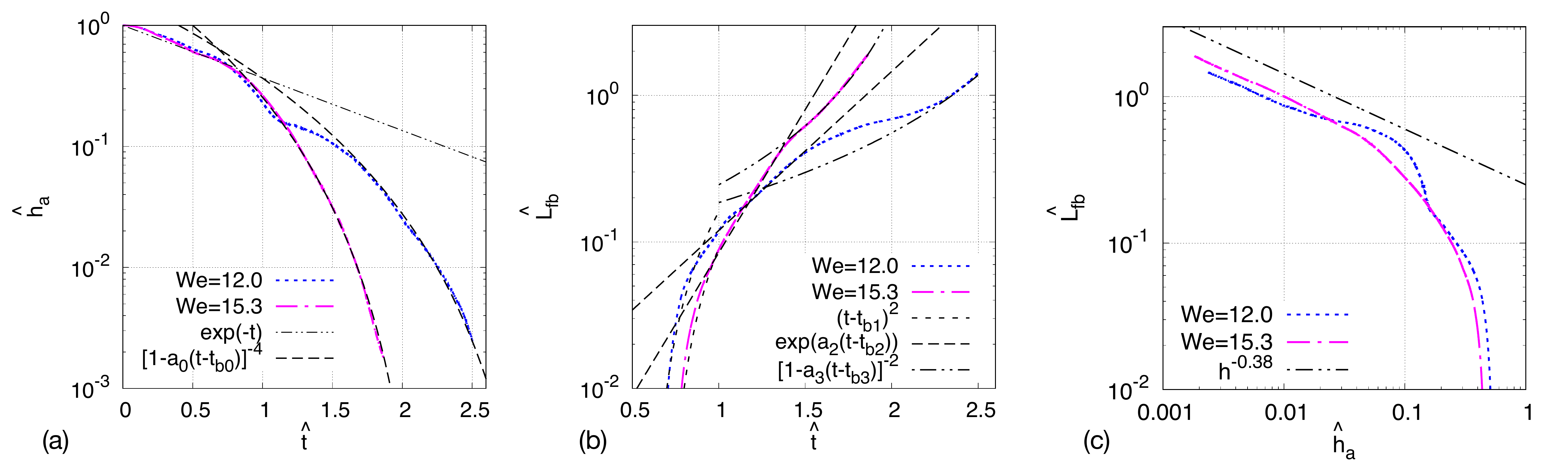}
\caption{Time evolutions of (a) sheet thickness $\hat{h}_a$ and (b) bag length $\hat{L}_{fb}$ for the cases $\text{We}=12.0$ and 15.3. (c) $\hat{L}_{fb}$ as a function of $\hat{h}_a$. }
\label{fig:bag_inflation}
\end{figure}

The model of \citet{Villermaux_2009a} predicts the sheet thickness for a disk decays exponentially over time, \ie, 
\begin{align}
	\hat{h}_a \sim \exp(-4\hat{t})\,. 
\end{align}
Then Eq.~\eqref{eq:bag_tip} can be integrated, yielding two asymptotic limits 
\begin{align}
	\hat{L}_{fb} \sim \hat{t}^2 &  \quad \mathrm{for}\ \hat{t}\ll 1\,, \label{eq:bag_length_VB_short}\\
	\hat{L}_{fb} \sim \exp(4 \hat{t}) & \quad  \mathrm{for}\ \hat{t}\gg 1\,,\label{eq:bag_length_VB_long}
\end{align}

In the model of \cite{Reyssat_2007a}, the bag was approximated as a spherical shell that extends uniformly in all direction. Also assuming the pressure difference inside and outside scales with the stagnation pressure, they predicted that 
\begin{align}
	\hat{L}_{fb} \sim (1-b \hat{t})^{-2}\,. \label{eq:bag_length_Reyssat}
\end{align}
Then through mass conservation, it can be shown that 
\begin{align}
	\hat{h}_a  \sim (\hat{L}_{fb})^{-2} \sim (1-b\hat{t})^{4}\,. 
	\label{eq:bag_thickness_Reyssat}
\end{align}

The simulation results for $\hat{h}_a$ for $\text{We}=12.0$ and 15.3 are shown in figure~\ref{fig:bag_inflation}(a). It can be observed that the early-time decrease of $\hat{h}_a$ in Phases I and II is exponential, similar to the model of \citet{Villermaux_2009a}. However, since at this time range, the drop is not exactly a disk, the expression for $\hat{h}_a$ should be corrected to
\begin{equation}
\hat{h}_a=\exp(-\hat{t}),
\end{equation}
instead of $\exp(-4\hat{t})$ in the original model. When the bag starts to inflate in the long-time, the decay of $\hat{h}_a$ switches to a power law, as predicted by the model of \cite{Reyssat_2007a}, \ie,
\begin{equation}
\hat{h}_a=[1-b(\hat{t}-\hat{t}_{b,0})]^{-4},, \label{eq:h_bag_inflate}
\end{equation}
where $b$ and $\hat{t}_{b,0}$ are model coefficients which need to be fitted based on the simulation results. In the figure, $b=0.37$ and 0.58, and $\hat{t}_{b,0}=0.4$ and 0.5, for $\text{We}=12.0$ and 15.3, respectively. The temporal evolutions of $\hat{h}_a$ for $\text{We}=12.0$ and 15.3 are quite similar until the end of Phase III ($\hat{t}\approx1.2$). Afterwards, $\hat{h}_a$ decays faster for $\text{We}=15.3$, which is mainly due to the long Phase IV for $\text{We}=12.0$, during which $\hat{h}_a$ decays more gradually.

The evolution of $\hat{L}_{fb}$ is shown in figure~\ref{fig:bag_inflation}(b). In the early time of bag inflation, $\hat{L}_{fb}$ increases with time following a power law, similar to Eq.~\eqref{eq:bag_length_VB_short}. We have to add a case dependent parameter $\hat{t}_{b1}$, \ie,
\begin{equation}
\hat{L}_{fb}=(\hat{t}-\hat{t}_{b1})^{2},
\end{equation}
where the fitted values for $\hat{t}_{b1}$ are 0.6 and 0.7 for $\text{We}=12.0$ and 15.3, respectively. In the intermediate term, when the drop deforms as a disk, the increase of $\hat{L}_{fb}$ is exponential, as predicted by the model of \citet{Villermaux_2009a} and Eq.~\eqref{eq:bag_length_VB_long}. It is found that
\begin{equation}
\hat{L}_{fb}=\exp[b_2(\hat{t}-\hat{t}_{b2})],
\end{equation}
where $b_2=2.5$ and 4.5, and $\hat{t}_{b2}=1.85$ and 1.55 for $\text{We}=12.0$ and 15.3, respectively. Similar trends have also been observed in the experiments \citep{Villermaux_2009a}. In the long time, when the bag inflates and deforms as a shell with decreasing thickness, it is observed that $\hat{L}_{fb}$ increases following another power law, similar to Eq.~\eqref{eq:bag_length_Reyssat} given by the model of \cite{Reyssat_2007a}, \ie,
\begin{equation}
\hat{L}_{fb}=[1-b_3(\hat{t}-\hat{t}_{b3})]^{-2},
\end{equation}
where $b_3=0.98$ and 1.5, and $\hat{t}_{b3}=2.35$ and 1.68 for $\text{We}=12.0$ and 15.3, respectively. During bag inflation, $\hat{L}_{fb}$ increases with decreasing $\hat{h}_a$, see figure~\ref{fig:bag_inflation}(c). The variation of $\hat{L}_{fb}$ for small $\hat{h}_a$ follows a power law, yet the power index is -0.38, instead of -0.5 given by Eq.~\eqref{eq:bag_thickness_Reyssat}. The agreement between the present simulation results with the theoretical scaling relations indicates that the early stage of bag inflation is well captured.

\subsection{Phase VI Bag rupture}
\label{sec:phase6}
\begin{figure}
\centering
\includegraphics[trim={0cm 0cm 0cm 0cm},clip,width=0.99\textwidth]{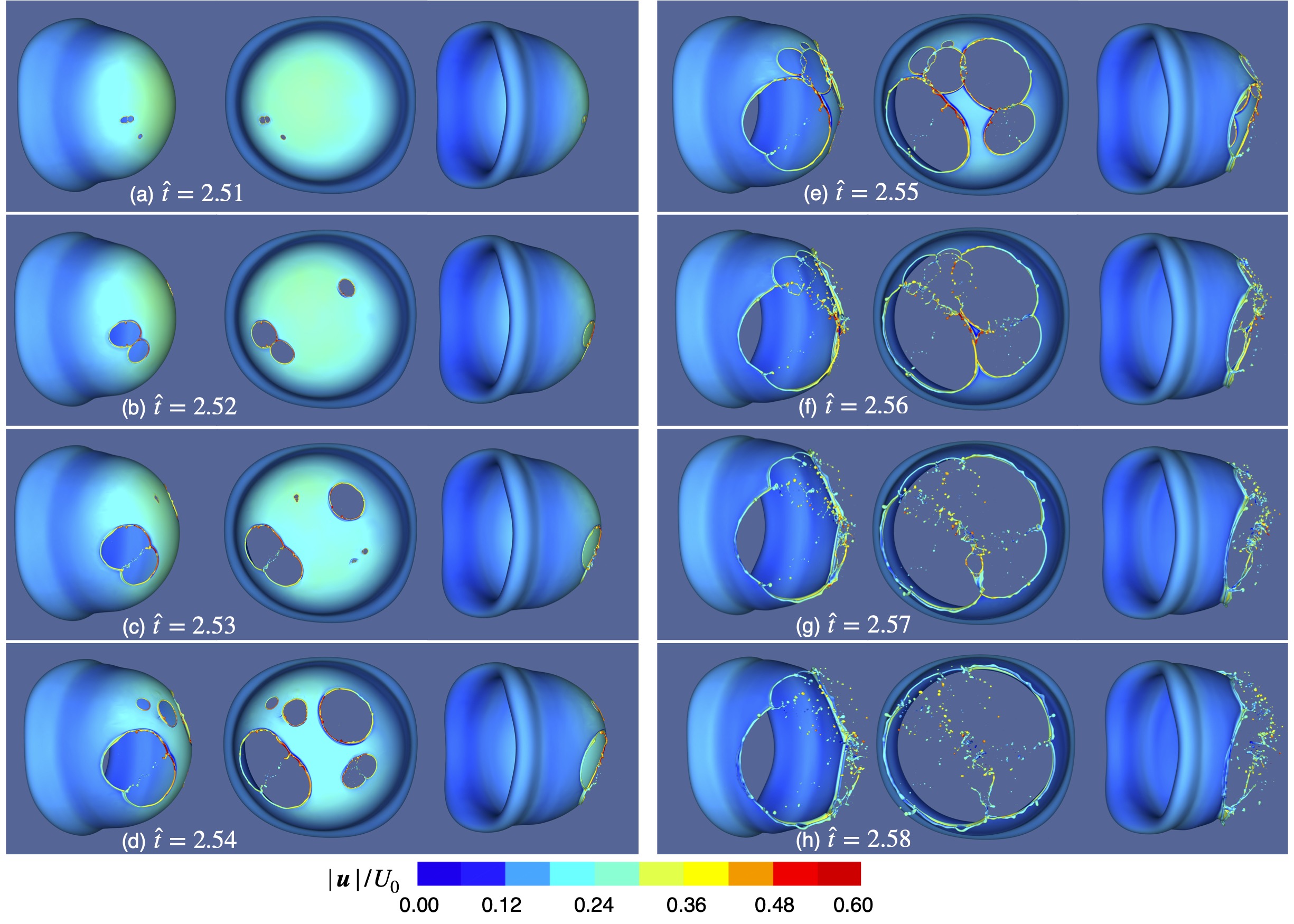}
\caption{Time evolution of the drop surface from $\hat{t}=t/\tau_d=2.51$-2.58 for $\text{We}=12.0$, showing the breakupof the bag due to appearance and merging of holes. The color on the interfaces represents the velocity magnitude. The three different views are shown for each time.  }
\label{fig:hole_dynamics}
\end{figure}

As the bag sheet thickness continues to decreases, the two interfaces will eventually pinch, forming a hole in the liquid sheet. Generally holes are formed at different locations at different times. The holes will expand as the rim retracts with the Taylor-Culick velocity \citep{Opfer_2014a, Ling_2017a, Agbaglah_2021a}. When different holes merge together, the liquid sheet rupture completely, producing numerous small filaments and droplets and the remaining rim. The bag rupture process for $\text{We}=12.0$ is shown in figure~\ref{fig:hole_dynamics}. 

\subsubsection{Effect of numerical breakup}
{
In an ideal scenario without any tiny bubbles or contaminants in the liquid sheet and with no thermal fluctuations, as considered in the present simulations, the two interfaces will eventually pinch, resulting in holes in the liquid sheet. Van der Waals forces become important to the interfacial dynamics when the sheet thickness is between about 10 to 100 nm and the interaction between van der Waals forces and surface tension typically dictate the sheet rupture and hole formation. The critical sheet thickness at which holes first appear due to van der Waals forces is referred to as the \emph{physical cutoff length}, which is in the order of tens of nanometers. While the rupture dynamics of a stationary free liquid sheet has been extensively studied \citep{Erneux_1993e,Ida_1996x}, the rupture of dynamic moving sheet is less understood and the corresponding critical sheet thickness could be significantly larger, as indicated in drop aerobreakup experiment by \citet{Opfer_2014a}.
} 
The minimum cell size in a 3D simulation, referred to as the \emph{numerical cutoff length}, is significantly larger than the physical counterpart, and therefore, the interface pinching occurs earlier in the simulation. When the liquid sheet thickness reduces to about two times the minimum cell size of the octree mesh, numerical error in VOF reconstruction will act as a perturbation on the interfaces, which eventually causes numerical pinching between the two interfaces of the sheet. The interface pinching will form small holes in the liquid sheet. The numerical breakup of the VOF method has both pros and cons. On one hand, it will automatically allow topology change, and no additional procedure is required as needed for other methods like the front-tracking method \citep{Lu_2018a}. On the other hand, the length scale for pinching to occur is related to mesh resolution. As a result, it will not produce mesh-independent results for liquid sheet breakup and hole formation. Recently, the manifold death method has been proposed by \citet{Chirco_2022a} to model sheet rupture. Thin sheets are detected by taking quadratic moments of an indicator obtained from the VOF function and then pinching is manually induced based on a user-defined cutoff length that is independent of the cell size. 
{
The manifold death method has been applied recently to simulate bag breakup by \citet{Chirco_2022a} and \citet{Tang_2022h}, and it was observed that the method reduces the influence of numerical breakup on hole formation. 
} 
Nevertheless, since the cutoff length scale in the manifold death method needs to be larger than the cell size, it will not produce ``more physical" results unless the mesh resolution is comparable to the physical length scale. The larger numerical cutoff length, which is about 1 \textmu m in the present simulation, will cause hole formation earlier than in reality, and as a result, the numerical simulation will not be able to capture the late stage of bag inflation and the bag rupture.
 
\begin{figure}
\centering
\includegraphics[trim={0cm 0cm 0cm 0cm},clip,width=0.99\textwidth]{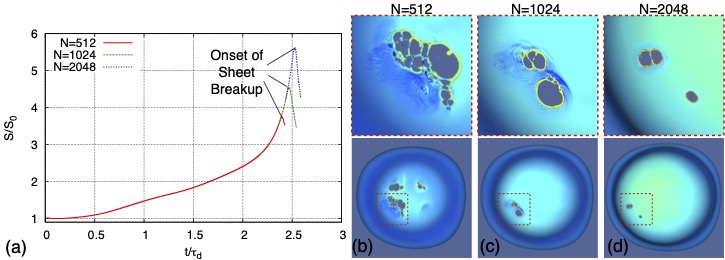}
\caption{Effect of mesh resolution on the bag bursting for $\text{We}=12.0$. (a) Time evolution of the drop surface area. (b)-(d) Drop surfaces when the holes are just formed for different mesh resolutions $N=512$, 1024, 2048, respectively.   }
\label{fig:bag_breakup_mesh}
\end{figure}

In spite of the limitations of the numerical simulation, the results at different resolutions obtained here are still useful for understanding the bag rupture dynamics, and the results can be used to extrapolate the unresolved stage of bag inflation. The evolution of the total surface area of the drop, normalized by its initial value, \ie, $S/S_0$, for $\text{We}=12.0$, is shown in figure~\ref{fig:bag_breakup_mesh}(a). During the interaction between the drop and the free stream, the gas kinetic energy is transferred to the surface energy of the drop, and thus the surface area increases over time. After holes are formed, the surface area will change to decrease due to the capillary expansion of the holes. Therefore, $S$ reaches the maximum value around the onset time for sheet breakup. As the mesh resolution increases from $N=512$ to 2048, the breakup is delayed, and we can resolve the bag until $S/S_0\approx 5.5$ with $N=2048$. According to figure~\ref{fig:bag_inflation}(a) and the fitted function Eq.~\eqref{eq:h_bag_inflate}, one can estimate the breakup time for a given physical cutoff length smaller than the minimum cell size. Then, based on the estimated breakup time, extrapolation from figure~\ref{fig:bag_breakup_mesh}(a) can be used to estimate the surface area when the bag breaks according to the physical cutoff length.

The bag morphology when holes are just formed is shown in figures~\ref{fig:bag_breakup_mesh}(b), (c), and (d) for $N=512$, 1024, and 2048, respectively. The holes formed due to the lack of resolution in VOF reconstruction typically exhibit irregular shapes. Yet after they are formed, surface tension tends to regularize the shape of the hole and to form a rim on the edge of the hole. For $N=512$, many small holes are formed simultaneously, and thus, holes start to interact with each other before the rim gets a chance to develop. With the most refined mesh, $N=2048$, only three small holes are initially formed, and the subsequent evolutions of rims and holes are well captured, see also figure~\ref{fig:hole_dynamics}.

\subsubsection{Hole dynamics and drop formation}
\begin{figure}
\centering
\includegraphics[trim={0cm 0cm 0cm 0cm},clip,width=0.99\textwidth]{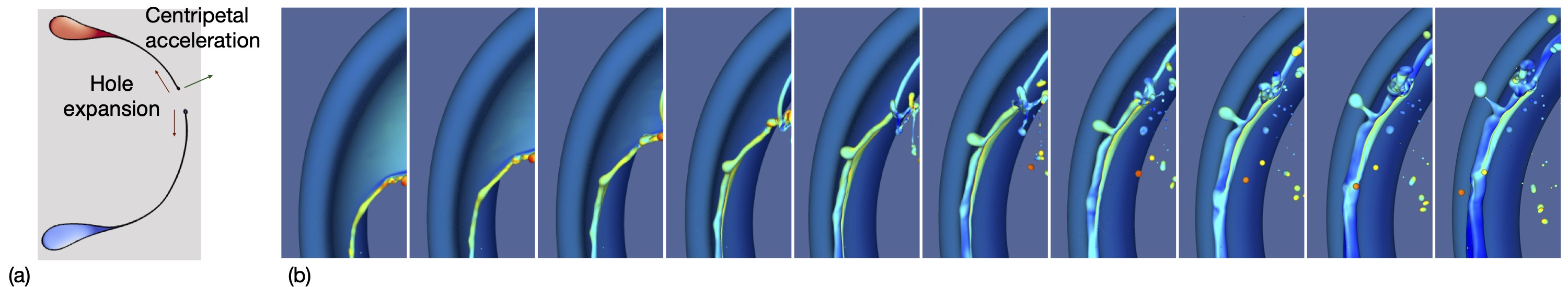}
\caption{Rayleigh-Taylor instability due to centripetal acceleration of the rim, forming fingers at the rim, which later detach to form children droplet. The results are for $\text{We}=12.0$ and $\hat{t}=2.534$-2.589 with an increment of 0.008. The color on the interfaces represents the velocity magnitude, see figure~\ref{fig:hole_dynamics} for the color scale. }
\label{fig:rim_RTI}
\end{figure}

After a hole is formed, the capillary retraction of the rim causes the hole to expand. The rim retraction velocity is the Taylor-Culick velocity, $U_{TC}=\sqrt{2 \sigma /\rho_l h}$, where $h$ is the sheet thickness near the rim. As the sheet thickness in the bag is quite uniform, the rim retraction velocity is approximately constant and agrees well with $U_{TC}$ \citep{Ling_2017a, Agbaglah_2021a}. Nevertheless, as the rim moves along the curved liquid sheet, it experiences a centripetal acceleration along the radial direction of the bag, see figure~\ref{fig:rim_RTI}(a). Along with the density difference, an RTI is triggered along the rim. The development of RTI leads to the formation of fingers normal to the rim pointing outward. The subsequent Rayleigh-Plateau instability of the fingers detaches droplets. The process of finger formation and drop detachment at the rim is shown in figure~\ref{fig:rim_RTI}(b). The color on the interface represents the velocity magnitude, from which the interfacial velocity variation along the rim can be recognized. Similar drop formation processes were also observed in bubble bursting \citep{Lhuissier_2012a}.

\begin{figure}
\centering
\includegraphics[trim={0cm 0cm 0cm 0cm},clip,width=0.99\textwidth]{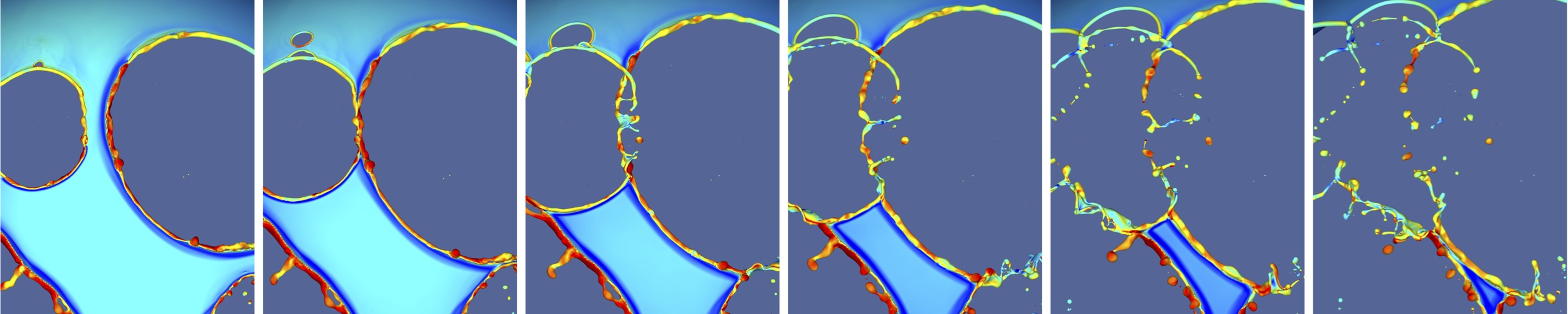}
\caption{Interaction between rims of different holes when the hole-merging occurs. The results are for $\text{We}=12.0$  and $\hat{t}=t/\tau_d=2.544$-2.562 with an increment of 0.004. The color on the interfaces represents the velocity magnitude, see figure~\ref{fig:hole_dynamics} for the color scale. }
\label{fig:rim_impingement}
\end{figure}

Figure~\ref{fig:rim_impingement} shows the interaction between two holes. It can be observed that when the rims of different holes collide, a new fish-shaped liquid lamella is formed, the disintegration of which produces a large number of small droplets. The hole-hole interaction has been recently investigated by \citet{Agbaglah_2021a} {and \citet{Tang_2022h}} through numerical simulation. The simulation results of \citet{Agbaglah_2021a} show that the collision of rims forms a rim instead of a lamella as observed here {and in the results of  \citet{Tang_2022h}}. The discrepancy is probably due to the fact that the sheet thicknesses considered by \citet{Agbaglah_2021a} is larger than those in the present cases. The small sheet thickness leads to a higher rim retraction velocity when they collide. This is also why the lamella formation was not observed in the present simulations with coarser meshes. \citet{Neel_2020e} suggested characterizing the rim collision dynamics using the Weber number $\text{We}_{h}$, defined based on $U_{TC}$ and rim radius. It can be shown that $\text{We}_{h}$ is proportional to the ratio between the holes' distance and the sheet thickness. When $\text{We}_{h}$ is sufficiently large, the rim collision will generate an expanding lamella with transverse modulation on the rim. The lamella will also break later to form a distribution of children droplets.

While the present simulations are useful in revealing the qualitative droplet formation mechanisms, the larger numerical cutoff length scale will not allow for a quantitative prediction of the statistics of the droplets formed from the bag rupture. Such a prediction will only be possible if one can use a mesh resolution down to the physical cutoff length, or if a subgrid sheet model can be developed to represent the unresolved sheet inflation and breakup dynamics. These tasks are relegated to future works.

\section{Turbulent wake and drop dynamics}
\label{sec:dynamics}
\subsection{Effect of drop deformation on the aerodynamic drag and lift}
\begin{figure}
\centering
\includegraphics[trim={0cm 0cm 0cm 0cm},clip,width=0.99\textwidth]{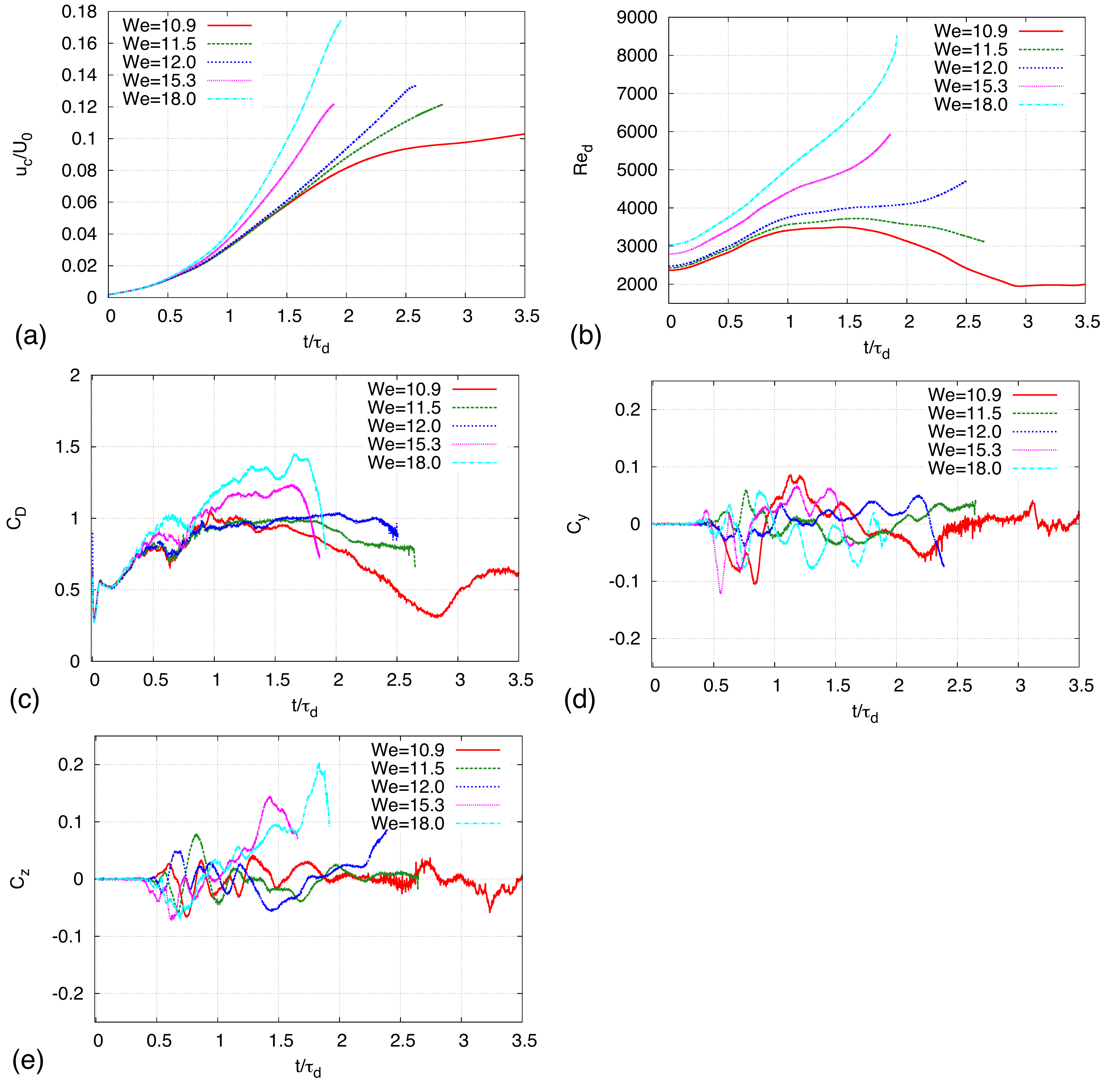}
\caption{Time evolutions of (a) streamwise velocity of the drop $u_c$, (b) drop Reynold number $Re_d$, (c) drag coefficient $C_D$, and (d, e) lift coefficients in the lateral $y$ and $z$ directions, for drops at different $\text{We}$. }
\label{fig:Cd_bag}
\end{figure}

In the aerobreakup process, the drop accelerates due to aerodynamic drag. As discussed in section \ref{sec:time_scales}, the unsteady contributions to the drag, triggered by the impulsive gas acceleration, to the drop velocity are generally small due to the low gas-to-liquid density ratio. The drop acceleration is mainly dictated by the quasi-steady drag corresponding to the relative velocity $U_0-u_d$. The shape deformation has a strong impact on the quasi-steady drag and the resulting drop acceleration. The temporal evolutions of the normalized drop velocity, $\hat{u}_d$, for different $\text{We}$ are shown in figure~\ref{fig:Cd_bag}(a). When $\text{We}$ increases, the drop deforms more significantly, and the drop velocity increases faster. It can be seen that the drop velocity reaches about 18\% of $U_0$ for $\text{We}=18.0$ when the drop breaks, compared to about 12\% for $\text{We}=15.3$.

The lateral radius $R$ is the characteristic length for the gas flow around the drop, based on which the instantaneous drop Reynolds number is defined as
\begin{equation}
	\text{Re}_d = \frac{\rho_g (U_0-u_d) 2R}{\mu_g}\, . 
\end{equation}
The evolutions of $\text{Re}_d$ for different $\text{We}$ are shown in figure~\ref{fig:Cd_bag}(b). At time zero, $\text{Re}_d=\text{Re}$. As time evolves, the differences in $\text{Re}_d$ for different cases are magnified. Though the relative velocity $U_0-u_d$ decreases over time, $\text{Re}_d$ increases, thanks to the rapid increase of $R$. For $\text{We}=18.0$, $\text{Re}_d$ can increase almost three times of its initial value. For the cases $\text{We}=10.9$ and 11.5, when $\hat{R}$ turns to decrease at later time, $\text{Re}_d$ also decreases. For all the cases considered, $\text{Re}_d$ varies between about 2000 to 8000. For a solid sphere, this range of $\text{Re}_d$ lies in the chaotic vortex shedding regime and subcritical turbulent wake regime \citep{Tiwari_2020a}. The shedding of the turbulent wake can be observed in figure~\ref{fig:wake_We15}(c). 

Since the drag coefficient $C_D$ is defined (see Eq.~\eqref{eq:Cd}) based on the instantaneous relative velocity and drop radius, the time variations of $U_0-u_d$ and $R$ actually do not contribute to changes in $C_D$ over time. While a constant $C_D$ is often assumed in previous studies \citep{Marcotte_2019a}, a non-monotonic time evolution of $C_D$ is observed here, as shown in figure\ref{fig:Cd_bag}(c), which is due to the complex drop shape deformation in different phases. Once again, we take the case $\text{We}=12.0$ as the representative example to explain the physics behind it. When the drop remains approximately a sphere at early time, $C_D\approx 0.5$, which is expected for a spherical drop with large density and viscosity contrast. Then, as the drop deforms into an ellipsoid with decreasing streamwise thickness in Phase I, $C_D$ increases until it reaches a local maximum. When the drop continues to deform into a disk in Phase II, $C_D$ increases again until it reaches a plateau value of about 1, which is similar to the typical $C_D$ for a thin, flat circular disk. When the disk deforms into a bag in Phases III and IV due to RTI, the change in the drop shape is relatively small, so $C_D$ increases slowly. When the bag inflates in Phase V, as the bag shape becomes more aligned with the gas flow, $C_D$ decreases. The amplitude of variation in $C_D$ is generally more profound for cases with larger $\text{We}$.

\subsection{Effect of transient development of gas flow}
The present configuration involves a sudden exposure of the drop to the free stream, which causes the boundary layer around the drop to take a finite time to develop, as shown in figure~\ref{fig:wake_We15}(c). This boundary layer development contributes to viscous-unsteady drag and the resulting large amplitude variation of $C_D$ near $\hat{t}=0$. Despite the high Reynolds numbers, the flow remains approximately axisymmetric in Phase I, and the transition to a turbulent wake does not occur until Phase II. The formation of the disk in Phase II enhances flow separation and accelerates the transition of the wake to turbulence. When the wake becomes turbulent, the gas pressure on the leeward side of the drop increases, as shown at $\hat{t}=0.79$ in figure~\ref{fig:wake_We15}(b). Consequently, the drag decreases around $\hat{t}\approx 0.6$ to 0.7, as shown in figure~\ref{fig:Cd_bag}(c). The breakdown of wake symmetry induces lift on the drop. The lift coefficients in the lateral $y$ and $z$ directions are initially zero since the flow around the drop is approximately axisymmetric, but $C_y$ and $C_z$ become nonzero and start to oscillate in time around $\hat{t}=0.4$, as shown in figure~\ref{fig:Cd_bag}(d) and (e). The low-frequency oscillation observed in the lift coefficients is related to the shedding of the wake. The wake features are consistent with the range of $\text{Re}$ for the present cases.

\section{Conclusions}
\label{sec:conclusions}
Detailed numerical simulations were performed in the present study to investigate the drop aerobreakup in the moderate Weber number ($\text{We}$) regime. We have considered an ideal configuration, in which an initially stationary and spherical drop is subjected to a uniform gas stream. The liquid and gas are taken to be water and air, and the drop size is fixed at 1.9 mm, following the experiment of \citet{Jackiw_2021a}. A parametric study is carried out by varying the gas stream velocity ($U_0$). Due to the interaction with the uniform gas stream, the drop deforms from a sphere to a forward bag with the opening facing upstream. When $\text{We}$ is sufficiently high, the bag is unstable and will eventually inflate and be pierced through by the gas stream. 

The numerical simulations were performed using the \emph{Basilisk} solver. The mass-momentum consistent volume-of-fluid method is used to captured the sharp gas-liquid interface. The computational domain is discretized using quad-tree/octree mesh and adaptive mesh refinement technique is employed to reduce the total number of cells. To resolve the bag inflation and rupture dynamics, the finest mesh used in the simulation is equivalent to 2048 cells across the initial drop diameter.  

The key findings of the present study are summarized as follows. 

\begin{itemize}
\item \emph{High-resolution 3D simulation is necessary to capture drop aerobreakup.} Both 2D axisymmetric and 3D simulations were performed using identical numerical methods and initial/boundary conditions. The converged 2D simulation results are shown to be significantly different from the 3D simulation and experimental results, because 2D simulations cannot resolve the turbulent wake, and as a result, will not be able to capture the correct bag shape and drop acceleration. Grid-refinement studies were performed for the 3D simulations, and the adaptive mesh with a minimum cell size equivalent to 512 cells across the initial drop diameter is sufficient to yield mesh-independent results for the drop shape, drag coefficient, and gas enstrophy. The 3D simulation results are compared against previous experiments, and good agreement is achieved.

\item \emph{Different phases in the drop morphological evolution are identified.} The non-monotonic evolution of the rate of change of the lateral drop radius ($dR/dt$) clearly shows different phases in the overall process, including {I) ellipsoid deformation}, {II) disk formation}, {III) disk deformation}, {IV) bag development}, {V) bag inflation}, and {VI) bag rupture}.

\item \emph{The asymptotic early-time drop dynamics are independent of $\text{We}$.} In the asymptotic limit of $t\to 0$, the evolutions of the drop shapes for different $\text{We}$ collapse. This $\text{We}$-independence in early-time drop dynamics is consistent with the time scale analysis and is further affirmed by the good agreement between drop velocities obtained by the present simulations and the inviscid compressible simulations in previous studies.

\item \emph{A new internal-flow deformation model is proposed for Phase I.} The drop deforms as an ellipsoid in Phase I, and the deformation is dictated by the internal flow, which is, in turn, driven by the stagnation flow near the windward pole. An improved internal-flow deformation model is proposed, which respects the $\text{We}$-independent asymptotic limit at time zero. The results of the present model show better agreement with the present simulation results, compared to previous models.

\item \emph{The disk is formed when the Laplace pressure at the edge is in balance with the stagnation pressure. } The drop deforms toward a disk with a rounded edge in Phase II. The edge rim of the disk is formed when $dR/dt$ reaches a local maximum, namely, $d^2R/dt^2=0$, and the Laplace pressure at the edge rim reaches a balance with the gas stagnation pressure on the windward surface.

\item \emph{Rayleigh-Taylor instability dictates the thinning of the disk center.} When the disk is accelerated along the streamwise direction, the windward surface is unstable while the leeward surface is stable. The development of the RTI becomes the dominant mechanism for the continuous drop lateral extension and the thinning of the disk center in Phase III. The surface tension at the edge rim resists the RTI development, and as a result, $dR/dt$ decreases in time.

\item \emph{Bag piercing does not guarantee bag fragmentation.} When the thin disk center curves downstream, the disk deforms to a forward bag. The bag can only be pierced through by the gas stream if the minimum thickness of the disk can decrease to be lower than a threshold, which is about 20\% of the original drop diameter. More importantly, when $\text{We}$ is just lower than $\text{We}_{cr}$, the bag can be pierced through but does not fragment into droplets, as the ring rim will retract toward the center and eventually will turn back to a big drop.

\item \emph{Evolutions of the bag length and sheet thickness follow similar power laws when the bag inflates.} In Phase V, the rapid inflation of the bag causes the rim lateral velocity $dR/dt$ to increase rapidly over time. The increase of the bag length and the decrease of the sheet thickness follow exponential functions initially and then switch to power laws. The agreement with the power-law scaling actually indicates that the sheet expands uniformly in all directions when the bag inflates. That is because the gas pressure is approximately uniform inside the bag.

\item \emph{Hole-hole interaction is important to the disintegration of the bag.} The numerical cutoff length is significantly larger than the physical counterpart, and thus the bags in the simulations break earlier than in reality. Nevertheless, the simulation results for the most refined mesh well capture the dynamics of holes. When two holes merge, the collision between the two rims forms a lamella first and then the lamella disintegrates into small droplets.

\item \emph{The drop morphological change has a strong impact on the drag coefficient.} The evolutions of the drop deformation and acceleration are closely coupled. Due to the low gas-to-liquid density ratio, the drop acceleration is mainly due to the quasi-steady drag. Although the relative velocity $U_0-u_d$ decreases in time, the instantaneous Reynolds number increases with $R$. The non-monotonic evolution of the drag coefficient $C_D$ is mainly caused by the change of drop shape. $C_D$ decreases over time as the bag inflates, and the flow separation is delayed.

\end{itemize}

\section*{Acknowledgement}
This research is supported by the ACS Petroleum Research Fund (\#62481-ND9) and the NSF (\#1942324). The XSEDE/ACCESS and Frontera programs have provided the computational resources that contributed to the simulation results reported in this paper. The Baylor High Performance and Research Computing Services (HPRCS) have been used for the simulation and data processing. The authors would also like to acknowledge St\'ephane Zaleski, Jacob McFarland, and Florence Marcotte for the helpful discussions. The present simulations were performed using the \emph{Basilisk} solver, which is made available by St\'ephane Popinet and other collaborators.
\section*{Declaration of Interests}
The authors report no conflict of interest.

%\bibliographystyle{jfm}
%\bibliography{refs}

\begin{thebibliography}{88}
\expandafter\ifx\csname natexlab\endcsname\relax\def\natexlab#1{#1}\fi
\def\au#1{#1} \def\ed#1{#1} \def\yr#1{#1}\def\at#1{#1}\def\jt#1{\textit{#1}}
  \def\bt#1{#1}\def\bvol#1{\textbf{#1}} \def\vol#1{#1} \def\pg#1{#1}
  \def\publ#1{#1}\def\arxiv#1{#1}\def\org#1{#1}\def\st#1{\textit{#1}}

\bibitem[Agbaglah(2021)]{Agbaglah_2021a}
{\sc \au{Agbaglah, G.~G.}} \yr{2021}  \at{Breakup of thin liquid sheets through
  hole--hole and hole--rim merging}.  \jt{J.~Fluid Mech.}  \bvol{911},
  \pg{A23}.

\bibitem[Apte {\em et~al.\/}(2003)Apte, Gorokhovski \& Moin]{Apte_2003a}
{\sc \au{Apte, S.V.}, \au{Gorokhovski, M.} \& \au{Moin, P.}} \yr{2003}
  \at{{LES} of atomizing spray with stochastic modeling of secondary breakup}.
  \jt{Int.~J.~Multiphase Flow}  \bvol{29},  \pg{1503--1522}.

\bibitem[Arrufat {\em et~al.\/}(2020)Arrufat, Crialesi-Esposito, Fuster, Ling,
  Malan, Pal, Scardovelli, Tryggvason \& Zaleski]{Arrufat_2020a}
{\sc \au{Arrufat, T.}, \au{Crialesi-Esposito, M.}, \au{Fuster, D.}, \au{Ling,
  Y.}, \au{Malan, L.}, \au{Pal, S.}, \au{Scardovelli, R.}, \au{Tryggvason, G.}
  \& \au{Zaleski, S.}} \yr{2020}  \at{A momentum-conserving, consistent,
  volume-of-fluid method for incompressible flow on staggered grids}.
  \jt{Comput.~Fluids}  \bvol{215},  \pg{104785}.

\bibitem[Balachandar(2009)]{Balachandar_2009a}
{\sc \au{Balachandar, S.}} \yr{2009}  \at{A scaling analysis for point particle
  approaches to turbulent multiphase flows}.  \jt{Int.~J.~Multiphase Flow}
  \bvol{35},  \pg{801--810}.

\bibitem[Blanco \& Magnaudet(1995)]{Blanco_1995v}
{\sc \au{Blanco, A.} \& \au{Magnaudet, J.}} \yr{1995}  \at{The structure of the
  axisymmetric high-reynolds number flow around an ellipsoidal bubble of fixed
  shape}.  \jt{Phys.~Fluids}  \bvol{7},  \pg{1265--1274}.

\bibitem[Burelbach {\em et~al.\/}(1988)Burelbach, Bankoff \&
  Davis]{Burelbach_1988s}
{\sc \au{Burelbach, J.~P.}, \au{Bankoff, S.~G.} \& \au{Davis, S.~H.}} \yr{1988}
   \at{Nonlinear stability of evaporating/condensing liquid films}.
  \jt{J.~Fluid Mech.}  \bvol{195},  \pg{463--494}.

\bibitem[Chang {\em et~al.\/}(2013)Chang, Deng \& Theofanous]{Chang_2013a}
{\sc \au{Chang, C.-H.}, \au{Deng, X.} \& \au{Theofanous, T.~G.}} \yr{2013}
  \at{Direct numerical simulation of interfacial instabilities: a consistent,
  conservative, all-speed, sharp-interface method}.  \jt{J.~Comput.~Phys.}
  \bvol{242},  \pg{946--990}.

\bibitem[Chirco {\em et~al.\/}(2022)Chirco, Maarek, Popinet \&
  Zaleski]{Chirco_2022a}
{\sc \au{Chirco, L.}, \au{Maarek, J.}, \au{Popinet, S.} \& \au{Zaleski, S.}}
  \yr{2022}  \at{Manifold death: a volume of fluid implementation of controlled
  topological changes in thin sheets by the signature method}.
  \jt{J.~Comput.~Phys.}  \bvol{467},  \pg{111468}.

\bibitem[Chou \& Faeth(1998)]{Chou_1998a}
{\sc \au{Chou, W.-H.} \& \au{Faeth, G.~M.}} \yr{1998}  \at{Temporal properties
  of secondary drop breakup in the bag breakup regime}.  \jt{Int.~J.~Multiphase
  Flow}  \bvol{24},  \pg{889--912}.

\bibitem[Dai \& Faeth(2001)]{Dai_2001a}
{\sc \au{Dai, Z.} \& \au{Faeth, G.~M.}} \yr{2001}  \at{Temporal properties of
  secondary drop breakup in the multimode breakup regime}.
  \jt{Int.~J.~Multiphase Flow}  \bvol{27},  \pg{217--236}.

\bibitem[Erneux \& Davis(1993)]{Erneux_1993e}
{\sc \au{Erneux, T.} \& \au{Davis, S.~H.}} \yr{1993}  \at{Nonlinear rupture of
  free films}.  \jt{Phys. Fluids A}  \bvol{5},  \pg{1117--1122}.

\bibitem[Flock {\em et~al.\/}(2012)Flock, Guildenbecher, Chen, Sojka \&
  Bauer]{Flock_2012a}
{\sc \au{Flock, A.~K.}, \au{Guildenbecher, D.~R.}, \au{Chen, J.}, \au{Sojka,
  P.~E.} \& \au{Bauer, H.-J.}} \yr{2012}  \at{Experimental statistics of
  droplet trajectory and air flow during aerodynamic fragmentation of liquid
  drops}.  \jt{Int.~J.~Multiphase Flow}  \bvol{47},  \pg{37--49}.

\bibitem[Gao {\em et~al.\/}(2013)Gao, Guildenbecher, Reu, Kulkarni, Sojka \&
  Chen]{Gao_2013a}
{\sc \au{Gao, J.}, \au{Guildenbecher, D.~R.}, \au{Reu, P.~L.}, \au{Kulkarni,
  V.}, \au{Sojka, P.~E.} \& \au{Chen, J.}} \yr{2013}  \at{Quantitative,
  three-dimensional diagnostics of multiphase drop fragmentation via digital
  in-line holography}.  \jt{Opt.~Lett.}  \bvol{38},  \pg{1893--1895}.

\bibitem[Guildenbecher {\em et~al.\/}(2017)Guildenbecher, Gao, Chen \&
  Sojka]{Guildenbecher_2017a}
{\sc \au{Guildenbecher, D.~R.}, \au{Gao, J.}, \au{Chen, J.} \& \au{Sojka,
  P.~E.}} \yr{2017}  \at{Characterization of drop aerodynamic fragmentation in
  the bag and sheet-thinning regimes by crossed-beam, two-view, digital in-line
  holography}.  \jt{Int.~J.~Multiphase Flow}  \bvol{94},  \pg{107--122}.

\bibitem[Guildenbecher {\em et~al.\/}(2009)Guildenbecher, L\'opez-Rivera \&
  Sojka]{Guildenbecher_2009a}
{\sc \au{Guildenbecher, D.~R.}, \au{L\'opez-Rivera, C.} \& \au{Sojka, P.~E.}}
  \yr{2009}  \at{Secondary atomization}.  \jt{Exp.~Fluids}  \bvol{46},
  \pg{371}.

\bibitem[Hadj-Achour {\em et~al.\/}(2021)Hadj-Achour, Rimbert, Gradeck \&
  Meignen]{Hadj-Achour_2021z}
{\sc \au{Hadj-Achour, M.}, \au{Rimbert, N.}, \au{Gradeck, M.} \& \au{Meignen,
  R.}} \yr{2021}  \at{Fragmentation of a liquid metal droplet falling in a
  water pool}.  \jt{Phys.~Fluids}  \bvol{33},  \pg{103315}.

\bibitem[Han \& Tryggvason(1999)]{Han_1999a}
{\sc \au{Han, J.} \& \au{Tryggvason, G.}} \yr{1999}  \at{Secondary breakup of
  axisymmetric liquid drops. i. acceleration by a constant body force}.
  \jt{Phys.~Fluids}  \bvol{11},  \pg{3650--3667}.

\bibitem[Han \& Tryggvason(2001)]{Han_2001a}
{\sc \au{Han, J.} \& \au{Tryggvason, G.}} \yr{2001}  \at{Secondary breakup of
  axisymmetric liquid drops. ii. impulsive acceleration}.  \jt{Phys.~Fluids}
  \bvol{13},  \pg{1554--1565}.

\bibitem[Harper {\em et~al.\/}(1972)Harper, Grube \& Chang]{Harper_1972a}
{\sc \au{Harper, E.~Y.}, \au{Grube, G.~W.} \& \au{Chang, I.-D.}} \yr{1972}
  \at{On the breakup of accelerating liquid drops}.  \jt{J.~Fluid Mech.}
  \bvol{52},  \pg{565--591}.

\bibitem[Hinze(1955)]{Hinze_1955a}
{\sc \au{Hinze, J.~O.}} \yr{1955}  \at{Fundamentals of the hydrodynamic
  mechanism of splitting in dispersion processes}.  \jt{AIChE J.}  \bvol{1},
  \pg{289--295}.

\bibitem[van Hooft {\em et~al.\/}(2018)van Hooft, Popinet, van Heerwaarden,
  van~der Linden, de~Roode \& van~de Wiel]{Hooft_2018a}
{\sc \au{van Hooft, J.~A.}, \au{Popinet, S.}, \au{van Heerwaarden, C.~C.},
  \au{van~der Linden, S. J.~A.}, \au{de~Roode, S.~R.} \& \au{van~de Wiel, B.
  J.~H.}} \yr{2018}  \at{Towards adaptive grids for atmospheric boundary-layer
  simulations}.  \jt{Bound.-Layer Meteor.}  \bvol{167},  \pg{421--443}.

\bibitem[Hsiang \& Faeth(1992)]{Hsiang_1992a}
{\sc \au{Hsiang, L.-P.} \& \au{Faeth, G.~M.}} \yr{1992}  \at{Near-limit drop
  deformation and secondary breakup}.  \jt{Int.~J.~Multiphase Flow}  \bvol{18},
   \pg{635--652}.

\bibitem[Hsiang \& Faeth(1995)]{Hsiang_1995a}
{\sc \au{Hsiang, L.-P.} \& \au{Faeth, G.~M.}} \yr{1995}  \at{Drop deformation
  and breakup due to shock wave and steady disturbances}.
  \jt{Int.~J.~Multiphase Flow}  \bvol{21},  \pg{545--560}.

\bibitem[Ida \& Miksis(1996)]{Ida_1996x}
{\sc \au{Ida, M.~P.} \& \au{Miksis, M.~J.}} \yr{1996}  \at{Thin film rupture}.
  \jt{Appl.~Math.~Lett.}  \bvol{9},  \pg{35--40}.

\bibitem[Jackiw \& Ashgriz(2021)]{Jackiw_2021a}
{\sc \au{Jackiw, I.~M.} \& \au{Ashgriz, N.}} \yr{2021}  \at{On aerodynamic
  droplet breakup}.  \jt{J.~Fluid Mech.}  \bvol{913},  \pg{A33}.

\bibitem[Jackiw \& Ashgriz(2022)]{Jackiw_2022a}
{\sc \au{Jackiw, I.~M.} \& \au{Ashgriz, N.}} \yr{2022}  \at{Prediction of the
  droplet size distribution in aerodynamic droplet breakup}.  \jt{J.~Fluid
  Mech.}  \bvol{940},  \pg{A17}.

\bibitem[Jain {\em et~al.\/}(2015)Jain, Prakash, Tomar \&
  Ravikrishna]{Jain_2015a}
{\sc \au{Jain, M.}, \au{Prakash, R.~S.}, \au{Tomar, G.} \& \au{Ravikrishna,
  R.~V.}} \yr{2015}  \at{Secondary breakup of a drop at moderate weber
  numbers}.  \jt{Proc.~R.~Soc.~Lond.~A Mat.}  \bvol{471},  \pg{20140930}.

\bibitem[Jain {\em et~al.\/}(2019)Jain, Tyagi, Prakash, Ravikrishna \&
  Tomar]{Jain_2019a}
{\sc \au{Jain, S.~S.}, \au{Tyagi, N.}, \au{Prakash, R.~S.}, \au{Ravikrishna,
  R.~V.} \& \au{Tomar, G.}} \yr{2019}  \at{Secondary breakup of drops at
  moderate weber numbers: Effect of density ratio and reynolds number}.
  \jt{Int.~J.~Multiphase Flow}  \bvol{117},  \pg{25--41}.

\bibitem[Jalaal \& Mehravaran(2014)]{Jalaal_2014a}
{\sc \au{Jalaal, M.} \& \au{Mehravaran, K.}} \yr{2014}  \at{Transient growth of
  droplet instabilities in a stream}.  \jt{Phys.~Fluids}  \bvol{26},
  \pg{012101}.

\bibitem[Jing \& Xu(2010)]{Jing_2010a}
{\sc \au{Jing, L.} \& \au{Xu, X.}} \yr{2010}  \at{Direct numerical simulation
  of secondary breakup of liquid drops}.  \jt{Chinese J.~Aeronaut.}  \bvol{23},
   \pg{153--161}.

\bibitem[Joseph {\em et~al.\/}(1999)Joseph, Belanger \& Beavers]{Joseph_1999a}
{\sc \au{Joseph, D.~D.}, \au{Belanger, J.} \& \au{Beavers, G.~S.}} \yr{1999}
  \at{Breakup of a liquid drop suddenly exposed to a high-speed airstream}.
  \jt{Int.~J.~Multiphase Flow}  \bvol{25},  \pg{1263--1303}.

\bibitem[Kekesi {\em et~al.\/}(2014)Kekesi, Amberg \& Wittberg]{Kekesi_2014a}
{\sc \au{Kekesi, T.}, \au{Amberg, G.} \& \au{Wittberg, L.~P.}} \yr{2014}
  \at{Drop deformation and breakup}.  \jt{Int.~J.~Multiphase Flow}  \bvol{66},
  \pg{1--10}.

\bibitem[Kulkarni \& Sojka(2014)]{Kulkarni_2014a}
{\sc \au{Kulkarni, V.} \& \au{Sojka, P.~E.}} \yr{2014}  \at{Bag breakup of low
  viscosity drops in the presence of a continuous air jet}.  \jt{Phys.~Fluids}
  \bvol{26},  \pg{072103}.

\bibitem[Lee \& Reitz(2001)]{Lee_2001a}
{\sc \au{Lee, C.~S.} \& \au{Reitz, R.~D.}} \yr{2001}  \at{Effect of liquid
  properties on the breakup mechanism of high-speed liquid drops}.
  \jt{Atomization Spray}  \bvol{11},  \pg{1--19}.

\bibitem[Lhuissier \& Villermaux(2012)]{Lhuissier_2012a}
{\sc \au{Lhuissier, H.} \& \au{Villermaux, E.}} \yr{2012}  \at{Bursting bubble
  aerosols}.  \jt{J.~Fluid Mech.}  \bvol{696},  \pg{5--44}.

\bibitem[Ling {\em et~al.\/}(2017)Ling, Fuster, Zaleski \&
  Tryggvason]{Ling_2017a}
{\sc \au{Ling, Y.}, \au{Fuster, D.}, \au{Zaleski, S.} \& \au{Tryggvason, G.}}
  \yr{2017}  \at{Spray formation in a quasiplanar gas-liquid mixing layer at
  moderate density ratios: A numerical closeup}.  \jt{Phys.~Rev.~Fluids}
  \bvol{2},  \pg{014005}.

\bibitem[Ling {\em et~al.\/}(2011)Ling, Haselbacher \& Balachandar]{Ling_2011a}
{\sc \au{Ling, Y.}, \au{Haselbacher, A.} \& \au{Balachandar, S.}} \yr{2011}
  \at{{Importance of unsteady contributions to force and heating for particles
  in compressible flows. Part 1: Modeling and analysis for shock-particle
  interaction}}.  \jt{Int.~J.~Multiphase Flow}  \bvol{37},  \pg{1026--1044}.

\bibitem[Ling {\em et~al.\/}(2013)Ling, Parmar \& Balachandar]{Ling_2013b}
{\sc \au{Ling, Y.}, \au{Parmar, M.} \& \au{Balachandar, S.}} \yr{2013}  \at{A
  scaling analysis of added-mass and history forces and their coupling in
  dispersed multiphase flows}.  \jt{Int.~J.~Multiphase Flow}  \bvol{57},
  \pg{102--114}.

\bibitem[Liu \& Reitz(1993)]{Liu_1993a}
{\sc \au{Liu, A.~B.} \& \au{Reitz, R.~D.}} \yr{1993}  \at{Mechanisms of
  air-assisted liquid atomization}.  \jt{Atomization Spray}  \bvol{3},
  \pg{55--75}.

\bibitem[Liu \& Reitz(1997)]{Liu_1997a}
{\sc \au{Liu, Z.} \& \au{Reitz, R.~D.}} \yr{1997}  \at{An analysis of the
  distortion and breakup mechanisms of high speed liquid drops}.
  \jt{Int.~J.~Multiphase Flow}  \bvol{23},  \pg{631--650}.

\bibitem[Lu \& Tryggvason(2018)]{Lu_2018a}
{\sc \au{Lu, J.} \& \au{Tryggvason, G.}} \yr{2018}  \at{Direct numerical
  simulations of multifluid flows in a vertical channel undergoing topology
  changes}.  \jt{Phys.~Rev.~Fluids}  \bvol{3}~(8),  \pg{084401}.

\bibitem[Magnaudet \& Mougin(2007)]{Magnaudet_2007d}
{\sc \au{Magnaudet, J.} \& \au{Mougin, G.}} \yr{2007}  \at{Wake instability of
  a fixed spheroidal bubble}.  \jt{J.~Fluid Mech.}  \bvol{572},  \pg{311--337}.

\bibitem[Marcotte \& Zaleski(2019)]{Marcotte_2019a}
{\sc \au{Marcotte, F.} \& \au{Zaleski, S.}} \yr{2019}  \at{Density contrast
  matters for drop fragmentation thresholds at low ohnesorge number}.
  \jt{Phys.~Rev.~Fluids}  \bvol{4},  \pg{103604}.

\bibitem[Maxey \& Riley(1983)]{Maxey_1983a}
{\sc \au{Maxey, M.~R.} \& \au{Riley, J.~J.}} \yr{1983}  \at{Equation of motion
  for a small rigid sphere in a nonuniform flow}.  \jt{Phys.~Fluids}
  \bvol{26},  \pg{883--889}.

\bibitem[Mei {\em et~al.\/}(1991)Mei, Lawrence \& Adrian]{Mei_1991a}
{\sc \au{Mei, R.}, \au{Lawrence, C.~J.} \& \au{Adrian, R.~J.}} \yr{1991}
  \at{Unsteady drag on a sphere at finite {R}eynolds number with small
  fluctuations in the free-stream velocity}.  \jt{J.~Fluid Mech.}  \bvol{233},
  \pg{613--631}.

\bibitem[Meng \& Colonius(2018)]{Meng_2018a}
{\sc \au{Meng, J.~C.} \& \au{Colonius, T.}} \yr{2018}  \at{Numerical simulation
  of the aerobreakup of a water droplet}.  \jt{J.~Fluid Mech.}  \bvol{835},
  \pg{1108--1135}.

\bibitem[Mikaelian(1990)]{Mikaelian_1990a}
{\sc \au{Mikaelian, K.~O.}} \yr{1990}  \at{{Rayleigh-Taylor} and
  {Richtmyer-Meshkov} instabilities in multilayer fluids with surface tension}.
   \jt{Phys.~Rev.~A}  \bvol{42},  \pg{7211}.

\bibitem[Mikaelian(1996)]{Mikaelian_1996i}
{\sc \au{Mikaelian, K.~O.}} \yr{1996}  \at{Rayleigh-taylor instability in
  finite-thickness fluids with viscosity and surface tension}.  \jt{Phys. Rev.
  E}  \bvol{54},  \pg{3676}.

\bibitem[Miksis {\em et~al.\/}(1981)Miksis, Vanden-Broeck \&
  Keller]{Miksis_1981a}
{\sc \au{Miksis, M.}, \au{Vanden-Broeck, J.-M.} \& \au{Keller, J.~B.}}
  \yr{1981}  \at{Axisymmetric bubble or drop in a uniform flow}.  \jt{J.~Fluid
  Mech.}  \bvol{108},  \pg{89--100}.

\bibitem[Mostert {\em et~al.\/}(2022)Mostert, Popinet \& Deike]{Mostert_2022g}
{\sc \au{Mostert, W.}, \au{Popinet, S.} \& \au{Deike, L.}} \yr{2022}
  \at{High-resolution direct simulation of deep water breaking waves:
  transition to turbulence, bubbles and droplets production}.  \jt{J.~Fluid
  Mech.}  \bvol{942}.

\bibitem[Neel {\em et~al.\/}(2020)Neel, Lhuissier \& Villermaux]{Neel_2020e}
{\sc \au{Neel, B.}, \au{Lhuissier, H.} \& \au{Villermaux, E.}} \yr{2020}
  \at{{`Fines'} from the collision of liquid rims}.  \jt{J.~Fluid Mech.}
  \bvol{893},  \pg{A16}.

\bibitem[Opfer {\em et~al.\/}(2014)Opfer, Roisman, Venzmer, Klostermann \&
  Tropea]{Opfer_2014a}
{\sc \au{Opfer, L.}, \au{Roisman, I.~V.}, \au{Venzmer, J.}, \au{Klostermann,
  M.} \& \au{Tropea, C.}} \yr{2014}  \at{Droplet-air collision dynamics:
  Evolution of the film thickness}.  \jt{Phys.~Rev.~E}  \bvol{89},
  \pg{013023}.

\bibitem[{O'Rourke} \& Amsden(1987)]{ORourke_1987a}
{\sc \au{{O'Rourke}, P.~J.} \& \au{Amsden, A.~A.}} \yr{1987}  \bt{The tab
  method for numerical calculation of spray droplet breakup}. {\em Tech.
  Rep.\/} Tech. Rep. LA-UR-2105.  \org{Los Alamos National Laboratory}.

\bibitem[Pai \& Subramaniam(2006)]{Pai_2006a}
{\sc \au{Pai, M.~G.} \& \au{Subramaniam, S.}} \yr{2006}  \at{Modeling
  interphase turbulent kinetic energy transfer in lagrangian-eulerian spray
  computations}.  \jt{Atomization Spray}  \bvol{16},  \pg{807--826}.

\bibitem[Park {\em et~al.\/}(2002)Park, Yoon \& Hwang]{Park_2002a}
{\sc \au{Park, J.-H.}, \au{Yoon, Y.} \& \au{Hwang, S.-S.}} \yr{2002}
  \at{Improved tab model for prediction of spray droplet deformation and
  breakup}.  \jt{Atomization Spray}  \bvol{12},  \pg{387--401}.

\bibitem[Pilch \& Erdman(1987)]{Pilch_1987a}
{\sc \au{Pilch, M.} \& \au{Erdman, C.~A.}} \yr{1987}  \at{Use of breakup time
  data and velocity history data to predict the maximum size of stable
  fragments for acceleration-induced breakup of a liquid drop}.
  \jt{Int.~J.~Multiphase Flow}  \bvol{13},  \pg{741--757}.

\bibitem[Popinet(2009)]{Popinet_2009a}
{\sc \au{Popinet, S.}} \yr{2009}  \at{An accurate adaptive solver for
  surface-tension-driven interfacial flows}.  \jt{J.~Comput.~Phys.}
  \bvol{228}~(16),  \pg{5838--5866}.

\bibitem[Ranger \& Nicholls(1969)]{Ranger_1969a}
{\sc \au{Ranger, A.~A.} \& \au{Nicholls, J.~A.}} \yr{1969}  \at{Aerodynamic
  shattering of liquid drops.}  \jt{{AIAA} J.}  \bvol{7},  \pg{285--290}.

\bibitem[Reyssat {\em et~al.\/}(2007)Reyssat, Chevy, Biance, Petitjean \&
  Quere]{Reyssat_2007a}
{\sc \au{Reyssat, E.}, \au{Chevy, F.}, \au{Biance, A.-L.}, \au{Petitjean, L.}
  \& \au{Quere, D.}} \yr{2007}  \at{Shape and instability of free-falling
  liquid globules}.  \jt{Europhys.~Lett.}  \bvol{80},  \pg{34005}.

\bibitem[Rimbert {\em et~al.\/}(2020)Rimbert, Escobar, Meignen, Hadj-Achour \&
  Gradeck]{Rimbert_2020o}
{\sc \au{Rimbert, N.}, \au{Escobar, S.~C.}, \au{Meignen, R.}, \au{Hadj-Achour,
  M.} \& \au{Gradeck, M.}} \yr{2020}  \at{Spheroidal droplet deformation,
  oscillation and breakup in uniform outer flow}.  \jt{J.~Fluid Mech.}
  \bvol{904},  \pg{A15}.

\bibitem[Sakakeeny {\em et~al.\/}(2021)Sakakeeny, Deshpande, Deb, Alvarado \&
  Ling]{Sakakeeny_2021b}
{\sc \au{Sakakeeny, J.}, \au{Deshpande, C.}, \au{Deb, S.}, \au{Alvarado, J.~L.}
  \& \au{Ling, Y.}} \yr{2021}  \at{A model to predict the oscillation frequency
  for drops pinned on a vertical planar surface}.  \jt{J.~Fluid Mech.}
  \bvol{928},  \pg{A28}.

\bibitem[Sakakeeny \& Ling(2020)]{Sakakeeny_2020a}
{\sc \au{Sakakeeny, J.} \& \au{Ling, Y.}} \yr{2020}  \at{Natural oscillations
  of a sessile drop on flat surfaces with mobile contact lines}.
  \jt{Phys.~Rev.~Fluids}  \bvol{5},  \pg{123604}.

\bibitem[Sakakeeny \& Ling(2021)]{Sakakeeny_2021a}
{\sc \au{Sakakeeny, J.} \& \au{Ling, Y.}} \yr{2021}  \at{Numerical study of
  natural oscillations of supported drops with free and pinned contact lines}.
  \jt{Phys.~Fluids}  \bvol{33},  \pg{062109}.

\bibitem[Scardovelli \& Zaleski(1999)]{Scardovelli_1999a}
{\sc \au{Scardovelli, R.} \& \au{Zaleski, S.}} \yr{1999}  \at{Direct numerical
  simulation of free-surface and interfacial flow}.  \jt{Annu.~Rev.~Fluid
  Mech.}  \bvol{31},  \pg{567--603}.

\bibitem[Sharma {\em et~al.\/}(2021)Sharma, Singh, Rao, Kumar \&
  Basu]{Sharma_2021c}
{\sc \au{Sharma, Shubham}, \au{Singh, Awanish~Pratap}, \au{Rao, S.~Srinivas},
  \au{Kumar, Aloke} \& \au{Basu, Saptarshi}} \yr{2021}  \at{Shock induced
  aerobreakup of a droplet}.  \jt{J.~Fluid Mech.}  \bvol{929}.

\bibitem[Stefanitsis {\em et~al.\/}(2017)Stefanitsis, Malgarinos, Strotos,
  Nikolopoulos, Kakaras \& Gavaises]{Stefanitsis_2017a}
{\sc \au{Stefanitsis, D.}, \au{Malgarinos, I.}, \au{Strotos, G.},
  \au{Nikolopoulos, N.}, \au{Kakaras, E.} \& \au{Gavaises, M.}} \yr{2017}
  \at{Numerical investigation of the aerodynamic breakup of diesel and heavy
  fuel oil droplets}.  \jt{Int.~J.~Heat Fluid Fl.}  \bvol{68},  \pg{203--215}.

\bibitem[Stefanitsis {\em et~al.\/}(2019{\natexlab{{\em a\/}}})Stefanitsis,
  Malgarinos, Strotos, Nikolopoulos, Kakaras \& Gavaises]{Stefanitsis_2019b}
{\sc \au{Stefanitsis, D.}, \au{Malgarinos, I.}, \au{Strotos, G.},
  \au{Nikolopoulos, N.}, \au{Kakaras, E.} \& \au{Gavaises, M.}}
  \yr{2019{\natexlab{{\em a\/}}}}  \at{Numerical investigation of the
  aerodynamic breakup of droplets in tandem}.  \jt{Int.~J.~Multiphase Flow}
  \bvol{113},  \pg{289--303}.

\bibitem[Stefanitsis {\em et~al.\/}(2019{\natexlab{{\em b\/}}})Stefanitsis,
  Strotos, Nikolopoulos, Kakaras \& Gavaises]{Stefanitsis_2019a}
{\sc \au{Stefanitsis, D.}, \au{Strotos, G.}, \au{Nikolopoulos, N.},
  \au{Kakaras, E.} \& \au{Gavaises, M.}} \yr{2019{\natexlab{{\em b\/}}}}
  \at{Improved droplet breakup models for spray applications}.
  \jt{Int.~J.~Heat Fluid Fl.}  \bvol{76},  \pg{274--286}.

\bibitem[Strotos {\em et~al.\/}(2016)Strotos, Malgarinos, Nikolopoulos \&
  Gavaises]{Strotos_2016a}
{\sc \au{Strotos, G.}, \au{Malgarinos, I.}, \au{Nikolopoulos, N.} \&
  \au{Gavaises, M.}} \yr{2016}  \at{Predicting droplet deformation and breakup
  for moderate weber numbers}.  \jt{Int.~J.~Multiphase Flow}  \bvol{85},
  \pg{96--109}.

\bibitem[Tang {\em et~al.\/}(2022)Tang, Adcock \& Mostert]{Tang_2022h}
{\sc \au{Tang, K.}, \au{Adcock, T.} \& \au{Mostert, W.}} \yr{2022} Bag film
  breakup of droplets in uniform airflows. ArXiv:2211.14518.

\bibitem[Tanner(1997)]{Tanner_1997a}
{\sc \au{Tanner, F.~X.}} \yr{1997}  \at{Liquid jet atomization and droplet
  breakup modeling of non-evaporating diesel fuel sprays}.  \jt{SAE
  Trans.~J.~Eng.}  \bvol{106},  \pg{127--140}.

\bibitem[Taylor(1949)]{Taylor_1949a}
{\sc \au{Taylor, G.~I.}} \yr{1949}  \at{The shape and acceleration of a drop in
  a high-speed air stream}.  \bt{In {\em The Scientific Papers of G. I. Taylor.
  Vol. III. Aerodynamics and the Mechanics of Projectiles and Explosions.\/}
  (ed. \ed{G.~K. Batchelor})}.  \publ{Cambridge Univ. Press.}

\bibitem[Theofanous(2011)]{Theofanous_2011a}
{\sc \au{Theofanous, T.~G.}} \yr{2011}  \at{Aerobreakup of newtonian and
  viscoelastic liquids}.  \jt{Annu. Rev. Fluid Mech.}  \bvol{43},
  \pg{661--690}.

\bibitem[Theofanous \& Li(2008)]{Theofanous_2008a}
{\sc \au{Theofanous, T.~G.} \& \au{Li, G.~J.}} \yr{2008}  \at{On the physics of
  aerobreakup}.  \jt{Phys.~Fluids}  \bvol{20},  \pg{052103}.

\bibitem[Theofanous {\em et~al.\/}(2004)Theofanous, Li \&
  Dinh]{Theofanous_2004a}
{\sc \au{Theofanous, T.~G.}, \au{Li, G.~J.} \& \au{Dinh, T.-N.}} \yr{2004}
  \at{Aerobreakup in rarefied supersonic gas flows}.  \jt{J.~Fluid
  Eng.-T.~ASME}  \bvol{126},  \pg{516--527}.

\bibitem[Theofanous {\em et~al.\/}(2007)Theofanous, Li, Dinh \&
  Chang]{Theofanous_2007a}
{\sc \au{Theofanous, T.~G.}, \au{Li, G.~J.}, \au{Dinh, T.-N.} \& \au{Chang,
  C.-H.}} \yr{2007}  \at{Aerobreakup in disturbed subsonic and supersonic flow
  fields}.  \jt{J.~Fluid Mech.}  \bvol{593},  \pg{131--170}.

\bibitem[Theofanous {\em et~al.\/}(2013)Theofanous, Mitkin \&
  Ng]{Theofanous_2013a}
{\sc \au{Theofanous, T.~G.}, \au{Mitkin, V.~V.} \& \au{Ng, C.~L.}} \yr{2013}
  \at{The physics of aerobreakup. iii. viscoelastic liquids}.
  \jt{Phys.~Fluids}  \bvol{25},  \pg{032101}.

\bibitem[Theofanous {\em et~al.\/}(2012)Theofanous, Mitkin, Ng, Chang, Deng \&
  Sushchikh]{Theofanous_2012a}
{\sc \au{Theofanous, T.~G.}, \au{Mitkin, V.~V.}, \au{Ng, C.~L.}, \au{Chang,
  C.~H.}, \au{Deng, X.} \& \au{Sushchikh, S.}} \yr{2012}  \at{The physics of
  aerobreakup. ii. viscous liquids}.  \jt{Phys.~Fluids}  \bvol{24},
  \pg{022104}.

\bibitem[Tiwari {\em et~al.\/}(2020{\natexlab{{\em a\/}}})Tiwari, Pal, Bale,
  Minocha, Patwardhan, Nandakumar \& Joshi]{Tiwari_2020a}
{\sc \au{Tiwari, S.~S.}, \au{Pal, E.}, \au{Bale, S.}, \au{Minocha, N.},
  \au{Patwardhan, A.~W.}, \au{Nandakumar, K.} \& \au{Joshi, J.~B.}}
  \yr{2020{\natexlab{{\em a\/}}}}  \at{Flow past a single stationary sphere, 1.
  experimental and numerical techniques}.  \jt{Powder Technol.}  \bvol{365},
  \pg{115--148}.

\bibitem[Tiwari {\em et~al.\/}(2020{\natexlab{{\em b\/}}})Tiwari, Pal, Bale,
  Minocha, Patwardhan, Nandakumar \& Joshi]{Tiwari_2020b}
{\sc \au{Tiwari, S.~S.}, \au{Pal, E.}, \au{Bale, S.}, \au{Minocha, N.},
  \au{Patwardhan, A.~W.}, \au{Nandakumar, K.} \& \au{Joshi, J.~B.}}
  \yr{2020{\natexlab{{\em b\/}}}}  \at{Flow past a single stationary sphere, 2.
  regime mapping and effect of external disturbances}.  \jt{Powder Technol.}
  \bvol{365},  \pg{215--243}.

\bibitem[Vanden-Broeck \& Keller(1980)]{Vanden_1980a}
{\sc \au{Vanden-Broeck, J.-M.} \& \au{Keller, J.~B.}} \yr{1980}
  \at{Deformation of a bubble or drop in a uniform flow}.  \jt{J.~Fluid Mech.}
  \bvol{101},  \pg{673--686}.

\bibitem[Villermaux \& Bossa(2009)]{Villermaux_2009a}
{\sc \au{Villermaux, E.} \& \au{Bossa, B.}} \yr{2009}  \at{Single-drop
  fragmentation determines size distribution of raindrops}.  \jt{Nature Phys.}
  \bvol{5},  \pg{697--702}.

\bibitem[Villermaux \& Bossa(2011)]{Villermaux_2011a}
{\sc \au{Villermaux, E.} \& \au{Bossa, B.}} \yr{2011}  \at{Drop fragmentation
  on impact}.  \jt{J.~Fluid Mech.}  \bvol{668},  \pg{412}.

\bibitem[Wert(1995)]{Wert_1995a}
{\sc \au{Wert, K.~L.}} \yr{1995}  \at{A rationally-based correlation of mean
  fragment size for drop secondary breakup}.  \jt{Int.~J.~Multiphase Flow}
  \bvol{21},  \pg{1063--1071}.

\bibitem[Williams \& Davis(1982)]{Williams_1982b}
{\sc \au{Williams, M.~B.} \& \au{Davis, S.~H.}} \yr{1982}  \at{Nonlinear theory
  of film rupture}.  \jt{J.~Colloid Interface Sci.}  \bvol{90}~(1),
  \pg{220--228}.

\bibitem[Yang {\em et~al.\/}(2016)Yang, Jia, Sun \& Wang]{Yang_2016a}
{\sc \au{Yang, W.}, \au{Jia, M.}, \au{Sun, K.} \& \au{Wang, T.}} \yr{2016}
  \at{Influence of density ratio on the secondary atomization of liquid
  droplets under highly unstable conditions}.  \jt{Fuel}  \bvol{174},
  \pg{25--35}.

\bibitem[Zhang {\em et~al.\/}(2020)Zhang, Popinet \& Ling]{Zhang_2020a}
{\sc \au{Zhang, B.}, \au{Popinet, S.} \& \au{Ling, Y.}} \yr{2020}  \at{Modeling
  and detailed numerical simulation of the primary breakup of a gasoline
  surrogate jet under non-evaporative operating conditions}.
  \jt{Int.~J.~Multiphase Flow}  \bvol{130},  \pg{103362}.

\bibitem[Zhao {\em et~al.\/}(2013)Zhao, Liu, Xu, Li \& Lin]{Zhao_2013a}
{\sc \au{Zhao, H.}, \au{Liu, H.-F.}, \au{Xu, J.-L.}, \au{Li, W.-F.} \& \au{Lin,
  K.-F.}} \yr{2013}  \at{Temporal properties of secondary drop breakup in the
  bag-stamen breakup regime}.  \jt{Phys.~Fluids}  \bvol{25},  \pg{054102}.

\end{thebibliography}

\end{document}